\documentclass[preprint]{ptephy_v1}
\preprintnumber{2303.02633}

\usepackage{graphicx,hyperref}

\usepackage[caption=false, justification=centerlast]{subfig}

\usepackage{ulem}

\usepackage{enumitem}
\usepackage{physics,bm}
\usepackage{amsmath,amsfonts,amssymb}
\usepackage{comment,here,color}
\usepackage{natbib}

\newcommand{\x}{\tilde{u}_{*}}
\newcommand{\y}{\tilde{h}'(u_{*})}
\newcommand{\ya}{\tilde{a}'(u_{*})}

\begin{document}

\title{Patchwork Conditions for Holographic Nonlinear Responses: A Computational Method for Electric Conductivity and Friction Coefficient}

\author{Shuta~Ishigaki}
\affil{Department of Physics, Shanghai University, 99 Shangda road, Shanghai 200444, China\email{shutaishigaki@shu.edu.cn}}

\author{Shin~Nakamura}
\affil{Department of Physics, Chuo University, Kasuga, Bunkyo-ku, Tokyo 112-8551, Japan\email{nakamura@phys.chuo-u.ac.jp}}

\author{Kazuaki~Takasan}
\affil{Department of Physics, University of Tokyo, 7-3-1 Hongo, Tokyo 113-0033, Japan\email{kazuaki.takasan@phys.s.u-tokyo.ac.jp}}

\begin{abstract}
	We propose a new method to compute nonlinear transport coefficients in holography, such as nonlinear DC conductivity and nonlinear friction coefficient. The conventional method can be applied only to the models whose action in the gravity dual has the ``square-root structure,'' i.e., the Dirac-Born-Infeld action of the probe D-branes or the Nambu-Goto action of the probe strings. Our method is applicable to a wider range of holographic models whose action does not have such a square-root structure. 
    We propose a condition to obtain regular physical configurations in the gravity dual in the form of two simultaneous equations, which we call the patchwork condition.
    Our method also enables us to estimate the effective temperature of the nonequilibrium steady states in a wider range of holographic models. We show that a general model exhibits different effective temperatures for different fluctuation modes.
\end{abstract}

\maketitle
\newpage

\section{Introduction}
\label{sec:introduction}

The AdS/CFT correspondence (the gauge/gravity duality) \cite{Maldacena:1998,Gubser:1998,Witten:1998}, that is also called the holographic approach, is a promising framework for computing physical quantities in strongly coupled field theories.
In this approach, the expectation values of the operators of the quantum field theory are computed by a classical theory of fields on a higher-dimensional curved spacetime, which is often called ``bulk.''

Usually, the GKP-Witten prescription\cite{Gubser:1998,Witten:1998} is employed to calculate the expectation value of operators in the AdS/CFT correspondence.
In this prescription, the bulk field is divided into normalizable mode and non-normalizable mode, with the non-normalizable mode corresponding to the external source and the normalizable mode corresponding to the expectation value of the operator conjugate to the source.
On the other hand, in terms of the second-order differential equation that the bulk field follows, the normalizable mode and the non-normalizable mode are independent solutions. Therefore, the magnitude of the normalizable mode can be set freely. This means that the expectation value is not determined and there is no predictive power. 
This problem is usually resolved by imposing regularity on the bulk field. If the magnitude of the normalizable mode, i.e., the expectation value of the operator, is arbitrarily assigned for a given value of the source, the bulk field will usually be singular.
However, if the magnitude of the normalizable mode is set to a ``right'' value, the bulk field will be regular. Thus, when regularity is achieved for the bulk field, the magnitude of the normalizable mode, i.e., the expectation value of the operator, is determined.

The above is a general guiding principle for calculating the expectation value of an operator in the AdS/CFT correspondence. However, for practical use, it is important to find an efficient way to determine the value of the normalizable mode that makes the bulk field regular.
In general terms, it is necessary to request that the bulk field be regular at all points in the bulk spacetime, and to achieve this, the regularity of the bulk field must be examined at all points in the bulk. However, this is quite a tedious and inefficient task.

So far, several efficient ways of finding regular solutions have been developed. 
In all of the formulations proposed so far, the regularity of the bulk field in the entire region of the bulk spacetime is achieved just by realizing the regularity at a specific point in the bulk spacetime. The location of this point in the bulk where regularity should be imposed is known in advance in the models studied so far.
For example, in the computation of the retarded Green's function, the normalizable mode is determined by imposing an ingoing-wave boundary condition on the bulk field at the event horizon of the bulk black hole \cite{Son:2002sd}. This is equivalent to imposing regularity on the bulk field on the event horizon in the ingoing Eddington-Finkelstein coordinates \cite{Yee:2009vw}.
Also, for example, in the calculation of nonlinear responses, such as the calculation of nonlinear frictional forces~\cite{Herzog:2006gh,Gubser:2006bz} or nonlinear electric conductivity~\cite{Karch:2007pd}, the magnitude of the normalizable modes has been determined by imposing a reality condition on the bulk field (or on the on-shell action). The bulk theory for these cases is given by the Nambu-Goto (NG) action or the Dirac-Born-Infeld (DBI) action.
In these models, the location of the special point is first determined,
and the expectation values of the transport coefficients are given by imposing a condition that ensures reality of the bulk filed, at this point.
This special point is known as an effective horizon for the worldvolume theory.
In terms of regularity, the correct expectation value is obtained by imposing regularity of the bulk field at the predetermined effective horizon~\cite{Kim:2011zd}.

On the other hand, there are cases where no formulation has been established for efficiently finding the regular solutions. For example, as described in this paper, such a case is when the action of the bulk field is non-NG or non-DBI type, i.e., without the square-root structure, and contains a nonlinear term. In this type of model, as we shall see later, the location where regularity should be imposed on the bulk is not predetermined. Therefore, we have to examine the regularity of the bulk field at all positions in the bulk.
In this paper, we propose a new prescription that efficiently imposes regularity on bulk fields even in such cases.

Such an efficient method of imposing regularity is not only beneficial in practical computations but also brings new physical insights.
In general, the calculation of transport coefficients
in nonlinear regions requires calculations on a non-equilibrium steady-state background where dissipation exists. This is a great challenge from the viewpoint of non-equilibrium statistical physics.
The effective horizon
is the horizon of the analog black hole on which the fluctuations of the bulk field reside.
Using this effective horizon, the effective temperature characterizing fluctuations in the non-equilibrium steady-state is naturally defined within the framework of the bulk theory \cite{Kim:2011qh,Sonner:2012if}.\footnote{The early studies on the effective temperature for the systems of the fundamental string are found in \cite{Gubser:2006nz, Casalderrey-Solana:2007ahi}. See also the related works for \cite{Gursoy:2010aa,Nakamura:2013yqa}. } 

However, the definition of the concept of effective temperature in a non-equilibrium steady state is itself a non-trivial problem in general. Thus, in more general models and settings, there may be instances where the effective horizon itself does not exist or cannot be defined. Therefore, it is also really a non-trivial question whether the prescription of assuming the existence of an effective horizon in advance and imposing regularity there is always valid.

We provide an answer to this question for a range of models in this paper. We find that, as far as the models we deal in this paper, the regular solution always has an effective horizon and has a notion of effective temperature. However, unlike the models with the NG-type or the DBI-type action, we also find that some model can have several different effective temperatures depending on the mode we consider.

In addition to the motivation from the non-equilibrium statistical physics perspective described above, the motivations for conducting this study are as follows.
\begin{enumerate}[label=\roman*)]
\item 
Even in the analysis of the models with the DBI action or the NG action, it is instructive to decompose the action into a sum of basic terms and to examine the contribution of each term. Understanding the role of each term tells us the information on the physical mechanism behind the phenomenon we study. Since the square-root structure of the action is broken after the decomposition, we need a method applicable to the models without the square-root structure. 
\item 
Many phenomenological models of holography, such as the holographic superconductors proposed in \cite{Hartnoll:2008kx,Hartnoll:2008vx,Gubser:2008zu,Gubser:2008wv,Roberts:2008ns}, do not have action with the square-root structure. Even though some phenomenological models exhibit only linear responses, it is worthwhile developing a computational method of nonlinear responses for their possible nonlinear extensions.
\item 
We also have a technical reason. For example, let us consider the models with non-Abelian global symmetries. We may construct a model with a non-Abelian global symmetry by considering multiple probe D-branes overlapped on top of each other in the dual geometry (see, for example, Refs. \cite{Ammon:2008fc,Ammon:2009fe}). In this case, we need to employ the non-Abelian DBI action \cite{Tseytlin:1997csa}.
However, in order to apply the symmetrized trace in the non-Abelian DBI action, we need to expand the action in the power of the worldvolume fields. The symmetrized trace must be treated for each term of the expanded action.
Furthermore, the validity of the non-Abelian DBI action has been confirmed to finite orders of the derivative expansion
\cite{Tseytlin:1997csa, Hashimoto:1997gm}. In order to study the models with non-Abelian DBI action, we need a method that is applicable to the models without the square-root structure.
\end{enumerate}

With the above motivations in mind, we present a new method to compute the nonlinear responses for the models without the square-root structure in the action.
In this paper, we focus on the models where the action contains only the derivative of the fields, such as the field-strength of the bulk gauge field, but not the field without the derivative.

We find that a characteristic behavior of the solutions to the equations of motion is the signal for physical regular solutions. We call this behavior a ``patchwork phenomenon.''
We propose two simultaneous equations associated with the patchwork phenomenon as a condition for the physical regular solution from which we obtain the nonlinear responses.
We believe that our approach provides a general treatment applicable to a broader range of systems.


Before closing this section, we would like to emphasize the importance of studying nonlinear responses from the perspective of applications of holography not only to statistical physics but also to condensed matter physics. Nonlinear responses have been widely used as essential tools to obtain information about electrons and ions in solids. For example, the second harmonic generation (SHG), which is the emission of $2\omega$-frequency light under the injection of $\omega$-frequency light, is a commonly used tool to probe the breaking of the inversion symmetry of electronic states or lattice structures~\cite{Boyd_book}. Nonlinear responses are also important for technological applications: photovoltaic devices, telecommunications, laser technologies, quantum computers, and so on. However, there still remain theoretical problems to be resolved in the theory of nonlinear responses. (i) One is the systematic treatment of non-perturbative effects. The standard knowledge of nonlinear optical responses has been established on the basis of perturbation theory with respect to the external field strength~\cite{Boyd_book}. Thus, the understanding of the non-perturbative effects is relatively poor and still under intensive investigation~\cite{Huttner2017}. (ii) Another problem is the many-body effects. Previous theories are mainly based on free-fermion models, which describe the electron dynamics in many solid-state materials well. Because of this, the quantum many-body theory of nonlinear responses is less established even in the perturbative regime. Indeed, recent studies have started to examine general properties of nonlinear responses in quantum many-body systems~\cite{Watanabe2020,Takasan2023,Michishita2021}. 
Thanks to the power of holographic techniques, it is possible that both problems can be resolved by the holographic approach. 
For example, 
the negative differential resistance (NDR)\footnote{This may also be called negative differential conductivity (NDC).} has been obtained theoretically~\cite{Nakamura:2010zd},
and a possible origin of the NDR was also revealed through this approach~\cite{Ishigaki:2020coe}. 
This approach has also been utilized in the study of nonequilibrium phase transitions~\cite{Nakamura:2012ae,Ali-Akbari:2013hba,Zeng:2018ero,Vahedi:2018gvn,Imaizumi:2019byu,Endo:2023vov} associated with nonperturbative responses to external fields.
In recent condensed matter experiments, the nonperturbative effects in high harmonic generation\footnote{High harmonic generation (HHG), which is the $n\omega$-frequency light emission under the injection of light with frequency $\omega$, is a nonlinear process in the interaction between electrons and the external electromagnetic field. 
The $n=2$ case of HHG is nothing but the SHG. In recent experiments, the large-$n$ (e.g., $n \approx 10$ or larger) regime has been extensively studied and the observation of non-perturbative effects (plateau and cut-off behavior) has been reported~\cite{Ghimire2019}.} and nonlinear responses in strongly correlated materials~\cite{Dzsaber2021} are attracting great attention and the importance of the above problems is expected to increase.
We believe that the development of flexible and useful holographic techniques for nonlinear responses will help to solve these problems and this study can be considered as one of the significant steps in this direction.

The organization of this paper is as follows.
The overview of our proposal is described in Sec.~\ref{sec:outline}. 
We revisit the conventional computational method in Sec.~\ref{sec:conventional-methods}. We show how the conventional method breaks down when the action in the gravity dual is neither the DBI type nor the NG type.
In Sec.~\ref{sec:proposal}, we propose a new method to compute the nonlinear transport coefficients efficiently. Section \ref{sec:higher-order} is devoted to demonstrating how our proposal works for the calculations of nonlinear conductivity in several examples. We show our method also works for the models with the DBI action, in Sec.~\ref{sec:DBI}. We describe how our method is applicable in the computation of nonlinear friction coefficients in Sec.~\ref{sec:NG}.
Computations of the effective temperatures based on our method are given in Sec.~\ref{sec:effective-temp}. 
We find models in which the effective temperature differs from one mode to another.
In Appendix \ref{sec:Kim-Pang}, we overview the method proposed in \cite{Kim:2011zd} with comments.
We discuss the result at $T=0$ in our model in Appendix \ref{sec:T=0}. We employ the ingoing Eddington-Finkelstein coordinates for the bulk spacetime in most parts of the main text. We exhibit the computations in the Schwarzschild coordinates in Appendix \ref{sec:Schwarzschild} for the convenience of the readers.

\section{Overview of our proposal}
\label{sec:outline}
Before moving into detailed discussions, let us provide an overview of our proposal in this section to clarify our basic idea.
We consider the nonlinear DC electric conductivity as a prototypical example of nonlinear responses. 

A computational method of nonlinear DC electric conductivity in holography was first proposed in \cite{Karch:2007pd}. The basic idea presented in \cite{Karch:2007pd} is that the conductivity is determined by requesting the reality condition: the gauge field, hence the on-shell action, has to be real in the entire region of the dual bulk spacetime. The same idea was employed in the holographic computation of nonlinear frictional force acting on a particle in a medium \cite{Herzog:2006gh,Gubser:2006bz}.
We call these methods the ``real-action method'' in this paper.

The real-action method is strongly based on the special circumstances that the action in the gravity dual is the DBI type or the NG type. 
The DBI action of a D$p$-brane has the following form:
\begin{equation}
	S = - \tau_{p} \int \dd[p+1]{\xi} e^{-\Phi}\sqrt{- \det(h_{ab} + F_{ab})},
	\label{Dp-action}
\end{equation}
where $\tau_{p}$ is the tension of the D$p$-brane, $\xi^a$ are the worldvolume coordinates, $\Phi$ is the dilaton field, $F_{ab} = \partial_a A_b - \partial_b A_a$ is the field-strength of the worldvolume $\mathrm{U}(1)$ gauge field $A_a$.\footnote{$2\pi\alpha^{\prime}=1$ in our convention.} $h_{ab}$ is the induced metric
\begin{equation}
	h_{ab} = g_{MN} \partial_a X^{M} \partial_b X^{N},
\end{equation}
where $g_{MN}$ is the metric of the target spacetime, and $X^{M}(\xi^{a})$ represents the configuration of the D$p$-brane.  
We employ the probe approximation.
The real-action method works by virtue of the ``square-root structure'' of the integrand of (\ref{Dp-action}). 
The Nambu-Goto action for fundamental strings, which we employ in the computation of frictional force, has the same structure.
We shall review how to compute the nonlinear conductivity in the conventional methods in Sec.~\ref{sec:conventional-methods}.

In the absence of the square-root structure, the real-action method does not work, as we shall show in detail in Sec.~\ref{sec:failure}. This is explicitly shown in the following example.\footnote{See also Sec.~\ref{sec:conventional-methods} for the details.}
Let us consider an action, in the gravity dual, which is obtained by truncation of the derivative expansion of the DBI action:
\begin{equation}
	S = - \tau_{p} \int \dd[p+1]{\xi} e^{-\Phi}\sqrt{- \det h_{ab}}
	\left(1-\frac{1}{4}F_{ab}F^{ab} + \cdots
	\right)
	\label{Dp-truncated-action},
\end{equation}
where $\cdots$ denotes higher-order contributions of $F_{ab}$ which we ignore.
The field-strength appears in the form of power series, and the ``square-root structure'' of the original DBI action has disappeared in this action. 
Since we keep the term of the second order of $F_{ab}$,
we expect that the expectation value of the current agrees between the models (\ref{Dp-truncated-action}) and (\ref{Dp-action}) to the linear order of $E$,
where $E$ is the electric field acting on the system.
However, one finds that the on-shell action of the model (\ref{Dp-truncated-action}) has no chance to be complex, as we shall show in Sec.~\ref{sec:failure}, even if we employ an unphysical solution for $A_{a}$.
Thus the reality condition for the model of (\ref{Dp-truncated-action}) does not fix the conductivity at all.
When we have higher-order terms of $F_{ab}$ in the action, we shall see in Sec.~\ref{sec:higher-order} that the on-shell action can be complex. However, the reality condition for the action gives only a constraint for the range of possible values of the conductivity, and it does not fix the conductivity uniquely.

In the present paper, we propose an alternative method to compute the nonlinear DC conductivity and nonlinear frictional constant that is applicable to the models without the square-root structure.
Our argument consists of two parts: the basic idea of requiring regularity of the solution, and the proposal of a simplified method to achieve the regularity.

Our basic idea is as follows. We require that the gauge-invariant scalar quantity be regular over the bulk spacetime to obtain the correct electric conductivity (or the correct frictional constant). For example, we impose regularity of $F_{ab}F^{ab}$ in the computation of conductivity: $F_{ab}F^{ab}$ has to be regular in the entire region of the bulk spacetime.\footnote{In general, we need to impose regularity of other gauge-invariant scalar quantities such as $\nabla_{c}(F_{ab}F^{ab})\nabla^{c}(F_{ab}F^{ab})$ as well. The precise meaning of the ``entire region of the bulk spacetime'' should also be clarified. The details will be given in Sec.~\ref{sec:higher-order}.} We call this condition the regularity condition. 
The details will be discussed in Sec. \ref{sec:regularity}.

In order to check the regularity of the given solution, we need to survey the entire region of the bulk spacetime. However, surveying the entire region of the bulk spacetime for all possible solutions is awkward.
We propose a short-cut method to realize the regularity condition.
Our proposal is that the correct conductivity is obtained by just solving the following simultaneous equations:\footnote{The corresponding equations for the computation of nonlinear friction coefficient are $\left.\frac{\partial^{2} {\cal L}(\xi^{\prime},u)}{\partial {\xi^{\prime}}^2}\right|_{u=u_{*}}=0$ and $\left.\frac{\partial^{2} {\cal L}(\xi^{\prime},u)}{\partial u \partial \xi^{\prime}}\right|_{u=u_{*}}=0$, where $\xi^{\prime}=\partial_{u}\xi$ and $\xi(u)$ represents the configuration of the string. See for the details, Sec.~\ref{sec:NG}. } 
\begin{eqnarray}
\left.\frac{\partial^{2} {\cal L}(F_{ui},u)}{\partial F_{ui}^2}\right|_{u=u_{*}}&=&0,
\label{saddle1}
\\
\left.\frac{\partial^{2} {\cal L}(F_{ui},u)}{\partial u \partial F_{ui}}\right|_{u=u_{*}}&=&0,
\label{saddle2}
\end{eqnarray}
where ${\cal L}$ is the Lagrangian density of the bulk theory, $u$ is the radial coordinate in the bulk spacetime, and $i$ is a spatial direction in which we have the current.
We call (\ref{saddle1}) and (\ref{saddle2}) the ``patchwork condition.''
The patchwork condition gives us possible combinations of $(u_{*}, F_{ui}(u_{*}))$ that satisfy (\ref{saddle1}) and (\ref{saddle2}). 
The expectation value of the current density $J(u, F_{ui})$ is given in terms of $F_{ui}(u)$ and $u$, but $J(u, F_{ui})$ is independent of $u$ by virtue of the equation of motion $\dv{u} J(u, F_{ui})=0$.
Therefore, we obtain the expectation value of the current density by substituting $u=u_{*}$ and $F_{ui}=F_{ui}(u_{*})$, that we obtain from the patchwork condition, into $J(u, F_{ui})$. See Sec. \ref{sec:patchwork} for the details.

The origin of the term ``patchwork condition'' is as follows. 
When the gauge-invariant scalar quantity becomes regular in the entire region of the bulk spacetime, we find a phenomenon that we call ``patchwork'' occurs: two singular (unphysical) solutions recombine at a point in the bulk (which we define $u=u_{*}$) to form a smooth regular physical solution. As far as the models we studied, the regularity condition is equivalent to the condition that this ``patchwork'' occurs. We find that the condition for this patchwork occurs is that the simultaneous equations (\ref{saddle1}) and (\ref{saddle2}) hold at some point ($u=u_{*}$) in the bulk.
We conjecture that the patchwork condition
is equivalent to the regularity condition. 

As discussed in Secs.~\ref{sec:proposal} and  \ref{sec:effective-temp}, we find that $u=u_{*}$ is the location of the {\em effective horizon} for the fluctuation of the bulk field around the background on-shell configuration. 
It has been proposed that the Hawking temperature associated with the effective horizon for the fluctuations gives an effective temperature that characterizes the nonequilibrium steady states we consider \cite{Gubser:2006nz, Casalderrey-Solana:2007ahi,Kim:2011qh,Sonner:2012if,Gursoy:2010aa,Nakamura:2013yqa}.
Our proposal enables us not only the computation of nonlinear conductivity but also this effective temperature for the wide range of models that do not have the square-root structure in the action.
We find, in Sec.~\ref{sec:effective-temp}, that different modes of fluctuation can have different effective temperatures when we consider general models without the square-root structure, while the models with the DBI action (or the NG action) have a unique common effective temperature regardless of the modes we consider.

We have two remarks on the relation between our proposal and the method proposed in earlier works.
The first is the relation between the real-action method and our method.
When we employ the real-action method for the models with the square-root structure, we obtain $u_*$ as a function of $E$ by solving Eq.~(\ref{eq:realcond-1}) in Sec.~\ref{sec:conventional-methods}.
On the other hand, our method argues that $u_*$ is determined from the simultaneous equations (\ref{saddle1}) and (\ref{saddle2}).
When we employ our method for the models with the square-root structure, Eq.~(\ref{saddle1}) decouples from Eq.~(\ref{saddle2}), and it reduces to Eq.~(\ref{eq:realcond-1}). Then $u_*$ is determined solely by Eq.~(\ref{saddle1}) without knowing $F_{ui}(u_{*})$ that is given by Eq.~(\ref{saddle2}).

This is however special for the models with the square-root structure.
For the models without the square-root structure, 
$u_{*}$ depends on $F_{ui}(u_{*})$ in general\footnote{
    This is the case for the models in Sec.~\ref{sec:higher-order} with higher powers of $F_{\mu \nu}F^{\mu \nu}$. For the models that show linear conductivity such as given in (\ref{eq:action-F1}), $u_{*}$ agrees with the location of the horizon of the bulk geometry and it is determined independently of the equations of motion of the bulk field within the probe approximation.}:
we need to determine $u_{*}$ and $F_{ui}(u_{*})$
simultaneously.
This is the reason why we need the two equations for determining $u_*$.

The second comment is about how the regularity condition has been applied to holographic calculations in previous studies. 
Requiring regularity of solutions is quite common in the gauge/gravity duality, and it is not a new idea. However, finding an efficient way to realize regular solutions in practice is significant for calculations in the gauge/gravity duality. Therefore, in past studies, conditions to realize regular solutions have been investigated empirically.
For example, it was found in Ref.~\cite{Babington:2003vm} that regular D-brane configurations can be achieved by requesting that the D-brane reaches the event horizon of the bulk geometry or the point where the compact manifold on which the D-brane wraps
collapses. One finds that the equation of motion for the D-brane configuration, which is a second-order differential equation, degenerates to a first-order differential equation at these points. Therefore, the first derivative of the worldvolume field is given as a function of the worldvolume field itself there: this provides a boundary condition.
The boundary condition given in these points of degeneration was used to efficiently achieve the regular solutions. (See, e.g., Ref.~\cite{Albash:2007bq}.\footnote{The same boundary condition has been used, e.g., in ref~\cite{Christensen:1998hg}.})
The method to request the regularity of the solutions at the point of degeneration was also used in \cite{Hashimoto:2014yza} for the systems out of equilibrium.
In Ref.~\cite{Hashimoto:2014yza}, the real-action method was not used, and the method of imposing regularity on the solution was adopted.
The same idea was employed in \cite{Kinoshita:2016lqd,Kinoshita:2017mio}, which is not in the context of the gauge/gravity duality, in the analysis of the energy-extraction mechanism from the black hole.


\section{Conventional method}
\label{sec:conventional-methods}

In this section, we review the conventional computational method of nonlinear conductivity in holography, which we refer to as the real-action method. Another method proposed in \cite{Kim:2011zd} is discussed in Appendix \ref{sec:Kim-Pang}.

We consider the D3-D7 model as an example of holographic realization of conductors \cite{Karch:2007pd}. The DBI action for the probe D7-brane is given as,
\begin{equation}
	S = - \tau_7 \int \dd[8]{\xi}\sqrt{- \det(g_{ab} + F_{ab})},
 \label{eq:DBIaction}
\end{equation}
where $\tau_7$ is the tension of the D7-brane,
$\xi^a$ are the 8-dimensional \ worldvolume coordinates.
$g_{ab}$ and $F_{ab}$ are the induced metric and the field-strength of the $\mathrm{U}(1)$ gauge field on the worldvolume, respectively.\footnote{Hereafter, depending on the context, $g_{ab}$ will denote either an induced metric or a background spacetime metric.}
The bulk geometry is the 5-dimensional AdS-Schwarzschild black hole times $\mathrm{S}^5$. The line element of the 5-dimensional AdS-Schwarzschild black hole is given by
\begin{equation}
	\dd{s}^2 = \frac{-f(u) \dd{t}^2 + \dd{\vec{x}}^2}{u^2} + \frac{\dd{u}^2}{u^2 f(u)},
	\label{eq:bulk_metric}	
\end{equation} 
where $f(u) = 1- u^4/u_h^4$. $u$ is the radial direction and $(t, \vec{x})$ is the (3+1)-dimensional coordinates on which the dual field theory is defined. The boundary of the bulk geometry is located at $u=0$, and $u=u_h$ is the location of the horizon of the black hole. The Hawking temperature $T$ is given by $T = (\pi u_h)^{-1}$.  
The line element of the unit 5-sphere, which we write $\dd{\Omega_5}$, is given by
\begin{equation}
	\dd{\Omega_5}^2 = \dd{\theta}^2 + \sin^2\theta \dd{\psi}^2 + \cos^2\theta \dd{\Omega_3}^2,
\end{equation}
where $\dd{\Omega_3}$ denotes the line element of the unit 3-sphere.
We employ the static gauge where the worldvolume coordinates are identical with the target-space coordinates $(t, \vec{x}, u, \Omega_3)$.
The $\mathrm{S}^{3}$ part is covered by the D7-brane. $\theta(u)$ and $\psi(u)$ describe the configuration of the D7-brane.
In this study, we consider the configuration $\theta = \psi =0$.
We employ the probe approximation throughout this paper where the background geometry is not modified by the back reaction.

We assume that a constant electric field $E$ is applied in the $x^{1}$-direction. Hereafter, we write $x^{1}$ as $x$ for simplicity. We employ the following ansatz for $A_{x}$
\begin{equation}
	A_x = - E t + a(u),
	\label{eq:ansatz_of_A_DBI}
\end{equation}
and we set the other components of the gauge field to zero.
Let us integrate the $\mathrm{S}^3$ part first in (\ref{eq:DBIaction}) to reduce it to the $(4+1)$-dimensional theory. 
The Lagrangian density after the integration in our convention is given by
\begin{equation}
	\mathcal{L} = - \mathcal{N}
	g_{xx}
	\sqrt{
		|g_{tt}|g_{xx}g_{uu}
		- g_{uu} F_{tx}^2 + |g_{tt}| F_{ux}^2
	},
	\label{eq:DBI_Lagrangian}
\end{equation}
where $\mathcal{N} = \tau_7 (2\pi^2)$ takes account of the volume of the unit $\mathrm{S}^{3}$.

The expectation value of current density $J$ in $x$-direction is obtained by the GKP-Witten prescription as\footnote{In this paper, we define the current density as $J =- \frac{1}{\mathcal{N}}\pdv{\mathcal{L}}{A_x'(u)}$ with including the factor $1/\mathcal{N}$ for simplicity.}
\begin{equation}
	J =- \frac{1}{\mathcal{N}}
	\pdv{\mathcal{L}}{A_x'(u)}
	= \frac{
		|g_{tt}|g_{xx} a'(u)
	}{
		\sqrt{|g_{tt}|g_{xx}g_{uu}  -E^2 g_{uu} + |g_{tt}|a'(u)^2}
	}.
\label{Jequation}
\end{equation}
Note that the equation of motion for $A_x$ is $\dv{u} \pdv{\mathcal{L}}{A_x'(u)}=0$, which is integrated to be $\pdv{\mathcal{L}}{A_x'(u)}={\rm constant}$. In this sense, $J$ is a constant of integration specified by the boundary condition when we solve the equation of motion. Therefore, the problem of obtaining the correct value of $J$ is that of identifying the proper condition we impose on the solution.

\subsection{Real-action method}
The computational method proposed in \cite{Karch:2007pd} requests the reality of physical quantities. We call this method the real-action method. The same computational method has also been employed in the computation of drag force \cite{Herzog:2006gh,Gubser:2006bz,Herzog:2006se,Caceres:2006as}.

The real-action method goes as follows. Eq.~(\ref{Jequation}) can be rewritten as
\begin{equation}
	- J^2 g_{uu} \xi(u) + |g_{tt}| \chi(u) a'(u)^2 = 0,
	\label{eq:eom_DBI}
\end{equation}
where we defined
\begin{equation}
	\xi(u) = |g_{tt}| g_{xx} - E^2,~~~\chi(u) = |g_{tt}|g_{xx}^2 - J^2.
\end{equation}
Then the solution to the equation of motion is given by
\begin{equation}
	a'(u) = J \sqrt{
		\frac{g_{uu}}{|g_{tt}|}\frac{\xi(u)}{\chi(u)}
	},
	\label{eq:hPrime_DBI}
\end{equation}
where we have selected the branch of $a'(u)>0$.
Since $g_{tt}(u_h)=0$, both of $\xi(u)$ and $\chi(u)$ are negative in the vicinity of $u=u_h$ whereas
$\xi(u)$ and $\chi(u)$ are positive at the boundary $u=0$: $\xi(u)$ and $\chi(u)$ change the sign somewhere between the boundary and the horizon.
In order to make $a'(u)$ real in the entire region of the bulk spacetime, we need to request that both $\xi(u)$ and $\chi(u)$ go across zero simultaneously at some point, which we define $u=u_*$.
Therefore, the reality condition for this model is expressed as the following simultaneous equations
\begin{eqnarray}
	\xi(u_{*}) &=& \left. \left(|g_{tt}| g_{xx} - E^2\right)\right|_{u=u_{*}}=0,
	\label{eq:realcond-1}\\
	\chi(u_{*})&=& \left. \left(|g_{tt}|g_{xx}^2 - J^2\right)\right|_{u=u_{*}}=0,
	\label{eq:realcond-2}
\end{eqnarray}
from which we obtain
\begin{equation}
	J= \pm\left.\sqrt{|g_{tt}|}g_{xx}\right|_{u=u_{*}}
 = \left.\sqrt{g_{xx}}\right|_{u=u_{*}} E,
\label{eq:cancel-2}
\end{equation}
where the sign of the right-hand side has been determined so that $J\cdot E>0$ in the last equality.

Eq.~(\ref{eq:realcond-1})
determines $u_{*}$ explicitly in terms of $E$. For the present model, $u_{*}$ is given by
\begin{equation}
	u_* = (1+e^2)^{-1/4} u_h,
	\label{eq:uStar_DBI}
\end{equation}
and then we obtain \begin{equation}
	j = e (1+e^2)^{1/4},
	\label{eq:current_DBI}
\end{equation}
where we have defined dimensionless electric field and current density by
\begin{equation}
	e \equiv u_h^2 E = \frac{E}{(\pi T)^2},\quad
	j \equiv u_h^3 J = \frac{J}{(\pi T)^3},
\end{equation}
respectively.
For later use, we expand $u_*$ and the current density in powers of $e$,
\begin{equation}
	u_*/u_h = 1 - \frac{1}{4} e^2 + \frac{5}{32} e^4 - \frac{15}{128}e^6 + \order{e^7}.
	\label{eq:ustar_DBI_expanded}
\end{equation}
\begin{equation}
	j = e + \frac{1}{4} e^3 - \frac{3}{32} e^5 + \frac{7}{128}e^7 + \order{e^8}.
	\label{eq:current_DBI_expanded}
\end{equation}

Now we have a comment on the point $u=u_*$.
The point $u=u_*$ corresponds to the effective horizon of an open-string metric \cite{Seiberg:1999vs,Gibbons:2000xe} given by $\gamma_{ab} = g_{ab} - F_{ac}g^{cd}F_{db}$.
In the present case, the components of the open-string metric are given by
\begin{equation}
\begin{aligned}
    \gamma_{ab}\dd{\xi^a}\dd{\xi^b}
    =&
    - \frac{\xi(u)}{g_{xx}}\dd{t}^2
    - 2 \frac{E h'(u)}{g_{xx}}\dd{t}\dd{u}
    +\frac{g_{xx}g_{uu}+h'(u)^2}{g_{xx}}\dd{u}^2\\
    &+\left(
        g_{xx} + \frac{E^2}{g_{tt}}
        +\frac{h'(u)^2}{g_{uu}}
    \right)\dd{x}^2
    + \frac{\dd{y}^2 + \dd{z}^2}{u^2} + \dd{\Omega_3}^2.
\end{aligned}
\end{equation}
We can evaluate the effective temperature \cite{Kim:2011qh,Sonner:2012if,Nakamura:2013yqa},
the Hawking temperature associated with the effective horizon, as
\begin{equation}
    T_{*} = \frac{\sqrt{2+3 e^2}}{\sqrt{2}(1+e^2)^{1/4}} T.
\label{eq:effectiveT_DBI}
\end{equation}

\subsection{Failure of real-action method}
\label{sec:failure}

The computational methods of conductivity presented in the previous subsection is strongly based on the ``square-root'' structure of the DBI action.
We demonstrate that the method in the previous subsection do not work in more general cases where the action does not have the ``square-root'' structure.%
\footnote{This has been pointed out in Ref. \cite{Kim:2011zd}.}%
\footnote{
In Ref.~\cite{Baggioli:2016oju}, models with higher-order field strength have been considered. However, the nonlinear conductivity has been computed by the real-action method only for the case where the action has a square-root structure.
}

Let us consider a toy model whose action in the gravity dual is given by
\begin{equation}
	S = \mathcal{N}\int \dd[4]{x}\dd{u} \sqrt{-g} \mathcal{F},
	\label{eq:action-F1}
\end{equation}
where
\begin{equation}
	\mathcal{F}(u) \equiv -\frac{1}{4} F_{\mu\nu} F^{\mu\nu}
	= \frac{1}{2}g^{xx} \left(
		|g^{tt}| E^2 - g^{uu} a'(u)^2
	\right).
	\label{eq:calF_def}
\end{equation}
We use the same metric as that is given in (\ref{eq:bulk_metric}) and the same ansatz as (\ref{eq:ansatz_of_A_DBI}) for the gauge field.
We have substituted them in the last line.
It is known that this model exhibits linear conductivity.
This is the simplest model to demonstrate the failure of 
the real-action method.

Now, the current density is given by
\begin{equation}
	J = -\frac{1}{\mathcal{N}}\pdv{\mathcal{L}}{A_x'(u)}= \sqrt{\frac{|g_{tt}|g_{xx}}{g_{uu}}} a'(u).
	\label{Jequation-F1}
\end{equation}
In contrast to (\ref{Jequation}), the relationship between $J$ and $a^{\prime}$ is linear, and $a^{\prime}$ has no chance to be complex as far as we choose the constant of integration $J$ to be real: the reality condition for the gauge field (hence for the on-shell action) just states $J$ has to be real, and it gives no further constraint for $J$. 
Therefore, the reality of the action itself cannot fix the physical value of the current density in this simplest case.
As we will see in the next section, imposing regularity is more useful than the reality in a general model.

\section{Regularity condition and patchwork condition}
\label{sec:proposal}

\subsection{Regularity condition}
\label{sec:regularity}
Let us see how the regularity condition works for the model of (\ref{eq:action-F1}).
Since this model exhibits linear conductivity, 
the conductivity can be calculated by a method using the Kubo formula without using the method we are about to propose.
However, we use this simplest model to demonstrate how our method works.
As we have mentioned in Sec.~\ref{sec:outline}, our regularity condition requests that ${\cal F}$ is regular in the entire region of the bulk spacetime.
In the Schwarzschild coordinates, we have
\begin{equation}
	\mathcal{F}(u) =
	\frac{1}{2}\frac{E^2 g_{xx} - J^2}{|g_{tt}| g_{xx}^2}.
\end{equation}
$\mathcal{F}(u)$ has a chance to be divergent only at the event horizon where $g_{tt}$ vanishes.\footnote{Note that the effective horizon coincides with the event horizon of the bulk geometry in this model.} In order to avoid the divergence of $\mathcal{F}(u)$, we need to set $E^2 g_{xx} - J^2=0$ at the horizon which gives the right answer
\begin{equation}
    J=\sqrt{g_{xx}}|_{u=u_{h}}E,
    \label{eq:linearcondInEF}
\end{equation}
provided that we choose the correct sign. When (\ref{eq:linearcondInEF}) holds, $\mathcal{F}=E^{2}u_{h}^{2}/4$ at $u=u_h$.

Since ${\cal F}$ is scalar, the regularity of ${\cal F}$ leads us to the same conclusion in the ingoing Eddington-Finkelstein coordinates. However, it is instructive to see what is going on there. 
The 5-dimensional AdS-Schwarzschild geometry in the ingoing Eddington-Finkelstein coordinates is given by
\begin{equation}
    \dd{s}^{2}=-\frac{f(u)}{u^2}\dd{\tau}^2-\frac{2}{u^2}\dd{\tau} \dd{u}+\frac{1}{u^2}\dd{\vec{x}}^2,
    \label{ingoingEF}    
\end{equation}
where $f(u)=1-u^4/u_h^4$. The $u$ coordinate is common between (\ref{eq:bulk_metric}) and (\ref{ingoingEF}), and $\dd{\tau}=\dd{t}-\frac{\dd{u}}{f(u)}$. 
Let us employ the following ansatz for the gauge field
\begin{equation}
A_{x}=-E\tau+h(u),
\label{eq:ansatz_EF}
\end{equation}
and we set the other components of the gauge field to zero. Note that $h(u)$ and $a(u)$ are different from each other.
In the ingoing Eddington-Finkelstein coordinates, ${\cal F}$ is expressed as
\begin{equation}
	\mathcal{F}(u) =
	\frac{1}{g_{\tau u}g_{xx}}h^{\prime}\left( E+\frac{g_{\tau\tau}}{2 g_{\tau u}}h^{\prime}\right).
\end{equation}
If $\frac{g_{\tau\tau}}{g_{\tau u}}h^{\prime}$ is nonzero at the horizon, $h^{\prime}$ has to diverge there. This leads ${\cal F}$ divergent at the horizon.
Therefore, regularity of ${\cal F}$ ensure $\frac{g_{\tau\tau}}{g_{\tau u}}h^{\prime}\to 0$ at the horizon.
Now, the current density is given as
\begin{equation}
    J=-\frac{1}{\mathcal{N}}\pdv{{\cal L}}{A_{x}^{\prime}}
    =\sqrt{g_{xx}}\left( E+\frac{g_{\tau\tau}}{g_{\tau u}} h^{\prime}\right).
    \label{eq:JequationInEF}
\end{equation}
As we have seen in (\ref{Jequation-F1}), the reality of $h^{\prime}$ does not give any constraint for $J$ except for the trivial statement, the reality of $J$.
However, if we request the regularity of ${\cal F}$,
the $h^{\prime}$-dependence of the right-hand side of (\ref{eq:JequationInEF}) vanishes at the horizon.
Then, $J$ is given by (\ref{eq:linearcondInEF}).
One finds that the idea given in the past studies mentioned in Sec.~\ref{sec:outline} of requesting the regularity of the solution ($h^{\prime}$ for this case) at the point of degeneration (event horizon in the present model) also works for this case.

\subsection{Patchwork condition}
\label{sec:patchwork}
\begin{figure}[htbp!]
	\centering
	\subfloat[]{\includegraphics[width=8cm]{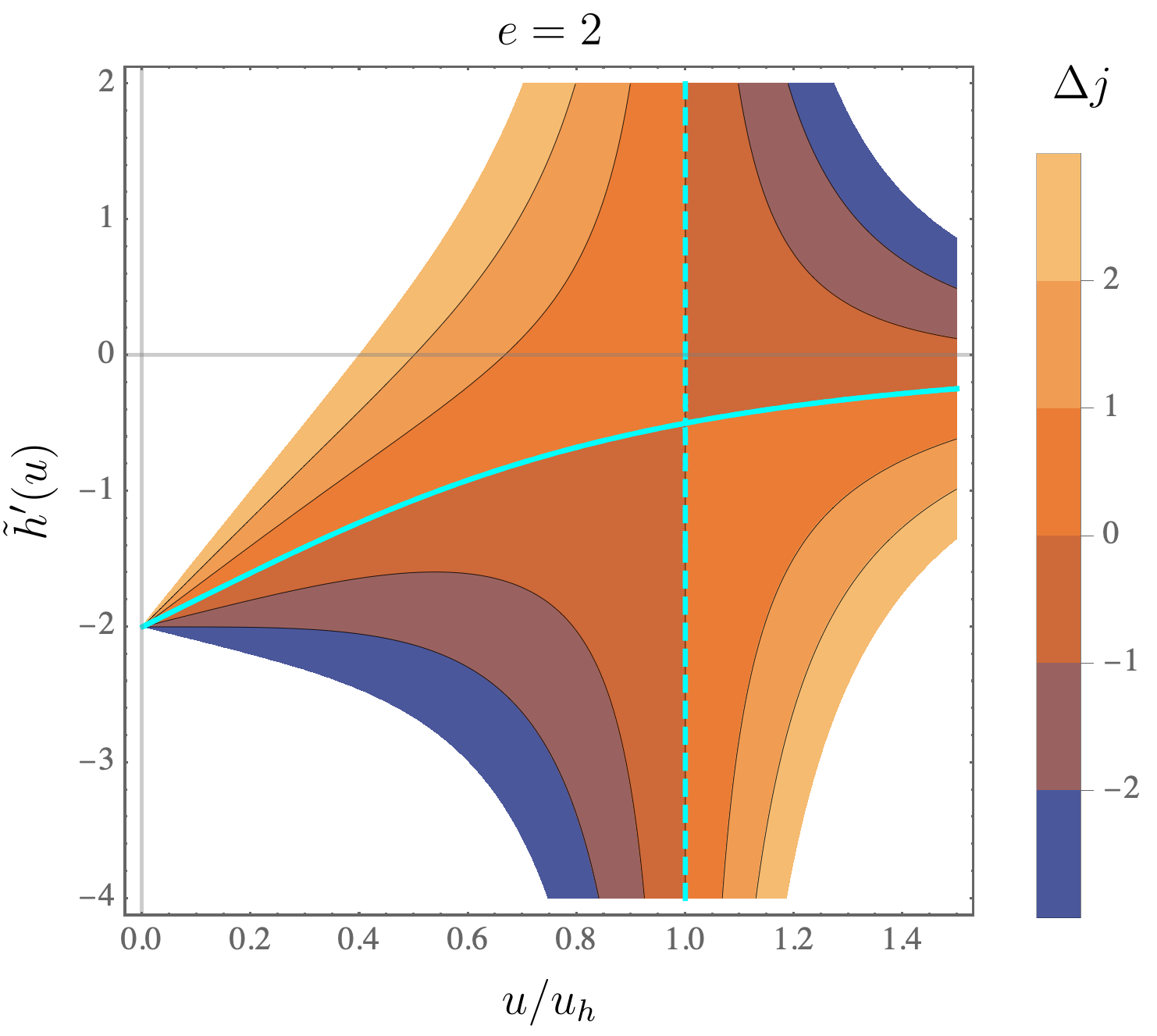}\label{fig:j_contours_F1-a}}
	\subfloat[]{\includegraphics[width=7.0cm]{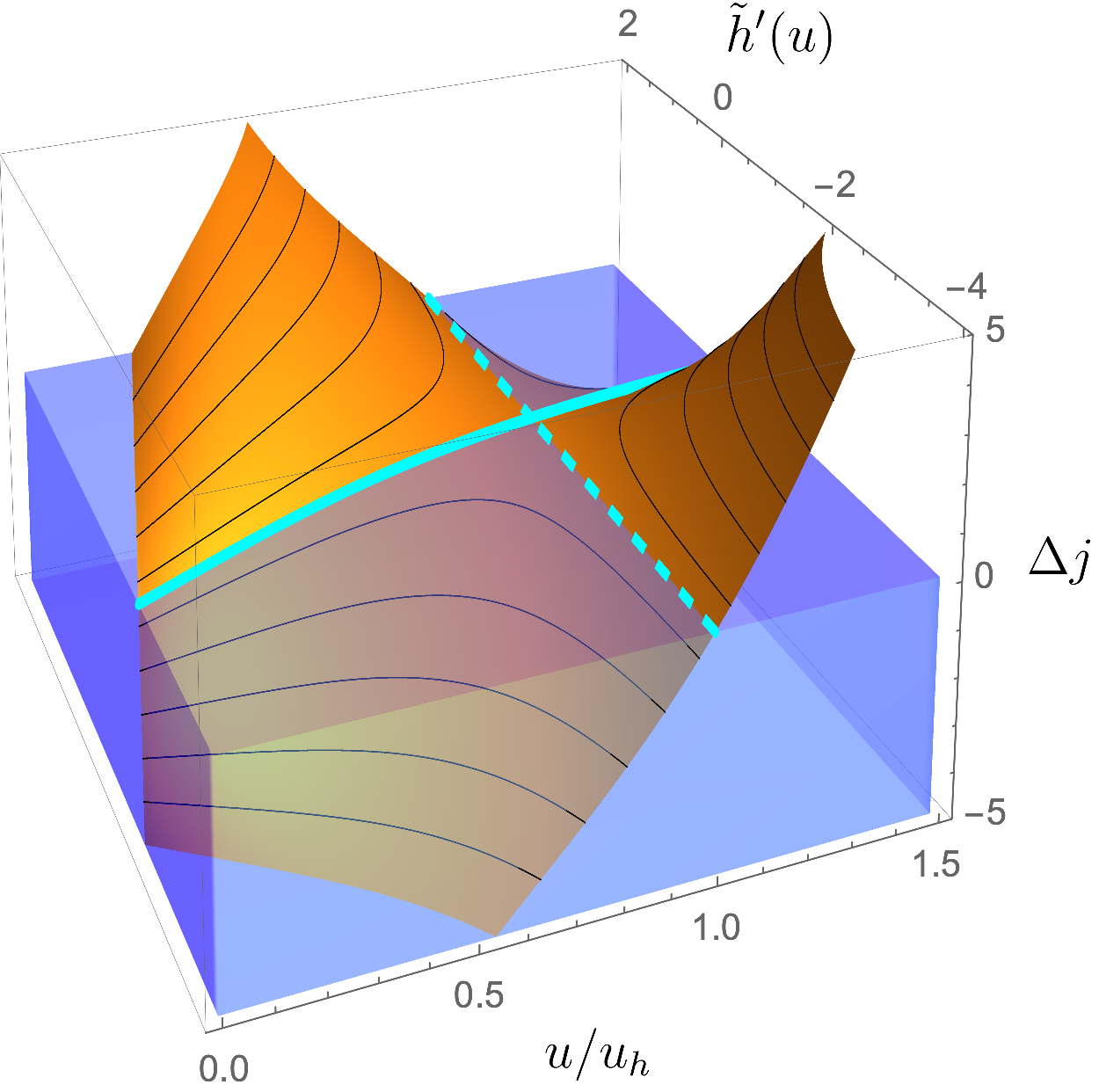}\label{fig:j_contours_F1-b}}
	\caption{
        (a)
        Contour plot of $J(u,h')$ for the model given by~\eqref{eq:action-F1}. Each contour represents the solution $h'(u)$ for given $J$. Here, $J$ and $h^{\prime}$ are rescaled as $j=u_h^3 J$ and $\tilde{h}'=u_h^2 h'$, respectively, and $\Delta j = j-j_\text{phys}$. The solid and dashed cyan curves are the contours at $j=j_{\text{phys}}$ that go through the patchwork point. The solid cyan curve is the regular physical solution and satisfies the proper boundary condition at $u=0$.
        (b)
        Surface plot of $J(u,h')$ for the model given by~\eqref{eq:action-F1}. 
        The orange surface represents $\Delta j$, while the top plane of the translucent blue box represents $\Delta j=0$.
        The solid and dashed cyan curves are the section of these two surfaces. The crossing point of these curves is the saddle point of the orange surface. Here, we set $e=2$ in these panels.
	}
	\label{fig:j_contours_F1}
\end{figure}

\begin{figure}[htbp!]
    \centering
    \includegraphics[width=8cm]{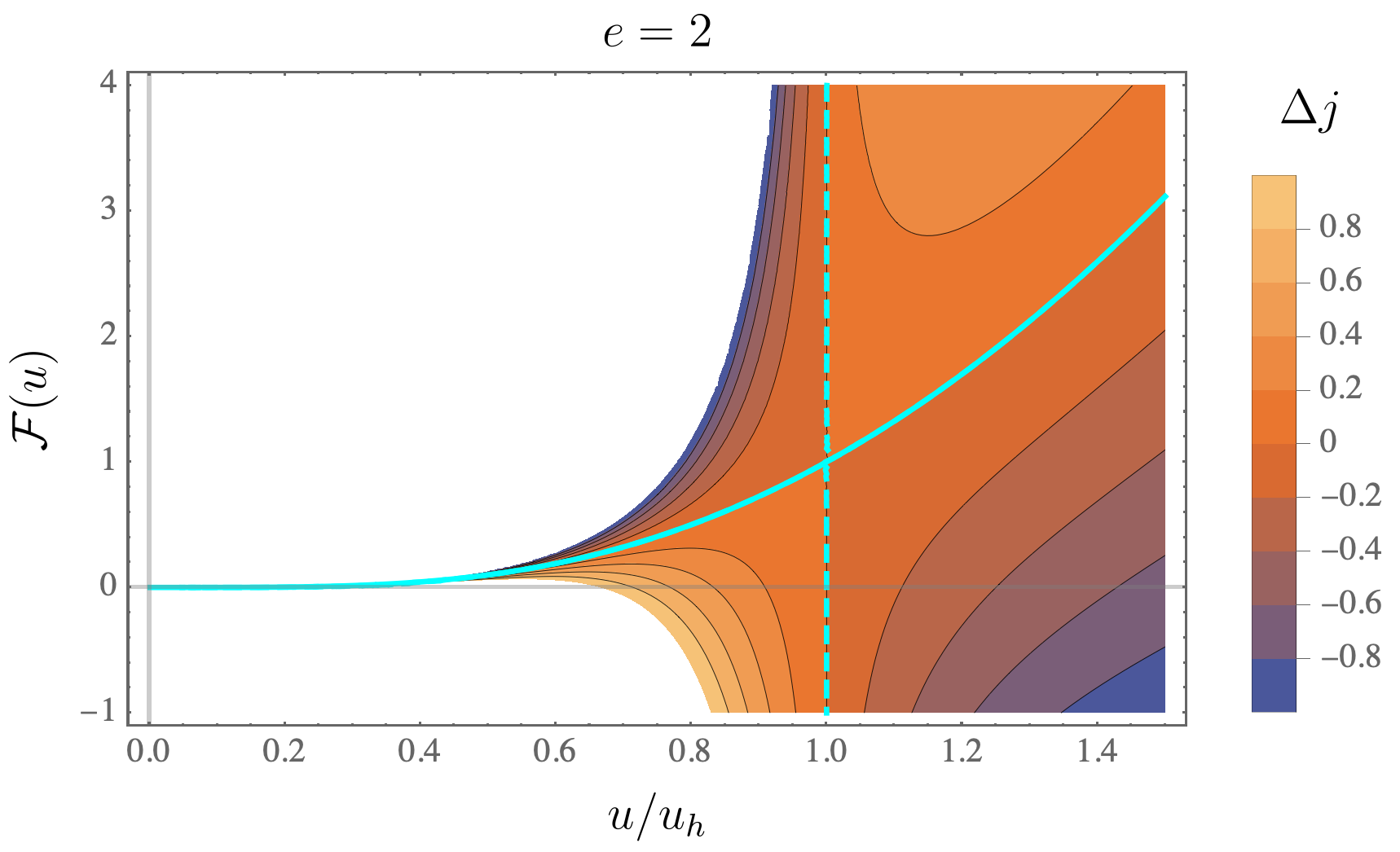}
    \caption{
    $\mathcal{F}(u)$ for the model given by~\eqref{eq:action-F1} with respect to the various values of $J$. The cyan curves correspond to those in Fig.~\ref{fig:j_contours_F1}. $\Delta j$ is defined as $\Delta j=u_{h}^{3}(J-J_{\text{phys}})$. Here, we set $e=2$.
    }
    \label{fig:calF-u_contours_F1}
\end{figure}

Let us consider the behavior of $h^{\prime}$ in more detail in the ingoing Eddington-Finkelstein coordinates.\footnote{If we employ the Schwarzschild coordinates in the model of (\ref{eq:action-F1}), the ``patchwork'' we explain in this section cannot be seen. The reason is that the point $u=u_{*}$ for this model is located at the event horizon. However, we can observe the ``patchwork'' in the models with higher-derivative terms given in Sec.~\ref{sec:higher-order} and Sec.~\ref{sec:DBI} even on the Schwarzschild coordinates (see Appendix \ref{sec:Schwarzschild}). This is because the points of $u=u_{*}$ for these models are located outside the event horizon where the Schwarzschild coordinates cover.} 
When $u\neq u_{h}$, $g_{\tau\tau}$ is nonzero and we obtain
\begin{eqnarray}
    h^{\prime}=\frac{g_{\tau u}}{g_{\tau\tau}}\frac{J-\sqrt{g_{xx}}E}{\sqrt{g_{xx}}}=\frac{1}{f(u)}\frac{J-\sqrt{g_{xx}}E}{\sqrt{g_{xx}}},
\end{eqnarray}
from (\ref{eq:JequationInEF}). 
Since $f(u)$ is a monotonically decreasing function that goes across zero at $u=u_h$, the behavior of $h^{\prime}$ can be categorized into the following three types.
\begin{enumerate}
    \item $\left(J-\sqrt{g_{xx}(u_h)}E \right) >0$\\
    $h^{\prime}|_{u\to u_h}=\infty$ when $u$ approaches $u_h$ from outside the horizon ($u<u_h$), whereas $h^{\prime}|_{u\to u_h}=-\infty$ when $u$ approaches $u_h$ from inside the horizon ($u>u_h$). $h^{\prime}(u)$ is discontinuous at $u=u_h$. One finds that ${\cal F}$ is divergent at the horizon.
    \item $\left(J-\sqrt{g_{xx}(u_h)}E \right) =0$\\
    We find $h^{\prime}|_{u\to u_h}=-E/4$ and finite. $h^{\prime}(u)$ is a smooth single-valued function of $u$ which goes across the horizon smoothly. One finds that ${\cal F}=E^{2} \frac{u^4 u_h^2}{2(u^2+u_h^2)}$ and is regular even at the horizon. ${\cal F}$ is singular at $u=\infty$, but this singularity is causally disconnected from the observer outside the horizon.
    \item $\left(J-\sqrt{g_{xx}(u_h)}E \right) <0$\\
    $h^{\prime}|_{u\to u_h}=-\infty$ when $u$ approaches $u_h$ from outside the horizon ($u<u_h$), whereas $h^{\prime}|_{u\to u_h}=\infty$ when $u$ approaches $u_h$ from inside the horizon ($u>u_h$). $h^{\prime}(u)$ is discontinuous at $u=u_h$. One finds that ${\cal F}$ is divergent at the horizon.
\end{enumerate}
The behaviour of $h^{\prime}(u)$ is shown in Fig.~\ref{fig:j_contours_F1}.
Figure \ref{fig:calF-u_contours_F1} shows the behaviour of $\mathcal{F}$ as well.
We defined $\tilde{h}'(u) = h'(u) u_h^2$.
Suppose that we keep $E$ fixed and change the value of $J$, which is a constant of integration, from $J>J_{\text{phys}}$ to $J<J_{\text{phys}}$, where $J_{\text{phys}}=\sqrt{g_{xx}(u_h)}E$ is the physical value.
When $J>J_{\text{phys}}$, the plot of $h^{\prime}(u)$ is separated into two curves that diverge at the horizon. These two curves approach each other and merge when $J$ goes to $J_{\text{phys}}$. 
$h^{\prime}(u)$ separates into two divergent curves again when $J$ goes lower than $J_{\text{phys}}$.
We call this behaviour of $h^{\prime}(u)$ at $J=J_{\text{phys}}$ ``patchwork.''
In this model, the regularity condition is equivalent to the condition that the ``patchwork'' occurs.

The condition that the patchwork occurs can be expressed by two equations.
To illustrate the behavior of the solution, we will use the analogy of a topographic map.
Suppose that $J(u,h^{\prime})$ is ``height of land'' at a location given by ``two-dimensional coordinates'' $(u,h^{\prime})$.
Then the curves we have in Fig.~\ref{fig:j_contours_F1}(a) show the contour map (the curves of equal height $J$) of the ``landform.'' Then we find that the location where patchwork occurs is the {\em saddle point} of the landform, and $J=J_{\text{phys}}$ is the height of the saddle point.
A point where contours cross is a saddle point in general.
Since the patchwork point we call is a point where contours of $J=J_{\text{phys}}$ cross, it is a saddle point obviously.
From this observation, we impose the following condition of the presence of a stationary point in the $u$--$h'$ plane:
\begin{eqnarray}
   \left. \frac{\partial J(u,h^{\prime})}{\partial u}\right|_{u=u_*}=0,
   \label{eq:patchJ-1}   \\
   \left. \frac{\partial J(u,h^{\prime})}{\partial h^{\prime}}\right|_{u=u_*}=0,
   \label{eq:patchJ-2}
\end{eqnarray}
where $u=u_*$ is the location of the saddle point.
In the present case $u_*=u_h$, but $u_*$ can be different from $u_h$ in more general models as we shall see later.
We also emphasize that eq.~(\ref{eq:patchJ-1}) is a condition for the partial derivative $\pdv{J}{u}$ rather than the ordinary derivative $\dv{J}{u}$.
It is because this condition arose from the presence of the saddle point in the parameter space of $(u,h')$.
Moreover, $\dv{J}{u}$ always vanishes because it is just the equation of motion of $A_{x}$.
Since $J=-\frac{1}{\mathcal{N}}\frac{\partial {\cal L}}{\partial A_{x}^{\prime}}$ and $A_{x}^{\prime}=h^{\prime}=F_{ux}$ in our convention, the foregoing two equations are rewritten as
\begin{eqnarray}
   \left. \frac{\partial^2 {\cal L}(F_{ux},u)}{\partial u \partial F_{ux}}\right|_{u=u_*}=0,
   \label{eq:patchL-1}   \\
   \left. \frac{\partial^2 {\cal L}(F_{ux},u)}{\partial F_{ux}^2}\right|_{u=u_*}=0,
   \label{eq:patchL-2}
\end{eqnarray}
which we have exhibited at (\ref{saddle1}) and (\ref{saddle2}) in Sec.~\ref{sec:outline}.
We call the simultaneous equations (\ref{eq:patchJ-1}) and (\ref{eq:patchJ-2}), or (\ref{eq:patchL-1}) and (\ref{eq:patchL-2}), as the ``patchwork condition.''%
\footnote{
    In the higher-order models such as we consider in Sec.~\ref{sec:higher-order}, Eq.~(\ref{eq:patchJ-2}) can also be understood as a condition that the equation of motion has multiple roots at $u=u_*$. The equation of motion is given as a polynomial of $h^{\prime}$, $P(h') \equiv J(u,h') - J=0$, in general. For example, the equation of motion (\ref{eq:eom_F2}) is a cubic equation of $h'(u)$.
    The discriminant of the polynomial, $\mathrm{disc}_{h'}(P)$, is given by the resultant of the polynomial and its derivative: $\mathrm{res}_{h'}(P,\partial_{h'}P)$.
    The resultant vanishes if and only if the equations $P=0$ and $\partial_{h'}P=0$ have a common root.
    $P=0$ is the equation of motion that has to be satisfied. Eq.~(\ref{eq:patchJ-2}) states that $\partial_{h'}P = \partial_{h'}J(u,h')=0$ has a root at $u=u_*$.
    Therefore, the discriminant vanishes at $u=u_*$, and the equation of motion has multiple roots at $u=u_*$. This is consistent since the patchwork point is the point where multiple roots exist. For the present model, however, the equation of motion (\ref{eq:JequationInEF}) is a linear equation of $h^{\prime}$, and this comment does not apply.
}

Note that the regularity condition for the present model has reduced to the simultaneous equations. Then, 
the conductivity can be obtained by solving the simultaneous equations. Indeed, (\ref{eq:patchJ-2}) gives $\left. \sqrt{g_{xx}}g_{\tau\tau}/g_{\tau u}\right|_{u=u_*}=0$ which yields $u_{*}=u_{h}$.\footnote{For the models with higher powers of $\mathcal{F}$, as we shall see in Sec.~\ref{sec:higher-order}, we need to solve both (\ref{eq:patchJ-1}) and (\ref{eq:patchJ-2}) to obtain $u_{*}$.} We obtain $h^{\prime}|_{u=u_{*}=u_{h}}=-E/4$ from (\ref{eq:patchJ-1}) and we find $\frac{g_{\tau\tau}}{g_\tau u}h^{\prime}$ vanishes at $u=u_{*}=u_h$. Then we obtain (\ref{eq:linearcondInEF}) from (\ref{eq:JequationInEF}). 

We can interpret the physical meaning of the patchwork condition as follows. 
Let us decompose $A_{\mu}$ as $A_{\mu}=\Bar{A}_{\mu}+a_{\mu}$ and consider the small fluctuation $a_{x}$ around the background solution $\Bar{A}_{x}$. We expand the Lagrangian density to the quadratic order of $a_{x}$ as
\begin{eqnarray}
{\cal L}=-\frac{1}{4}\sqrt{-\gamma}f_{\mu x}f_{\nu x}\gamma^{\mu\nu}\gamma^{xx}+\cdots,
\end{eqnarray}
where $f_{\mu\nu}=\partial_{\mu}a_\nu-\partial_\nu a_\mu$.
Here, $\gamma_{\mu\nu}$ denotes the effective metric that governs the dispersion relation of the small fluctuation $a_{x}$. $\gamma$ is the determinant of the effective metric. Although the effective metric agrees with the metric of the bulk geometry in the present model, it depends on $E$ in general when the original Lagrangian contains higher powers of $\mathcal{F}$. In the case of the DBI theory, the effective metric is given by the open-string metric \cite{Seiberg:1999vs,Gibbons:2000xe}.
We obtain
\begin{eqnarray}
\frac{\partial^{2} {\cal L}}{\partial f_{\mu x}\partial f_{\rho x}}
=-\frac{1}{2}\sqrt{-\gamma}\gamma^{\mu\rho}\gamma^{xx}.
\label{eq:effective-m}
\end{eqnarray}
Note that the left-hand side of (\ref{eq:effective-m}) is $\frac{\partial^{2} {\cal L}}{\partial F_{\mu x}\partial F_{\rho x}}$ evaluated at the given background.
We assume that $\sqrt{-\gamma}\gamma^{xx}$ is non-vanishing (Assumption 1). Then (\ref{eq:patchL-2}) is understood as the condition for $\gamma^{uu}|_{u=u_*}=0$ which means that $u=u_{*}$ is the location of the ``effective horizon'': the small fluctuation $a_{x}$ cannot propagate from inside the effective horizon $u>u_*$ to outside the effective horizon $u<u_*$ beyond $u=u_{*}$.

The other condition (\ref{eq:patchL-1}) is understood as follows.
The equation of motion of $A_{x}$ is written as
\begin{eqnarray}
0={\dv{u}}\frac{\partial {\cal L}}{\partial F_{ux}}
=\left(\frac{\partial}{\partial u}+\dv{F_{ux}}{u} \frac{\partial }{\partial F_{ux}} \right)\frac{\partial {\cal L}}{\partial F_{ux}}.
   \label{eq:eom_Ax}
\end{eqnarray}
Let us assume that $\dv{F_{ux}}{u}\frac{\partial^2 {\cal L}}{\partial F_{ux}^2}$ vanishes at $u=u_*$ (Assumption 2).
Then (\ref{eq:patchL-1}) is equivalent to the equation of motion (\ref{eq:eom_Ax}) at the location of the effective horizon $u=u_*$ where (\ref{eq:patchL-2}) holds.
Recall that (\ref{eq:patchL-1}) is the condition for the partial derivative with respect to $u$.

When Assumption 1 and Assumption 2 hold, the patchwork condition (and hence the regularity condition) requests the {\em presence of the effective horizon} on the background solution. 
As far as the authors have checked, all the models we deal with in this paper satisfy these assumptions at $u=u_*$. 
This suggests that the concept of effective temperature in nonequilibrium steady states exists in a wide range of holographic models not only for the models whose action has the square-root structure.


Now we have a comment on the coordinate system we employ. If we use the Schwarzschild coordinates, the patchwork phenomenon
is not observed for the field $a'(u)$ explicitly, in the present model linear in $\mathcal{F}$. The reason is as follows. The patchwork phenomenon occurs in such a way that the regular solution penetrates the effective horizon from outside the effective horizon into the inside of it. In the case of the present model, the location of the effective horizon coincides with that of the event horizon of the bulk geometry. Since the Schwarzschild coordinates cover only outside the event horizon, we cannot observe the recombination of the solutions where a solution inside the horizon and that outside the horizon merge into a single regular solution. However, we can work on the Schwarzschild coordinates in the cases of more general models we present in Sec.~\ref{sec:higher-order} and in Sec.~\ref{sec:DBI} where the effective horizon, the point where the patchwork occurs, is located outside the event horizon. In this case, the inside of the effective horizon is at least partly covered by the Schwarzschild coordinates. Even though we work on the ingoing Eddington-Finkelstein coordinates in Sec.~\ref{sec:higher-order} and in Sec.~\ref{sec:DBI}, we also present how it works on the Schwarzschild coordinates in Appendix \ref{sec:Schwarzschild}.

\section{Higher-order models}
\label{sec:higher-order}
We demonstrate that the patchwork condition we have proposed in Sec.~\ref{sec:patchwork} works in more general models that contain higher powers of $\mathcal{F}$ but without the square-root structure.
The bulk geometry is the AdS-Schwarzschild black hole whose metric is given by 
(\ref{ingoingEF}). 

\subsection{$\mathcal{F}^2$ model}\label{subsec:F2-model}
Now, we consider the following action involving a nonlinear term.
\begin{equation}
	S = \mathcal{N}\int\dd[4]{x}\dd{u}\sqrt{-g}\left(
		\mathcal{F} + \frac{c}{2}\mathcal{F}^2
	\right),
	\label{eq:action_F2}
\end{equation}
where $c$ is a real constant.%
\footnote{
In general, nonlinear Maxwell theories can be unstable, and a physical range of the coupling constant would be restricted.
In this study, however, we do not consider this problem.
}
$\mathcal{F}$ is given by the first equality of Eq.~(\ref{eq:calF_def}).
We will refer to this model as $\mathcal{F}^2$ model.
We consider the finite temperature cases here.\footnote{The results for $T=0$ is given in Appendix \ref{sec:T=0}.}

Let us employ the ingoing Eddington-Finkelstein coordinates on which the metric is given by (\ref{ingoingEF}).
The Lagrangian density with our ansatz (\ref{eq:ansatz_EF}) is explicitly written as
\begin{equation}
\begin{aligned}
	\mathcal{L} =&
	\frac{\mathcal{N}}{2}\sqrt{g_{\tau u}^2 g_{xx}}\Big[
		-
		\left(
			2 g^{\tau u} F_{\tau x}
			+ g^{uu} F_{ux}
		\right) F_{u x}\\
		&+
		\frac{c}{4}g^{xx}\left(
			2 g^{\tau u} F_{\tau x}
			+ g^{uu} F_{ux}
		\right)^2 F_{ux}^2
	\Big].
\end{aligned}
	\label{eq:Lagrangian_F2}
\end{equation}
The equation of motion for $A_x$ is given by
\begin{equation}
\begin{aligned}
	J = & - \frac{1}{\mathcal{N}}\pdv{\mathcal{L}}{F_{ux}}\\
	=& \sqrt{g_{\tau u}^2 g_{xx}}
	( - E g^{\tau u} + g^{uu} h'(u))\\
	&\times\left[
		1 - \frac{c}{2} g^{xx}  h'(u) (-2 E g^{\tau u} + g^{uu}h'(u))
	\right]\\
	\equiv& J(u, h'(u)),
\end{aligned}
\label{eq:eom_F2}
\end{equation}
where $J$ is an integration constant.
This is a cubic equation of $h'(u)$ that has three roots.
The explicit expressions of these roots are very complicated, and we do not exhibit them here.

Let us see the asymptotic behaviors of the three roots.
In the vicinity of the boundary of the bulk geometry, the solutions have series expansions, as follows:
\begin{eqnarray}
    h'(u) &=&
	-E + J u + \order{u^2},
 	\label{eq:hPrime_at_zero-1}\\
	h'(u) &=&
	\pm \frac{\sqrt{2/c}}{u^2} - E - \frac{J}{2} u + \order{u^2}.
	\label{eq:hPrime_at_zero-2}
\end{eqnarray}
Only the first solution (\ref{eq:hPrime_at_zero-1}) matches with our boundary condition $h(u)=0$ at $u\to0$.
On the other hand, the solutions in the vicinity of the event horizon are expanded as follows:
\begin{eqnarray}
	h'(u) &=&
	u_h^{-2} \frac{e - j}{c e^2} + \order{1-u/u_h},
 	\label{eq:hPrime_at_uh-1}\\
    h'(u) &=&
	u_h^{-2}\left[
        - \frac{e}{2 (1 - u/u_h)} - \frac{4 e + 3 c e^3 + 4 j}{4 c e^2}
    \right] \nonumber\\
    && + \order{1-u/u_h},
	\label{eq:hPrime_at_uh-2}\\
    h'(u) &=&
    u_h^{-2}\left[
    - \frac{e}{4 (1 - u/u_h)}  - \frac{3 c e^3 - 16 j}{8 c e^2}
    \right] \nonumber\\
    && + \order{1-u/u_h}
    \label{eq:hPrime_at_uh-3}.
\end{eqnarray}
Note that we have rescaled $E$ and $J$ as $e=u_{h}^{2}E$ and $j=u_{h}^{3}J$. We also obtain
\begin{eqnarray}
	\mathcal{F}(u) &=&
	- \frac{e - j}{c e}+ \order{1-u/u_h},
 	\label{eq:scF(uh)_F2-1}\\
    \mathcal{F}(u) &=&
	- \frac{e + j}{c e}+ \order{1-u/u_h},
	\label{eq:scF(uh)_F2-2}\\
    \mathcal{F}(u) &=&
    \frac{e^2}{8 (1- u/u_h)} - \frac{5}{16}e^2+ \order{1-u/u_h},
    \label{eq:scF(uh)_F2-3}
\end{eqnarray}
respectively.
If we choose the first two solutions (\ref{eq:hPrime_at_uh-1}) and (\ref{eq:hPrime_at_uh-2}), $\mathcal{F}(u)$ that is given by (\ref{eq:scF(uh)_F2-1}) and (\ref{eq:scF(uh)_F2-2}), respectively, remain finite at $u=u_h$ whatever combinations of $j$ and $e$ we have at non-zero value of $e$: the regularity of $\mathcal{F}(u)$ at $u=u_h$ alone does not fix the conductivity at all.\footnote{Note that it turns out later that $u=u_h$ is no longer the location of the effective horizon.}

\begin{figure}[htbp!]
	\centering
	\subfloat[]{\includegraphics[width=8.0cm]{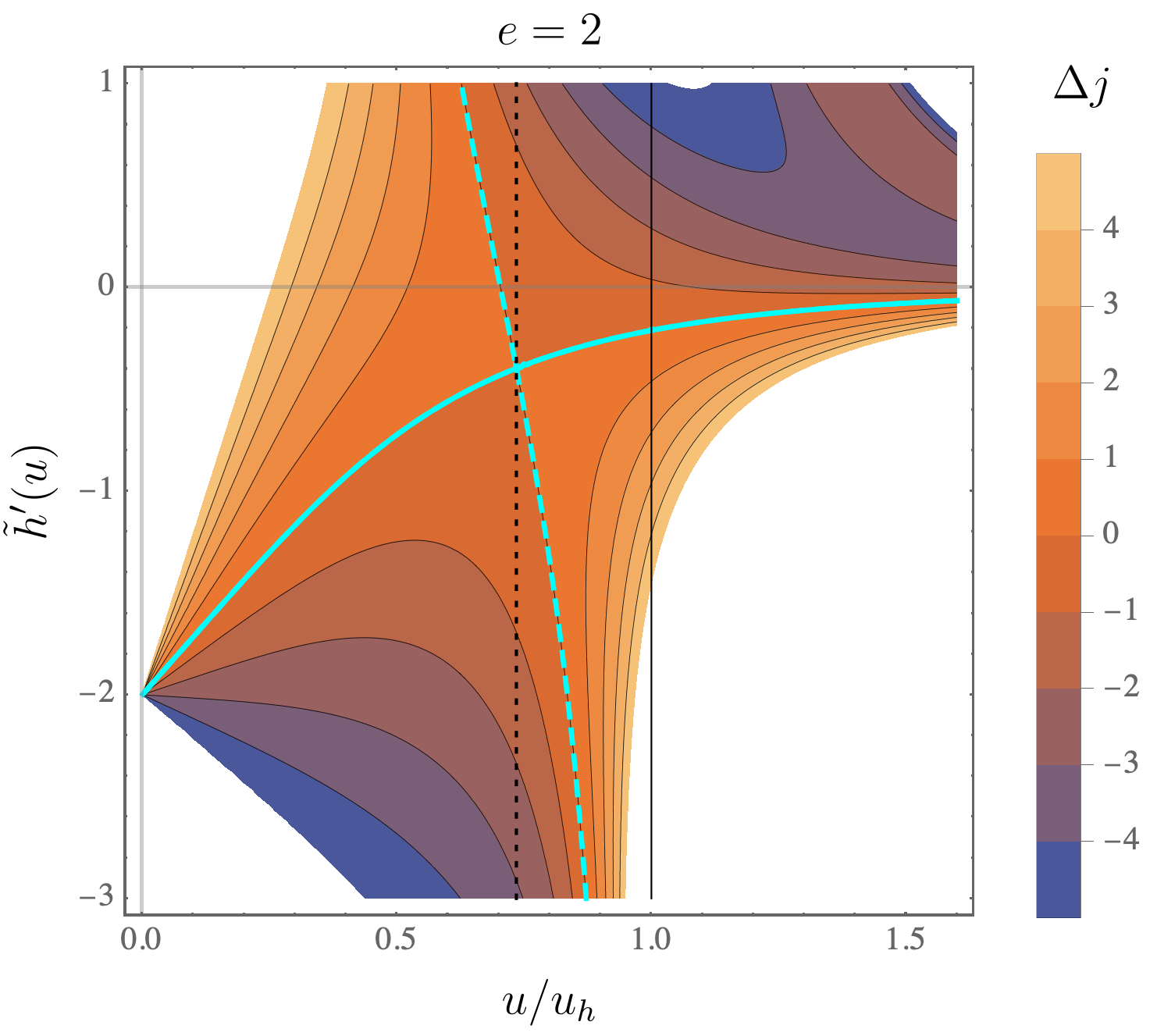}\label{fig:j_contours_F2-a}}
	\subfloat[]{\includegraphics[width=7.0cm]{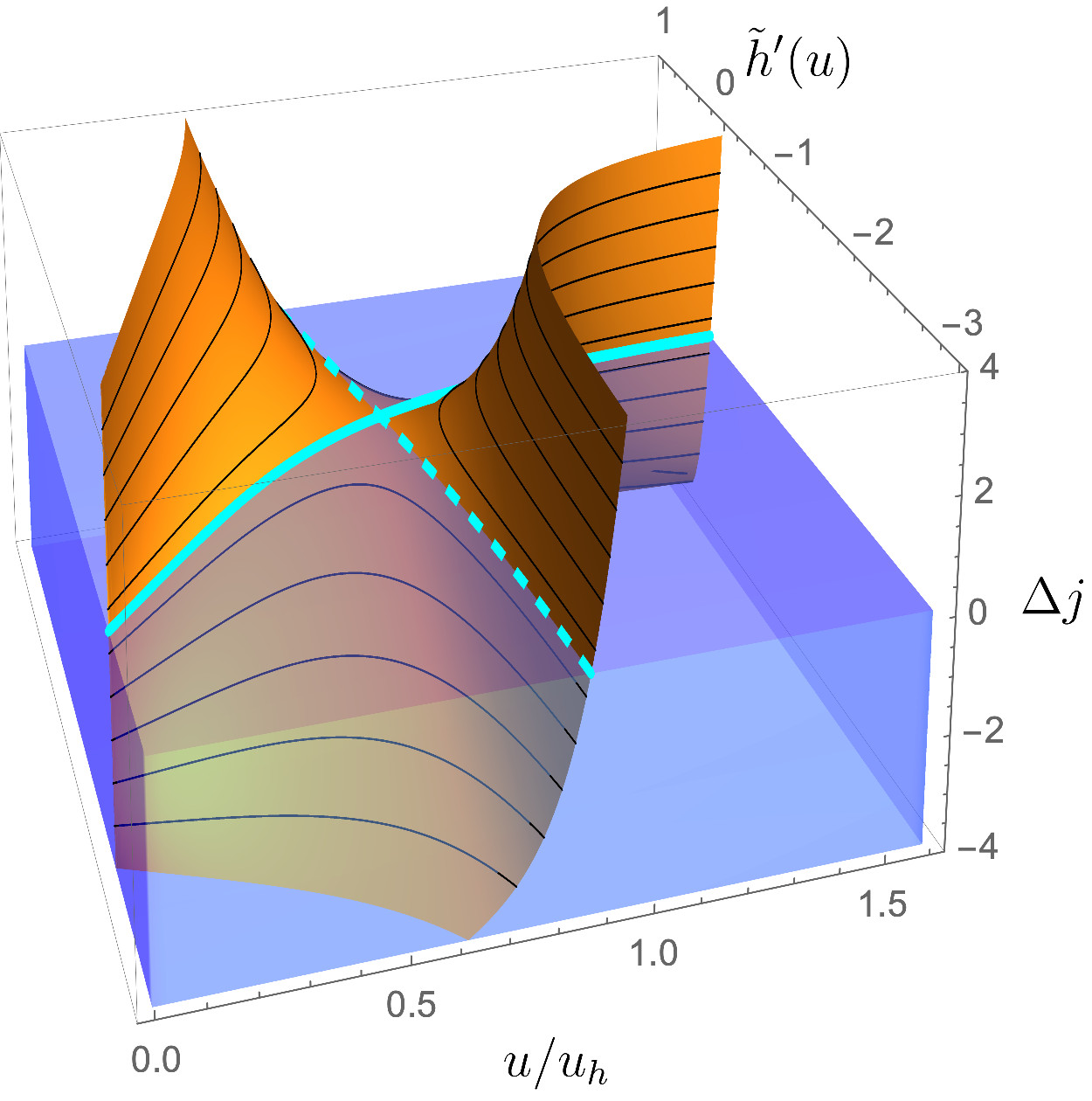}\label{fig:j_contours_F2-b}}
	\caption{
        (a)
        Contour plot of $J(u,h')$ for the $\mathcal{F}^2$ model [Eq.~\eqref{eq:action_F2}]. Each contour represents the solution $h'(u)$ for given $J$. Here, $J$ and $h^{\prime}$ are rescaled as $j=u_h^3 J$ and $\tilde{h}'=u_h^2 h'$, respectively, and $\Delta j = j-j_\text{phys}$. The solid and dashed cyan curves are the contours at $j=j_{\text{phys}}=2.8424$ that go through the patchwork point. The solid cyan curve is the regular physical solution and satisfies the right boundary condition at $u=0$.
        The vertical dotted and solid lines indicate the locations of the saddle point (effective horizon) $u=u_{*}$ and the black hole horizon $u=u_h$, respectively.
        (b)
        Surface plot of $J(u,h')$ for the $\mathcal{F}^2$ model [Eq.~\eqref{eq:action_F2}].
        The orange surface represents $\Delta j$, while the top plane of the translucent blue box represents $\Delta j=0$.
        The solid and dashed cyan curves are the section of these two surfaces. The crossing point of these curves is the saddle point of the orange surface. Here, we set $e=2$ and $c=1$ in these panels.
	}
	\label{fig:j_contours_F2}
\end{figure}

\begin{figure}[htbp!]
	\centering
	\includegraphics[width=8.0cm]{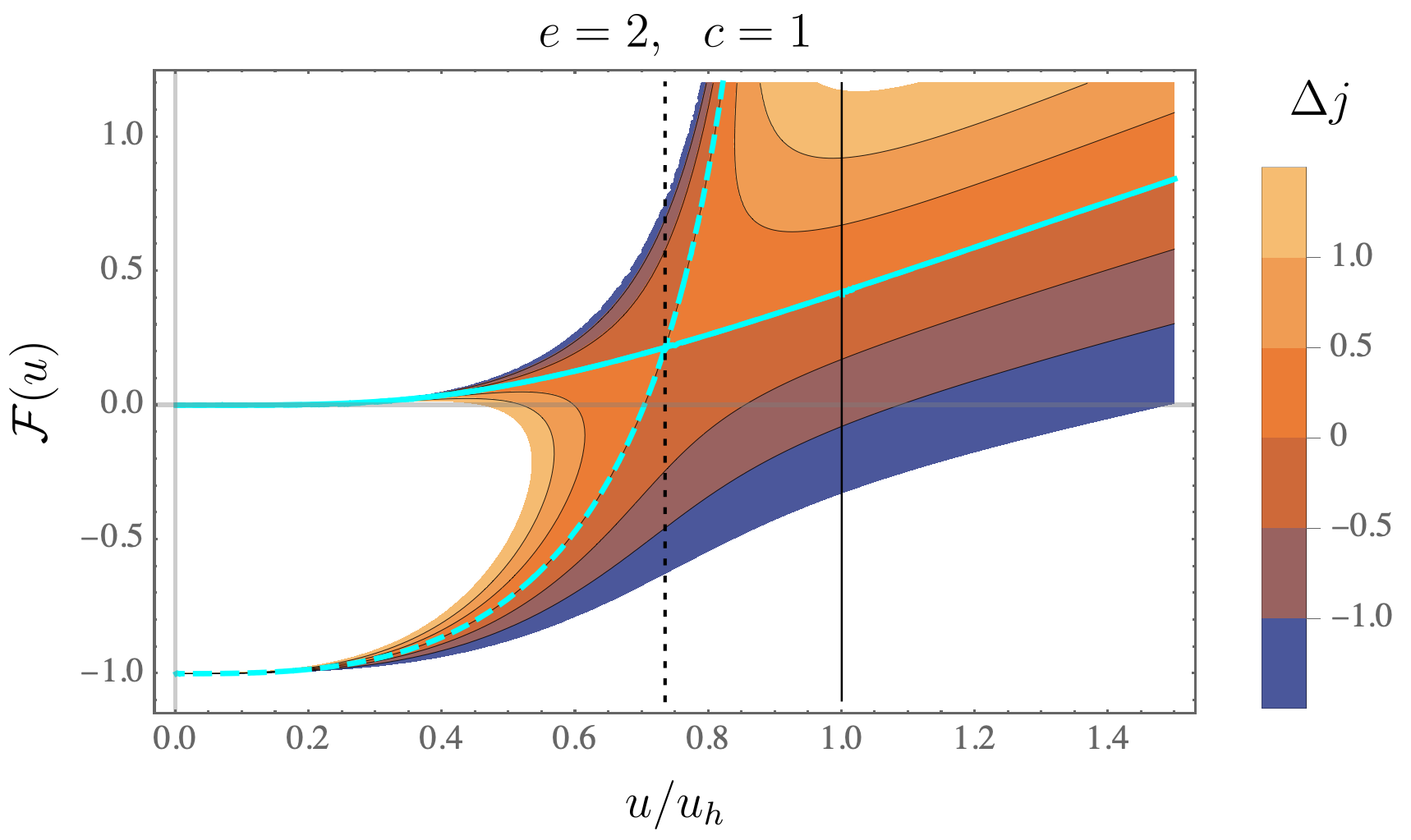}
	\caption{
		$\mathcal{F}(u)$ for the $\mathcal{F}^2$ model [Eq.~\eqref{eq:action_F2}] with respect to the various values of $j$.
        The cyan curves correspond to those in Fig.~\ref{fig:j_contours_F2}.
        $\Delta j$ is defined as $\Delta j=u_{h}^{3}(J-J_{\text{phys}})$.
		The vertical dotted and solid lines indicate the locations of the saddle point $u=u_{*}$ and the black hole horizon $u=u_h$, respectively.
        Here, we set $e=2$ and $c=1$.
	}
	\label{fig:calF-u_contours_F2}
\end{figure}

Let us survey the behavior of the solutions in the entire region of $0\le u \le u_h$.
The numerical results are shown in Fig.~\ref{fig:j_contours_F2}.
In Fig.~\ref{fig:j_contours_F2}, $h'(u)$ as a function of $u$ is given for various values of $j$,
for $e=2$ and $c=1$. We find that the patchwork occurs when $j=j_{\text{phys}}=2.8424$. The solutions with $j=j_{\text{phys}}$ are indicated by the cyan solid curve and cyan dotted curve in Fig.~\ref{fig:j_contours_F2} and in Fig.~\ref{fig:calF-u_contours_F2}. 
In Fig.~\ref{fig:calF-u_contours_F2}, we find that ${\cal F}(u)$ is regular in the entire region of the bulk spacetime for the cyan solid curve. Furthermore, it shows the desired behavior (\ref{eq:hPrime_at_zero-1}) at $u=0$ as is indicated in Fig.~\ref{fig:j_contours_F2}. Therefore, the cyan solid curve is the physical solution.
We also find that $h'(u)$ for the physical solution is finite at $u=u_h$ whereas that for the cyan dotted curve goes to $-\infty$ (see Fig.~\ref{fig:j_contours_F2}(a)). Therefore, we can identify that (\ref{eq:hPrime_at_uh-1}) and hence (\ref{eq:scF(uh)_F2-1}) correspond to the physical solution when $j$ is chosen to $j=j_{\text{phys}}$ for given $e$.

The location of the patchwork (we call it the patchwork point) is the saddle point of the ``landform'' in Fig.~\ref{fig:j_contours_F2}(b) if we regard $j=u_h^3 J(u,h'(u))$ as the ``height.'' 
The location of the saddle point $(u_*, h'(u_*))$ is given by the patchwork condition.
Eqs.~(\ref{eq:patchJ-1}) and (\ref{eq:patchJ-2}) for the present case are explicitly given as
\begin{gather}
	c \left(2 e^2-6 e (\tilde{u}_*^4-1) \y +3 (\tilde{u}_*^4-1)^2 \y^2\right)
	+ 2 - \frac{2}{\tilde{u}_*^4} = 0,\\
	6 \y \left(c e^2+1\right) + 
	3 c e (3-7 \tilde{u}_*^4) \y^2 + 
	c (11 \tilde{u}_*^4-3) (\tilde{u}_*^4-1) \y^3 +
	 \frac{2 e+2 \y}{\tilde{u}_*^4} = 0,
\end{gather}
respectively, where $\x \equiv u_{*}/u_h$ and $\y = u_h^2 h'(u_{*})$.
These equations have three different solutions for $\x^4$ obtained as
\begin{eqnarray}
   \x^4&=& \frac{2}{2-c e^2}, \label{eq:F2first_ustar}\\
   \x^4&=& \frac{3 c e^2+8 +\sqrt{9 c^2 e^4+72 c e^2+16}}{2(2 - c e^2)},\label{eq:F2second_ustar}\\
   \x^4&=& \frac{3 c e^2+8 -\sqrt{9 c^2 e^4+72 c e^2+16}}{2(2 - c e^2)}.\label{eq:F2third_ustar}
\end{eqnarray}
Meanwhile, the equations admit two choices for $\y$ in terms of $\x$
\begin{equation}
    \y = - \frac{
        3 c e \x^2 \pm \sqrt{3 c (2 + (c e^2 - 2)\x^4)}
    }{3 c \x^2 (1 - \x^4)}.
    \label{eq:hp_F2}
\end{equation}
There are naively six choices for the set of $\x^4$ and $\y$, but there are actually five choices because $\y$ is degenerate when we choose (\ref{eq:F2first_ustar}).
We can determine the physical choice by imposing that the results reduces to those in the linear theory under taking $e\to0$, i.e., $\lim_{e\to0} \x = 1$ and $\lim_{e\to0}\y = 0$.
In the current case, only the choice of eq.~(\ref{eq:F2third_ustar}) and the positive sign for the double sign in (\ref{eq:hp_F2}) agrees with the desired behavior.
For this reason, we choose (\ref{eq:F2third_ustar}) then $\x$ is given by
\begin{equation}
	\x = \left(
		\frac{
            3 c e^2 + 8
            - \sqrt{9 c^2 e^4+72 c e^2+16}
        }{2(2 - c e^2)}
	\right)^{1/4}.
	\label{eq:uStar_F2}
\end{equation}
The current density is obtained by substituting these results into Eq.~(\ref{eq:eom_F2}) evaluated at $u=u_*$:
\begin{equation}
	j = \frac{\left(\x^4 \left(c e^2-2\right)+2\right)^{3/2}}{3 \sqrt{3} \sqrt{c} \x^3 \left(1 - \x^4\right)}.
	\label{eq:J_F2}
\end{equation}
Our choice correctly leads the result of $j>0$ for $e>0$, which corresponds to choosing the positive sign for the double sign in (\ref{eq:hp_F2}).
The location where patchwork occurs in Fig.~\ref{fig:j_contours_F2} and the value of $j_{\text{phys}}$ we numerically found are given by substituting $e=2$ and $c=1$ to the above equations.

Let us check the validity of our computations by comparing them with the results from the DBI theory (\ref{eq:DBI_Lagrangian}).
The action of the DBI theory can be expanded as
\begin{equation}
\begin{aligned}
    S =& \tau_7 \int \dd[8]{\xi} \sqrt{-g}\Big[
        - 1 - \frac{1}{4} F_{ab}F^{ab}\\
        &- \frac{1}{32}\left(
            (F_{ab}F^{ab})^2 - 4 F_{ab}F^{bc}F_{cd}F^{da}
        \right)
    \Big] + \cdots
\end{aligned}
    \label{eq:DBI_expansion}
\end{equation}
If we employ the ansatz (\ref{eq:ansatz_EF}), the each term contributes as
\begin{align}
    F_{ab}F^{ab} =& 
    2 h'(u) g^{xx} (-2 g^{\tau u} E + g^{uu} h'(u)),\\
    F_{ab}F^{bc}F_{cd}F^{da} =&
    2 h'(u)^2 (g^{xx})^2 (- 2 g^{\tau u}E  + g^{uu} h'(u))^2,
\end{align}
respectively.
Dropping the constant term, the Lagrangian density of (\ref{eq:DBI_expansion}) is written as 
\begin{equation}
\begin{aligned}
    {\cal L}=&
    \frac{\mathcal{N}}{2}\sqrt{g_{\tau u}^2 g_{xx}}\Big[
		- 2
		\left(
			- 2 g^{\tau u} E
			+ g^{uu} h'(u)
		\right) F_{u x}\\
		&+
		\frac{1}{4}g^{xx}\left(
			- 2 g^{\tau u} E
			+ g^{uu} h'(u)
		\right)^2 h'(u)^2
	\Big]
\end{aligned}
\end{equation}
that agrees with Eq.~(\ref{eq:Lagrangian_F2}) with $c=1$.\footnote{Note that this equivalence does not hold if we include the fluctuations around the background solutions. See, for the details, Sec.~\ref{sec:effective-temp}.} Therefore, our results for $c=1$ need to agree with those of the DBI theory to the sub-leading order of the expansion with respect to the electric field $e$.
Eqs.~(\ref{eq:uStar_F2}) and (\ref{eq:J_F2}) are expanded as
\begin{equation}
	\x = 1 - \frac{1}{4} c e^2
	+ \frac{11}{32} c^2 e^4 - \frac{99}{128} c^3 e^6+ \order{e^7},
\end{equation}
and
\begin{equation}
	j = e + \frac{1}{4} c e^3  - \frac{3}{16} c^2 e^5 + \order{e^6},
	\label{eq:current_F2}
\end{equation}
respectively. Indeed, they agree with (\ref{eq:ustar_DBI_expanded}) and (\ref{eq:current_DBI_expanded}) to the sub-leading order of the small-$e$ expansion when $c=1$.

\subsection{$\mathcal{F}^3$ model}\label{subsec:F3-model}
Let us add an $\mathcal{F}^3$ term to the action of the $\mathcal{F}^2$ model with $c=1$:
\begin{equation}
	S = \mathcal{N} \int\dd[4]{x}\dd{u} \sqrt{-g}\left[
		\mathcal{F} + \frac{1}{2} \mathcal{F}^2
		+ b \frac{1}{2}\mathcal{F}^3
	\right],
 \label{eq:action-F3}
\end{equation}
where $b$ is a real constant.
We will refer to the model as $\mathcal{F}^3$ model.
If we set $b=1$, this action equals the truncated DBI action to the 6th order of $F_{\mu\nu}$ with the ansatz of (\ref{eq:ansatz_EF}) (when we do not consider the fluctuations).
We employ the ingoing Eddington-Finkelstein coordinates.
The equation of motion for $A_x$ is explicitly given by
\begin{equation}
\begin{aligned}
	J =& - \frac{1}{\mathcal{N}}\pdv{\mathcal{L}}{A_x'(u)}\\
	=& \sqrt{g_{\tau u}^2 g_{xx}}(- E g^{\tau u} + g^{uu} h'(u))
	\left(1 + \mathcal{F} + \frac{3}{2}b \mathcal{F}^2\right)\\
	\equiv& J(u,h'),
\end{aligned}
\end{equation}
within the ansatz (\ref{eq:ansatz_EF}).
In this case, the equation of motion is a quintic equation of $h'(u)$.
Let us examine the behavior of the solutions numerically since there is no algebraic formula for the roots of the quintic equation.

\begin{figure}[htbp!]
	\centering
	\subfloat[]{\includegraphics[width=8.0cm]{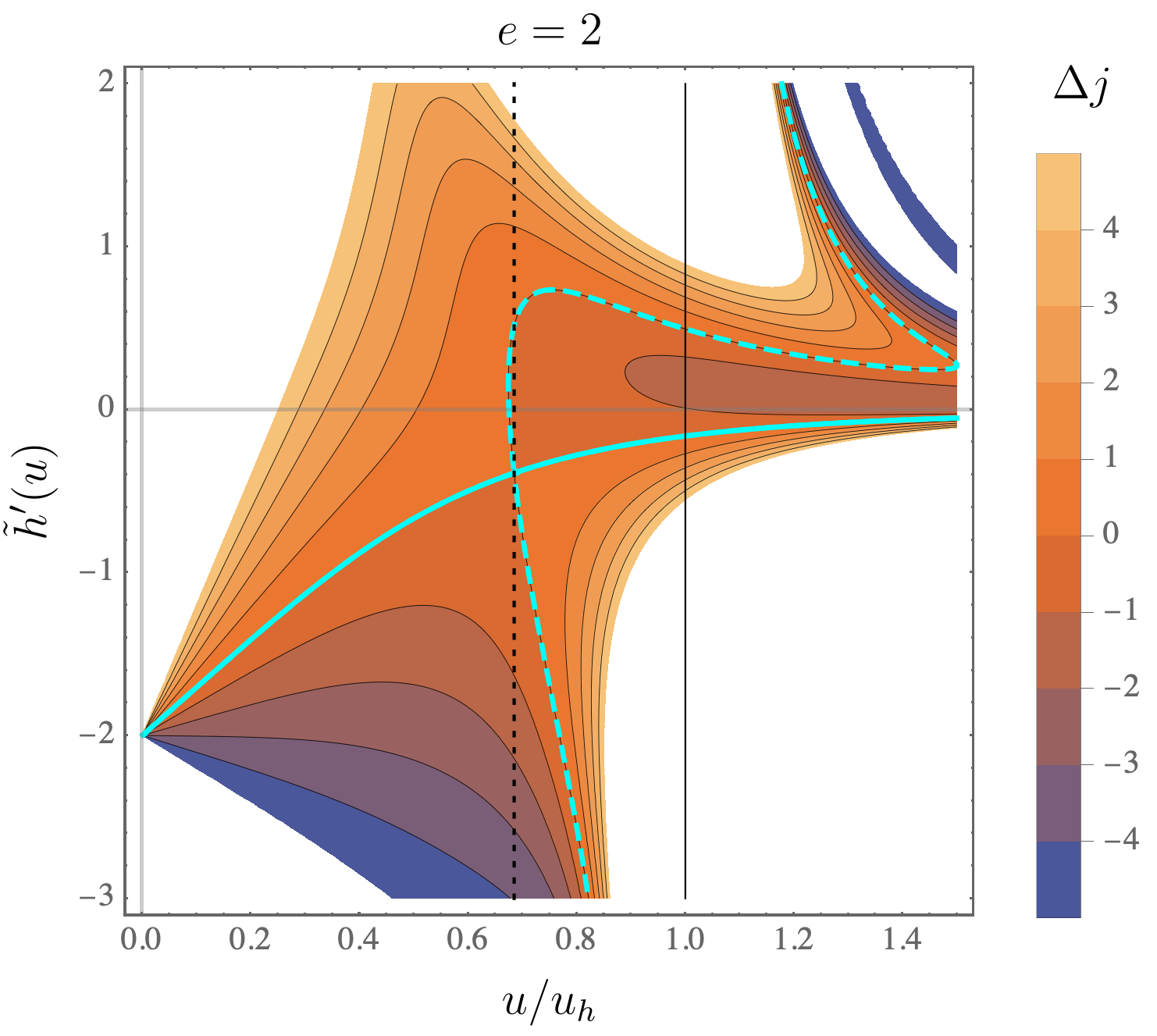}\label{fig:j_contours_F3-a}}
	\subfloat[]{\includegraphics[width=7.0cm]{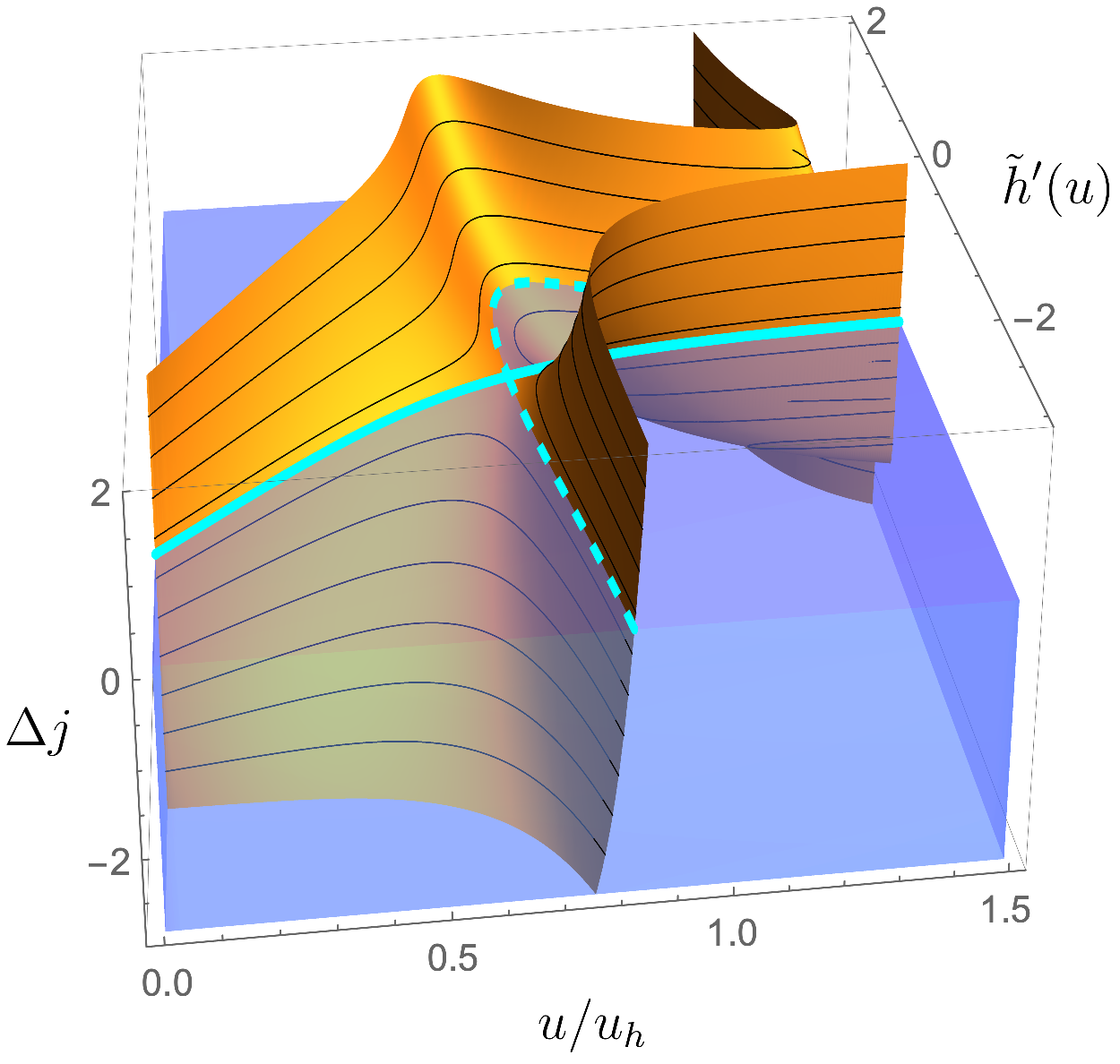}\label{fig:j_contours_F3-b}}
	\caption{
        (a)
        Contour plot of $J(u,h')$ for the $\mathcal{F}^3$ model [Eq.~\eqref{eq:action-F3}]. Each contour represents the solution $h'(u)$ for given $J$. Here, $J$ and $h^{\prime}$ are rescaled as $j=u_h^3 J$ and $\tilde{h}'=u_h^2 h'$, respectively, and $\Delta j = j-j_\text{phys}$. The solid and dashed cyan curves are the contours at $j=j_{\text{phys}}=2.9621$ that go through the patchwork point. The solid cyan curve is the regular physical solution and satisfies the right boundary condition at $u=0$.
        The vertical dotted and solid lines indicate the locations of the saddle point (effective horizon) $u=u_{*}$ and the black hole horizon $u=u_h$, respectively.
        (b)
        Surface plot of $J(u,h')$ for the $\mathcal{F}^3$ model [Eq.~\eqref{eq:action-F3}].
        The orange surface represents $\Delta j$, while the top plane of the translucent blue box represents $\Delta j=0$.
        The solid and dashed cyan curves are the section of these two surfaces. The crossing point of these curves is the saddle point of the orange surface. Here, we set $e=2$ and $b=1$ in these panels.
	}
	\label{fig:j_contours_F3}
\end{figure}
Let us consider the case where $b=1$ and $e=2$, for example. We find the patchwork condition formally gives four different values of $u_{*}^{4}$, but one is imaginary and two are negative. Only the physical value of $u_{*}$ we obtain is $u_{*}=0.68475 u_h$. For this value of $u_{*}$, the patchwork condition gives two values of $h^{\prime}$ which are $h^{\prime}=-0.389243 u_h^{-2}$ and $h^{\prime}=-4.73799 u_h^{-2}$. If we use the former, we obtain $j=2.9621$ while the latter gives $j=-2.9621$. We choose $j>0$. As a result, we obtain $j_{\text{phys}}=2.9621$ with the location of the patchwork $(u_{*}, h^{\prime})=(0.68475 u_h, -0.389243 u_h^{-2})$.
Figure \ref{fig:j_contours_F3} shows the solutions for various values of $j$ with $b=1$ and $e=2$. We find that the patchwork indeed occurs when $j=j_{\text{phys}}$ at $(u_{*}, h^{\prime})=(0.68475 u_h, -0.389243 u_h^{-2})$.

Although we could not obtain exact analytic representations of $u_{*}$ and $h^{\prime}(u_{*})$, we get their approximate representations by employing the small-$e$ approximation. We expand $u_{*}$ and $h^{\prime}(u_{*})$ as $u_{*}=\sum_{k}u_{(k)}e^{k}$, $h^{\prime}=\sum_{k}h^{\prime}_{(k)}e^{k}$, respectively, and substitute them into the patchwork condition. Then we obtain the coefficients $u_{(k)}$ and $h^{\prime}_{(k)}$ order by order. The results with arbitrary $b$ are as follows.
\begin{gather}
	\tilde{u}_* =
	1 - \frac{e^2}{4} + \frac{11 - 6 b}{32} e^4 + \frac{99}{128}(b - 1) e^6 + \order{e^8},
 \\
 \tilde{h}^{\prime}(u_{*})= -\frac{1}{4}e+\frac{3}{32}e^{3}+\frac{9b-26}{128}e^{5} + \order{e^7}.
\end{gather}
We have taken the physical choice of $\tilde{u}_{*}$ and $\tilde{h}'(u_*)$ in the same manner as we did in the $\mathcal{F}^2$ model.
The current density is obtained as
\begin{equation}
	j = e + \frac{e^3}{4} +
	\frac{3}{32} (b-2) e^5 + \order{e^7}.
	\label{eq:current_F3}
\end{equation}
Eq.~(\ref{eq:current_F3}) agrees with Eq.~(\ref{eq:current_DBI_expanded}) in the D3-D7 model to the order of $e^5$ with $b=1$.
If we set $b=0$, the coefficient of $e^5$ in Eq.~(\ref{eq:current_F3}) becomes $(-1)\times 3/16$ which agrees with the coefficient of $e^5$ in Eq.~(\ref{eq:current_F2}) in the $\mathcal{F}^2$ model with $c=1$.
Interestingly, the order of $e^5$ contribution of $j$ vanishes when $b=2$.

\begin{figure}[htbp!]
	\centering
	\includegraphics[width=8.0cm]{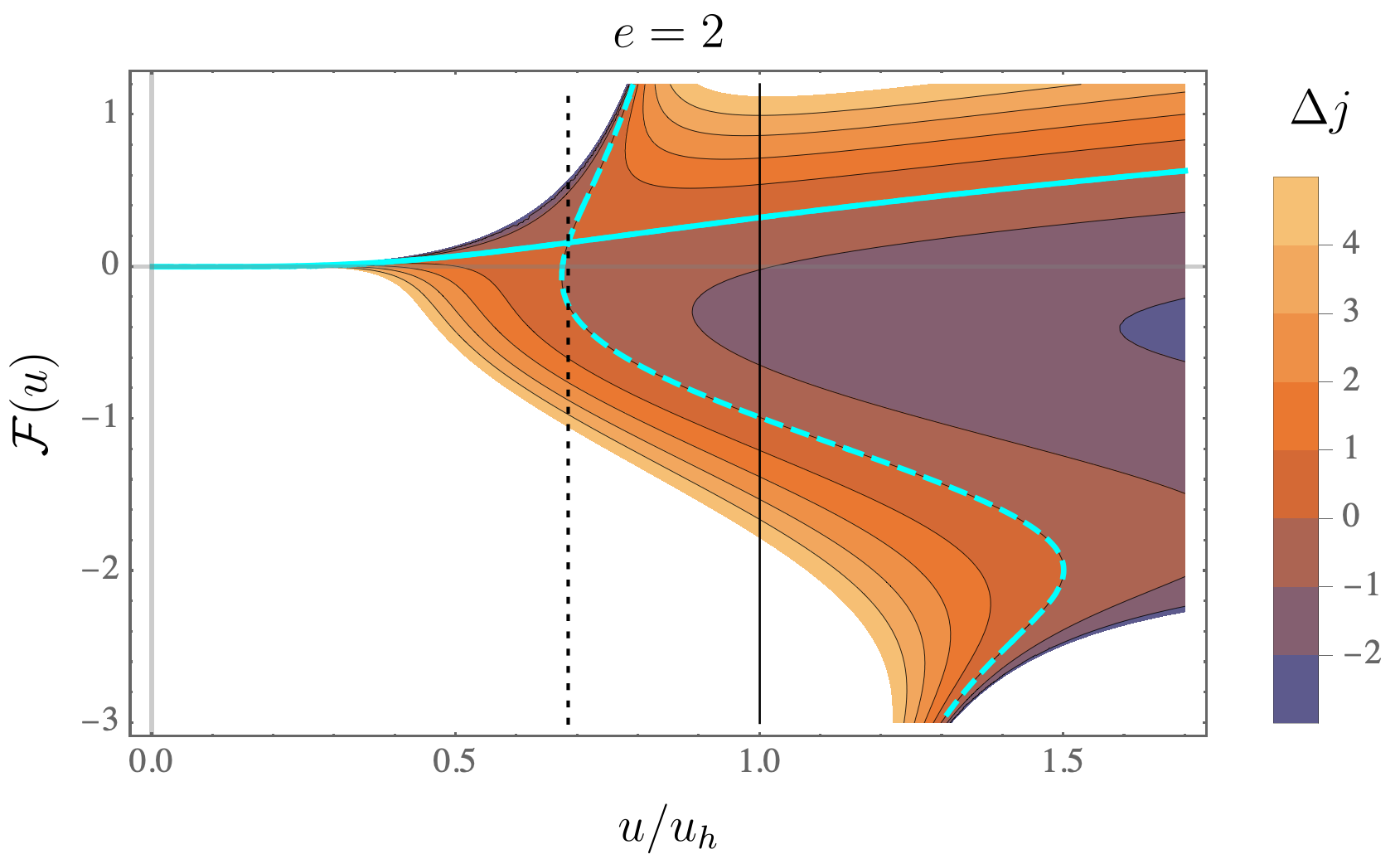}
	\caption{
        $\mathcal{F}(u)$ for the $\mathcal{F}^3$ model [Eq.~\eqref{eq:action-F3}] with respect to the various values of $j$.
        The cyan curves correspond to those in Fig.~\ref{fig:j_contours_F3}.
        $\Delta j$ is defined as $\Delta j=u_{h}^{3}(J-J_{\text{phys}})$.
		The vertical dotted and solid lines indicate the locations of the saddle point $u=u_{*}$ and the black hole horizon $u=u_h$, respectively.
        Here, we set $e=2$ and $b=1$.
	}
	\label{fig:calF-u_contours_F3}
\end{figure}

We have a comment on the regularity of the solutions in this model. Figure \ref{fig:calF-u_contours_F3} shows $\mathcal{F}$ as functions of $u$ for various values of $j$. One finds that there are solutions that satisfy the right boundary condition at $u=0$ and $\mathcal{F}$ is regular for $0\le u \le u_{h}$. They start from $\mathcal{F}=0$ at $u=0$ and penetrate the event horizon of the bulk geometry with negative but finite $\mathcal{F}$ without encountering the ``patchwork.'' In this sense, these solutions are regular outside the event horizon. However, we find that $\partial_{u} \mathcal{F}$ of these solutions diverge at a point of $u>u_{h}$: $(\nabla_\mu \mathcal{F})(\nabla^\mu \mathcal{F})$, in terms of a scalar quantity, is singular inside the event horizon. We exclude these solutions because of the following reason.

Let us consider the effective metric $\gamma_{ab}$ on these solutions. We define the singular point as $u=u_{\text{sing}}$ where $(\nabla_\mu \mathcal{F})(\nabla^\mu \mathcal{F})$ diverges.
Since these solutions reach the singular point without encountering the patchwork phenomenon, $\gamma^{uu}<0$ throughout the region of $0\le u <u_{\text{sing}}$ including the location of the event horizon. We find that other components of the effective metric do not change the sign and $\det \gamma_{ab}>0$ in $0\le u \le u_{\text{sing}}$. Therefore we regard that the singularity at $u_{\text{sing}}$ is not hidden by any (effective) horizon from the viewpoint of the fluctuations governed by the effective metric, and we discard these solutions.\footnote {In general, the speed of propagation of signals can be modified in the theories with higher-derivative terms, and it can even be superluminal. When it is the case, the causal structure has to be discussed based on the fastest mode, but not the null hypersurface of the bulk geometry. See, for example, \cite{Izumi:2014loa} for the discussions in the Gauss-Bonnet gravity.}

It is interesting to refer to the results for the $\mathcal{F}^2$ model at $T=0$ given in Appendix \ref{sec:T=0}. The $\mathcal{F}^2$ model yields singularity at $u=\infty$ in the presence of $J$ and $E$. There is no event horizon in the bulk since $T=0$. However, the singularity is hidden by the effective horizon in this case.

\section{DBI model}\label{sec:DBI}
For comparison, let us consider the model with the DBI action (\ref{eq:DBIaction}).
The Lagrangian density after the integration of the $\mathrm{S}^3$ part in the ingoing Eddington-Finkelstein coordinates in our convention is explicitly given by
\begin{equation}
	\mathcal{L} = - \mathcal{N}
	g_{xx}
	\sqrt{
		-2E h^{\prime}g_{\tau u} +g_{xx}g_{\tau u}^{2}-g_{\tau\tau} {h^{\prime}}^{2} 
	},
\label{eq:DBI_Lagrangian_EF}
\end{equation}
where $\mathcal{N}$ is the same as that in (\ref{eq:DBI_Lagrangian}).

The current density $J$ in $x$-direction is obtained as
\begin{equation}
\begin{aligned}
	J =& - \frac{1}{\mathcal{N}}
	\pdv{\mathcal{L}}{A_x'(u)}\\
	=& \frac{
		g_{xx} g_{\tau\tau}\left( h'(u)+E g_{\tau u}/g_{\tau \tau} \right)
	}{
		\sqrt{-g_{\tau \tau} \left( h'(u)+E g_{\tau u}/g_{\tau \tau} \right)^{2}
  +\left(E^2+g_{\tau \tau}g_{xx}\right) g_{\tau u}^{2}/g_{\tau \tau}}
	}.
\end{aligned}
\label{eq:J_DBI}
\end{equation}
The condition (\ref{eq:patchJ-2}) gives
\begin{eqnarray}
 \left.(E^2+g_{\tau \tau}g_{xx})\right|_{u=u_{*}}=0.
 \label{eq:patchwork2_DBI_EF}
\end{eqnarray}
Note that $u_{*}$ can be determined solely by (\ref{eq:patchwork2_DBI_EF}) for the present case.
$h'(u_{*})$ can be determined by solving the condition corresponding to (\ref{eq:patchJ-1}) together with (\ref{eq:patchwork2_DBI_EF}). However, we do not need to do it in the present model. If we substitute (\ref{eq:patchwork2_DBI_EF}) into (\ref{eq:J_DBI}), $h'(u)+E g_{\tau u}/g_{\tau \tau}$ is cancelled in (\ref{eq:J_DBI}) at $u=u_*$ up to the sign, and we obtain
\begin{equation}
    J=\pm \left.g_{xx}\sqrt{-g_{\tau \tau}}\right|_{u=u_{*}}=\left.\sqrt{g_{xx}}\right|_{u=u_{*}}E,
\end{equation}
where we have chosen the sign in such a way that $J>0$ for $E>0$ in the last equality.\footnote{Similar cancellation of the worldvolume field has been seen in the computations of Refs.~\cite{Hashimoto:2014yza,Ishigaki:2021vyv} and was utilized in \cite{Endo:2023vov} in the presence of finite charge density.} This agrees with what we have obtained in (\ref{eq:cancel-2}). Of course, if we work in the Schwarzschild coordinates, we obtain (\ref{eq:realcond-1}) instead of (\ref{eq:patchwork2_DBI_EF}), and (\ref{eq:cancel-2}) is reproduced.

\begin{figure}[htbp!]
	\centering
	\subfloat[]{\includegraphics[width=8.0cm]{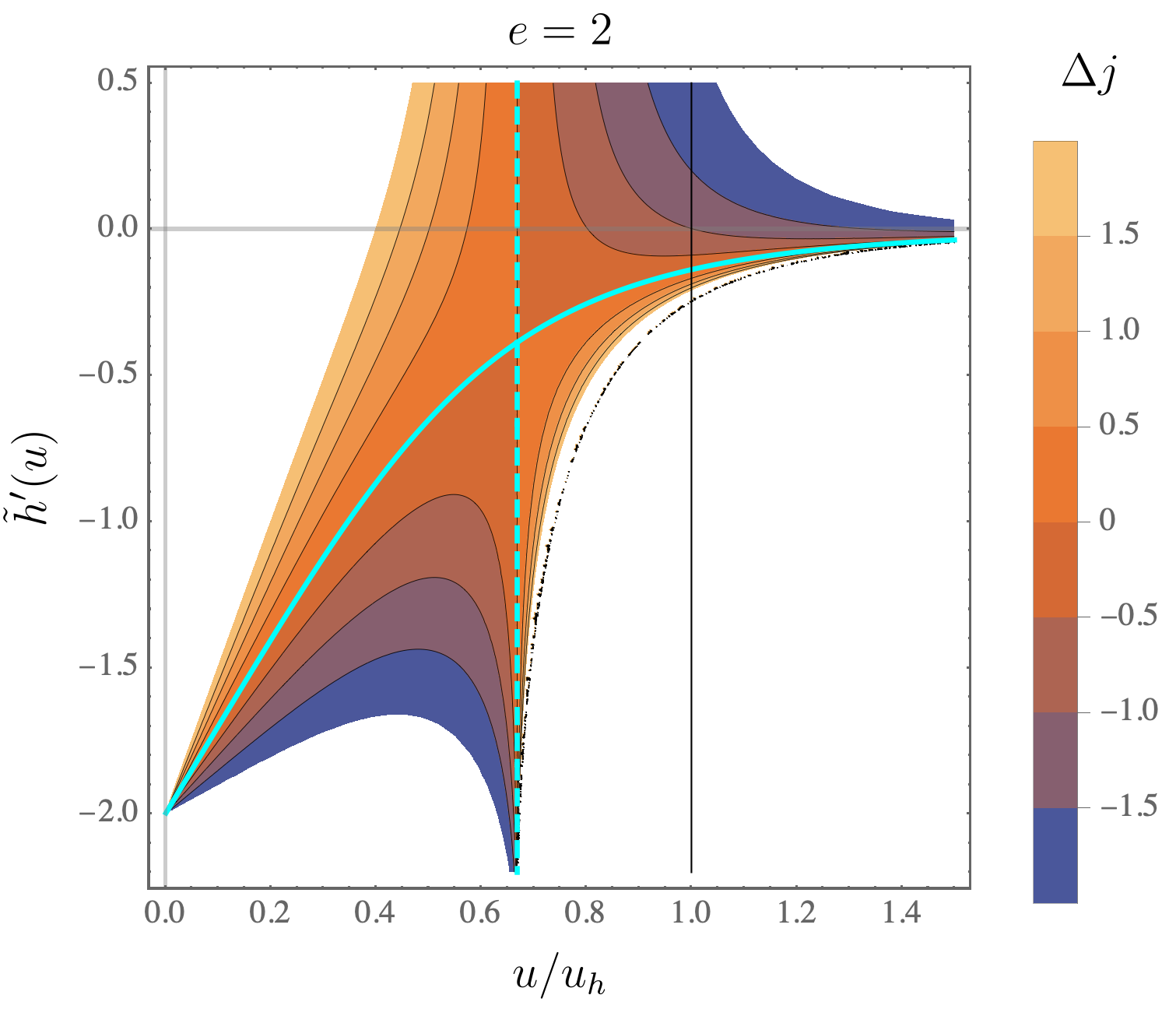}\label{fig:j_contours_DBI-a}}
	\subfloat[]{\includegraphics[width=7.0cm]{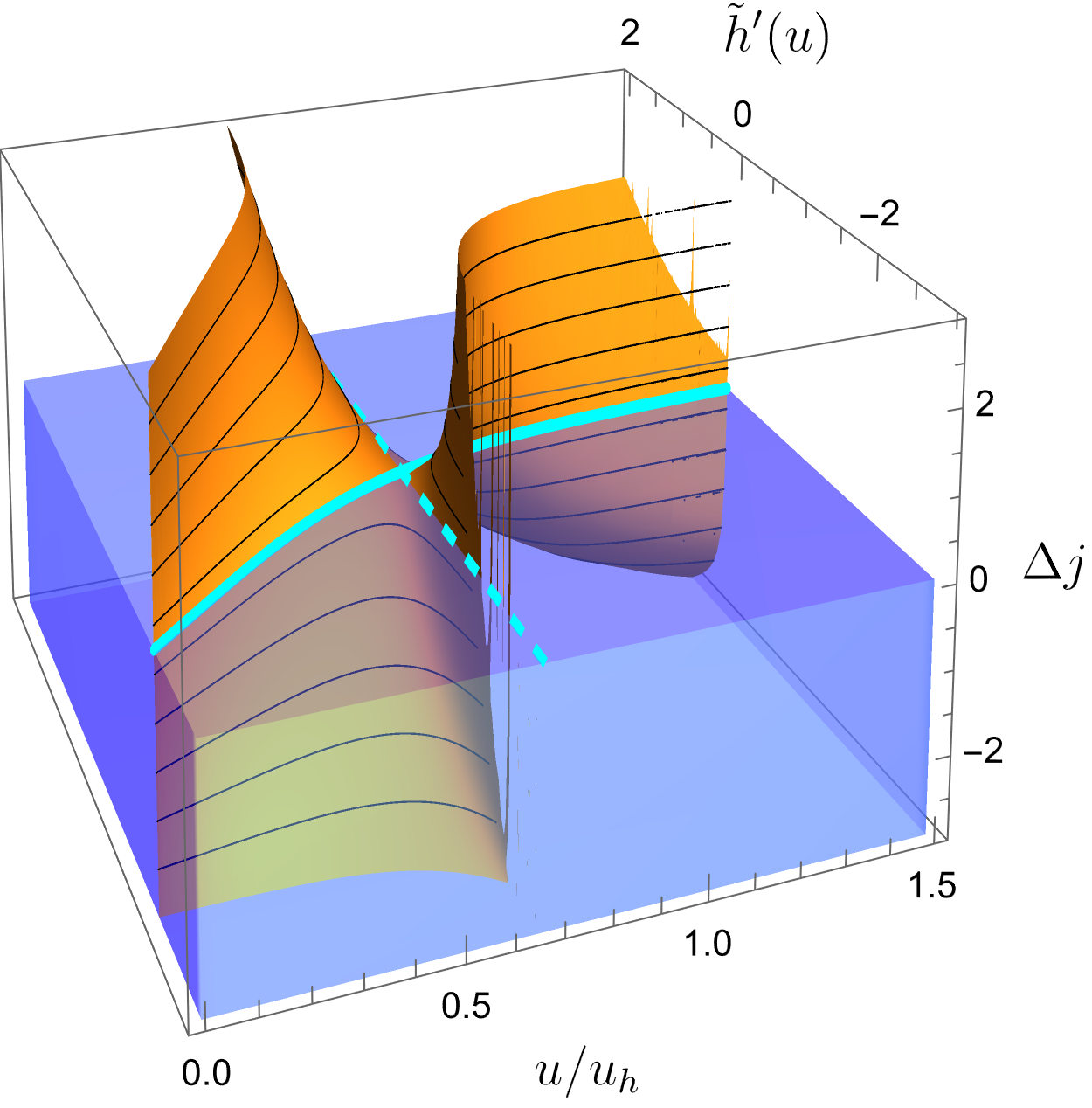}\label{fig:j_contours_DBI-b}}
	\caption{
		(a)
        Contour plot of $J(u,h')$ for the DBI model [Eq.~\eqref{eq:DBIaction}]. Each contour represents the solution $h'(u)$ for given $J$. Here, $J$ and $h^{\prime}$ are rescaled as $j=u_h^3 J$ and $\tilde{h}'=u_h^2 h'$, respectively, and $\Delta j = j-j_\text{phys}$. The solid and dashed cyan curves are the contours at $j=j_{\text{phys}}=2.9907$ that go through the patchwork point. The solid cyan curve is the regular physical solution and satisfies the right boundary condition at $u=0$.
        The vertical solid line indicates the black hole horizon $u=u_h$. Note that $u=u_*$ coincides with the cyan dashed line.
        (b)
        Surface plot of $J(u,h')$ for the DBI model [Eq.~\eqref{eq:DBIaction}].
        The orange surface represents $\Delta j$, while the top plane of the translucent blue box represents $\Delta j=0$.
        The solid and dashed cyan curves are the section of these two surfaces. The crossing point of these curves is the saddle point of the orange surface. Here, we set $e=2$ in these panels.
        }
	\label{fig:j_contours_DBI}
\end{figure}

\begin{figure}[htbp!]
	\centering
	\includegraphics[width=8.0cm]{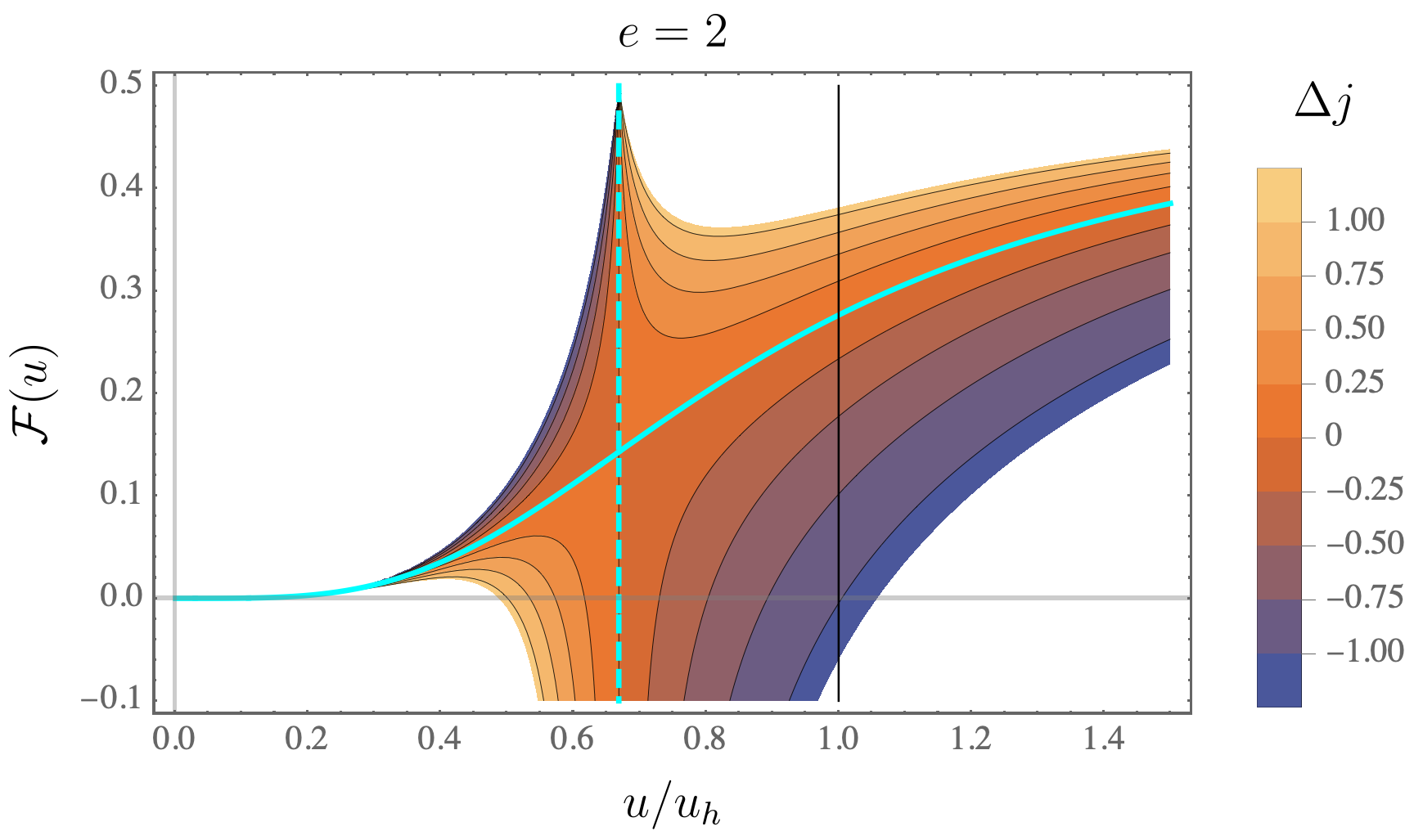}
	\caption{
        $\mathcal{F}(u)$ for the DBI model [Eq.~\eqref{eq:DBIaction}] with respect to the various values of $j$.
        The cyan curves correspond to those in Fig.~\ref{fig:j_contours_DBI}.
        $\Delta j$ is defined as $\Delta j=u_{h}^{3}(J-J_{\text{phys}})$.
		The vertical solid line indicates the locations of the black hole horizon $u=u_h$. Here, we set $e=2$.
	}
	\label{fig:calF-u_DBI}
\end{figure}

Let us see the patchwork indeed occurs for the regular solution, explicitly.
Figure \ref{fig:j_contours_DBI} shows various solutions $h'(u)$ as functions of $u$ for various values of $j$.
The cyan curve corresponds to the solution with $j=j_{\text{phys}}$, where $j_{\text{phys}}$ is the value of $j$ obtained from Eq.~(\ref{eq:current_DBI}). We find that the patchwork occurs when $j=j_{\text{phys}}$. This is only the solution that is smoothly connected from $u=0$ to $u=u_h$.
Figure \ref{fig:calF-u_DBI} shows $\mathcal{F}(u)$
for various $j = j_{\text{phys}} + \Delta j$.
We can see that $\mathcal{F}(u)$ is regular only when $j=j_{\text{phys}}$.

\section{Nonlinear friction coefficient}
\label{sec:NG}
In this section, we apply our computational method to the calculation of the nonlinear friction coefficient. We employ the model where a Nambu-Goto string is dragged in the AdS-Schwarzschild black hole geometry (\ref{eq:bulk_metric}) in $5$-dimensions \cite{Herzog:2006gh,Gubser:2006bz}.%
\footnote{
    We work in the Schwarzschild coordinates in this section.
}
Let us call this model the NG model for short.
The action of the test string is
\begin{equation}
    S_{\text{NG}} = - \frac{1}{2\pi\alpha'}\int\dd[2]{\sigma}\sqrt{-\det h_{\alpha\beta}},
\end{equation}
where $h_{\alpha\beta} = g_{\mu\nu}\partial_{\alpha}x^{\mu}\partial_{\beta}x^{\nu}$ is the induced metric and $\alpha, \beta$ denote indices of worldsheet coordinates.
We take the worldsheet coordinates as $\sigma^{\alpha} = (t, u)$ and employ an ansatz $\vec{x} = (v t + \xi(u), 0, 0)$ for the description of the configuration of the string.
$v$ is the velocity of the endpoint of the string at the boundary of the bulk geometry.
The action becomes
\begin{equation}
    S_{\text{NG}} = - \frac{1}{2\pi\alpha'}\int\frac{\dd{t}\dd{u}}{u^2}
    \sqrt{1 + f(u) x'(t,u)^2 - \frac{\dot{x}(t,u)^2}{f(u)}},
    \label{eq:Nambu-Goto_with_ansatz}
\end{equation}
where the dot represents the differentiation with respect to $t$.
We define the Lagrangian density by $S_{\text{NG}} = \frac{1}{2\pi\alpha'}\int\dd{t}\dd{u} \mathcal{L}_{\text{NG}}$.
The system has a constant of motion for $u$-coordinate which is given by
\begin{equation}
    \pi_{\xi} \equiv
    - \pdv{\mathcal{L}_{\text{NG}}}{x'(t,u)}
    = \frac{f(u) \xi'(u)}{u^2 \sqrt{1 + f(u) \xi'(u)^2 - v^2/f(u)}}.
    \label{eq:pi_xi}
\end{equation}
Now the patchwork condition is given by the following two simultaneous equations:
\begin{eqnarray}
\left.\pdv{\pi_{\xi}(u,\xi')}{u}\right|_{u_*} &=&0,
\label{eq:patchwork-pi-1}\\
\left.\pdv{\pi_{\xi}(u,\xi')}{\xi'(u)}\right|_{u_*} &=&0. 
\label{eq:patchwork-pi-2}
\end{eqnarray}
The condition (\ref{eq:patchwork-pi-2}) is explicitly given by
\begin{equation}
    0=\left.\pdv{\pi_{\xi}}{\xi'(u)}\right|_{u_*} =
    \frac{f(u_*) - v^2}{
        u_*^2\left[
            1 + f(u_*) \xi'(u_*)^2 - v^2/f(u_*)
        \right]^{3/2}
    }.
\label{eq:pw_pi_2}    
\end{equation}
Note that (\ref{eq:pw_pi_2}), which is $f(u_*)-v^2 = 0$, solely gives $u_{*}$ for this case as we have seen for the DBI theory. 
The locus of the effective horizon is given by $u_*^4/u_h^4 = 1-v^2$.
We can also obtain $\xi(u_*)$ by solving (\ref{eq:patchwork-pi-1}) together with (\ref{eq:patchwork-pi-2}), but we do not need to do it for the present case. 
Substituting $f(u_*)-v^2 = 0$ into (\ref{eq:pi_xi}), $\xi'$ is cancelled up to the sign in (\ref{eq:pi_xi}), and we obtain $\pi_{\xi}=\pm u_{*}^{-2} v$ which is explicitly given as
\begin{equation}
    \pi_{\xi} = u_h^{-2} \frac{v}{\sqrt{1-v^2}},
    \label{eq:pixi}
\end{equation}
where we have chosen the sign so that $\pi_{\xi}>0$ for $v>0$.
This quantity corresponds to the drag force acting on the system
of velocity $v$.
The result completely agrees with those in Refs.~\cite{Herzog:2006gh,Gubser:2006bz}.
Note that the factor of $2\pi\alpha'$ is absent from our definition of $\pi_\xi$.

Now, we consider a truncated action that is obtained by the derivative expansion of (\ref{eq:Nambu-Goto_with_ansatz}) to the order of $(\partial_{\alpha} x)^4$. We call this model the $(\partial x)^4$-truncated model.
The relevant part of the Lagrangian density is given by
\begin{equation}
    \mathcal{L}_{\text{NG}}^{(4)}
    =
    - \frac{1}{u^2}\left[
        \frac{1}{2}\left(
            f {x'}^2 - \frac{\dot{x}^2}{f}
        \right)
        - \frac{1}{8}\left(
          f {x'}^2 - \frac{\dot{x}^2}{f}
        \right)^2
    \right].
    \label{eq:delx4-model}
\end{equation}
The constant of motion is given by
\begin{equation}
    \pi_{\xi} \equiv - \pdv{\mathcal{L}_{\text{NG}}^{(4)}}{x'}
    =
    \frac{\xi'(u)}{2 u^2}\left(
        2 f(u) + v^2 - f(u)^2 \xi'(u)^2
    \right).
    \label{eq:pi_truncated}
\end{equation}
We obtain the patchwork condition as
\begin{gather}
    \left.\pdv{\pi_{\xi}}{\xi'}\right|_{u_*}
    = \frac{1}{2 u_*^2}\left(
        v^2 + 2 f - 3 f^2 \xi'(u_*)^2
    \right) = 0,\\
    \begin{aligned}
    \left.\pdv{\pi_{\xi}}{u}\right|_{u_*}
    =& \frac{\xi'(u_*)}{u_*^3}\left(
        f (f - u_* f') \xi'(u_*)^2
        + u_* f' - 2 f - v^2
    \right)\\
    =& 0.
    \end{aligned}
\end{gather}
In this case, we need to solve both of the above equations to obtain $u_*$ and $\xi'(u_*)$.
The suitable solution is
\begin{equation}
    \frac{u_*^4}{u_h^4} =
    \frac{v^2 + 2}{3 v^2 + 2},\quad
    \xi'(u_*) =
    \frac{\sqrt{3 v^4 + 8 v^2 + 4}}{2 v}.
\end{equation}
Evaluating (\ref{eq:pi_truncated}) at $u=u_*$ with these results, we obtain
\begin{equation}
    \pi_{\xi} = u_h^{-2}\left( v + \frac{v^3}{2} \right).
    \label{eq:pixi_truncated}
\end{equation}
The result agrees with (\ref{eq:pixi}) to the order of $v^3$.

\begin{figure}[htbp!]
	\centering
	\subfloat[]{\includegraphics[width=8.0cm]{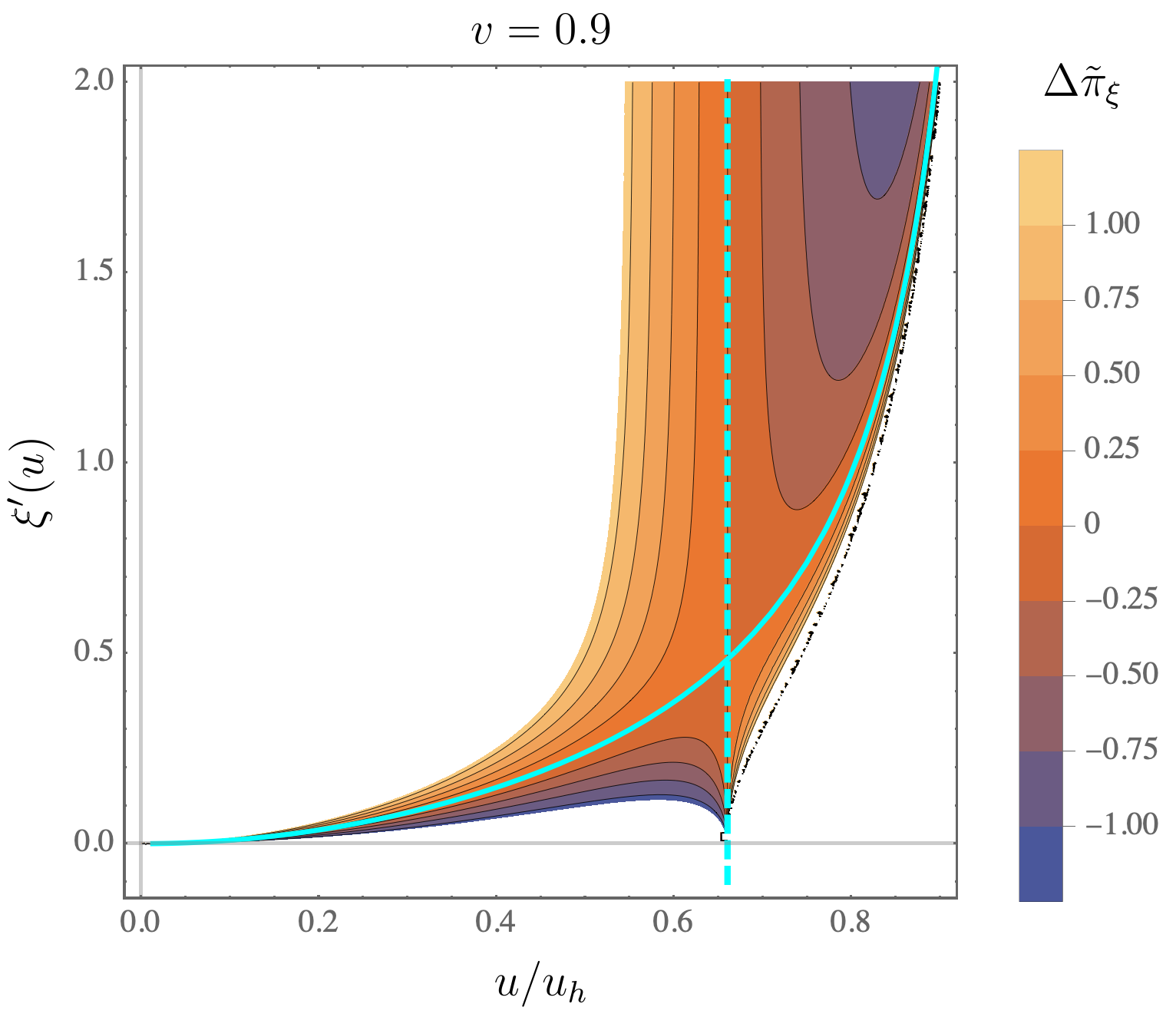}\label{fig:patchwork_NG-action-a}}
 	\subfloat[]{\includegraphics[width=7.5cm]{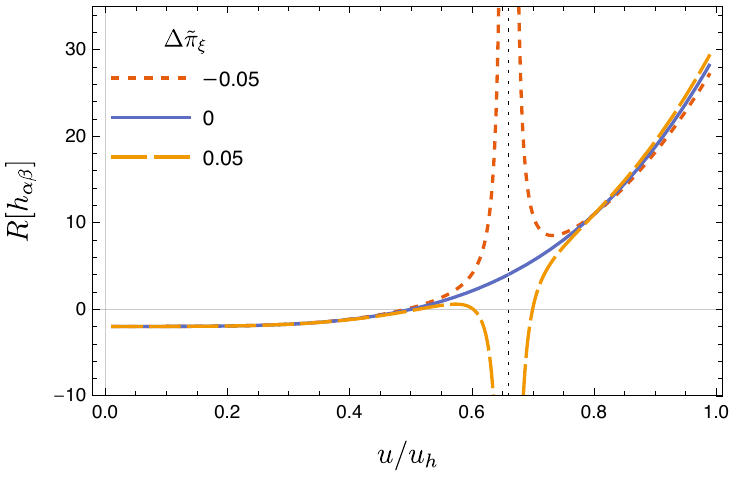}\label{fig:patchwork_NG-action-b}}
	\caption{
        Description of the patchwork phenomenon for the NG model. 
        (a)
        $\xi'(u)$ vs. $u/u_h$ for various values of $\Delta\tilde{\pi}_\xi \equiv u_h^2(\pi_{\xi} - \pi_{\text{phys}})$.
        The vertical dashed cyan line is located at $u=u_*$.
        The solid cyan curve is the regular physical solution of $\xi'(u)$.
        (b)
        Scalar curvature constructed from $h_{\alpha\beta}$ as a function of $u/u_h$ for various $\Delta\tilde{\pi}_{\xi}$.
        The vertical dotted line shows the location of $u=u_*$.
        The curve is smoothly connected at $u=u_*$ only when $\Delta\tilde{\pi}_{\xi}=0$, i.e.\ $\pi_{\xi} = \pi_{\text{phys}}$. In these panels, we set $v=0.9$.
	}
	\label{fig:patchwork_NG-action}
\end{figure}

\begin{figure*}[htbp!]
	\centering
 	\subfloat[]{\includegraphics[width=8.0cm]{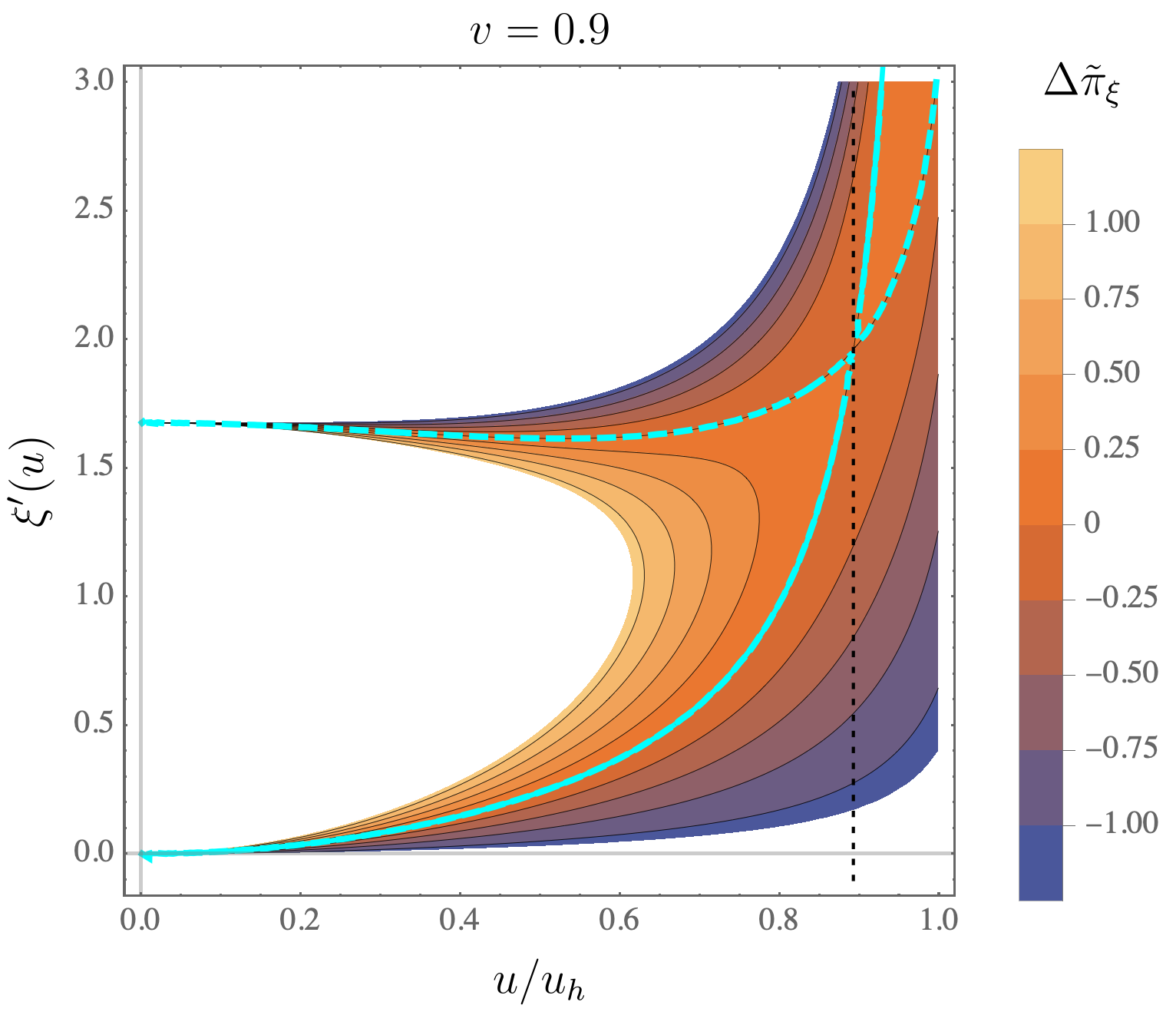}\label{fig:patchwork_dx4-model-a}}\\
	\subfloat[]{\includegraphics[width=7.5cm]{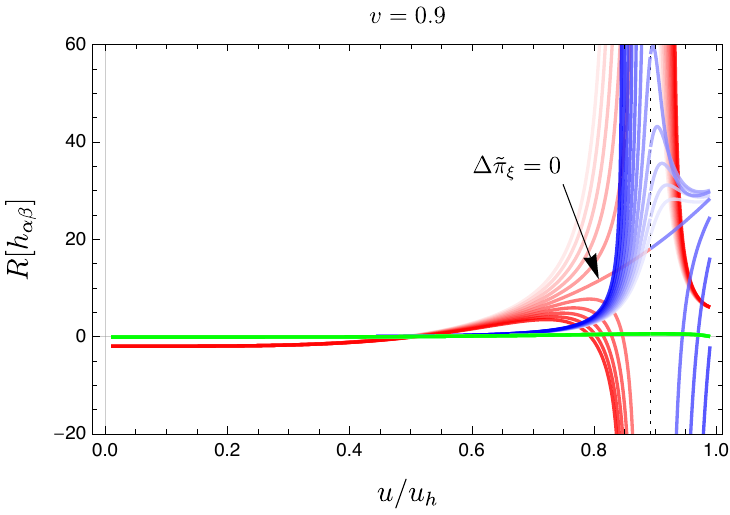}\label{fig:patchwork_dx4-model-b}}
    \subfloat[]{\includegraphics[width=7.5cm]{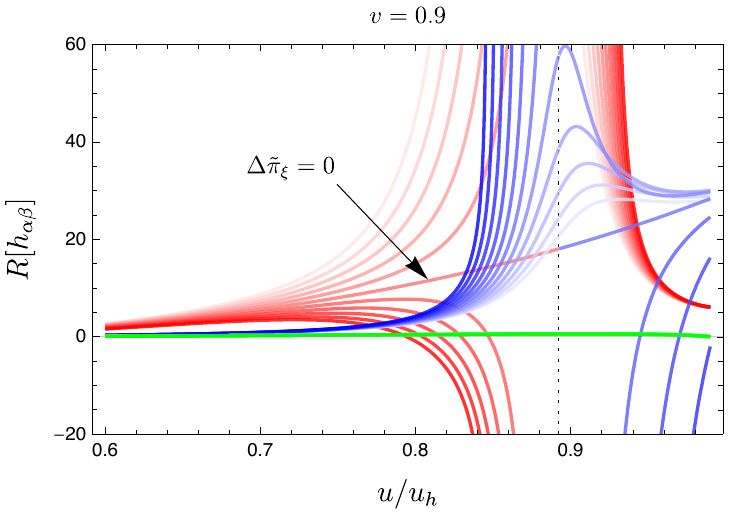}\label{fig:patchwork_dx4-model-c}}
	\caption[]{
        Description of the patchwork phenomenon for the $(\partial x)^4$-truncated model.
        (a)
        $\xi'(u)$ vs. $u/u_h$ for various values of $\Delta\tilde{\pi}_\xi \equiv u_h^2(\pi_{\xi} - \pi_{\text{phys}})$.
        The vertical dotted line indicates the location of $u=u_*$.
        The solid cyan curve is the regular physical solution of $\xi'(u)$.
        (b)
        Scalar curvature as a function of $u/u_h$ for various $\pi_{\xi}$.
        The red, blue, and green families of curves correspond to the three roots $\xi'(u)$ of Eq.~(\ref{eq:pi_truncated}).
        The brightness of each curve represents the value of $\Delta\tilde{\pi}_{\xi}$ in such a way that the brighter curves have lower $\Delta\tilde{\pi}_{\xi}$.
        The range of $\Delta\tilde{\pi}_{\xi}$ is $[-0.5, 0.5]$.
        The vertical dotted line indicates the location of $u=u_*$.
        (c)
        The enlarged view of the bottom left panel around $u=u_*$.
        In all the panels, we set $v=0.9$.
	}
	\label{fig:patchwork_dx4-model}
\end{figure*}

We expect that the patchwork condition (\ref{eq:patchwork-pi-1}) and (\ref{eq:patchwork-pi-2}) realize a regular configuration.
Figure \ref{fig:patchwork_NG-action}(a) for the NG model and Fig.~\ref{fig:patchwork_dx4-model}(a) for the $(\partial x)^4$-truncated model show profiles of $\xi'(u)$ for several values of $\Delta \tilde{\pi}_{\xi}\equiv u_h^2(\pi_{\xi} - \pi_{\text{phys}})$, where $\pi_{\text{phys}}$ is $\pi_{\xi}$ that satisfies the patchwork condition.
$\pi_{\text{phys}}$ is explicitly given by (\ref{eq:pixi}) and (\ref{eq:pixi_truncated}) for the NG model and the $(\partial x)^4$-truncated model, respectively.
The cyan curves that realize the patchwork phenomenon corresponds to the solutions with $\Delta \tilde{\pi}_{\xi}=0$.
$\xi'(u)$, which is not a gauge-invariant scalar quantity, is divergent at $u=u_h$ because of the artifact of the coordinate singularity of the Schwarzschild-AdS$_5$ geometry.

An appropriate quantity to see the regularity is the scalar curvature $R$ constructed by the induced metric $h_{\alpha\beta}$ \cite{Kim:2011zd}.\footnote{
$x_{\mu} \nabla_{\alpha}\nabla^{\alpha} x^{\mu}$ also detects the regularity.
Note that $\nabla_{\alpha} x^{\mu} \nabla^{\alpha} x_{\mu} = {\delta^\alpha}_\alpha=2$ 
does not work for this purpose.}
Figure \ref{fig:patchwork_NG-action}(b) shows the scalar curvature as a function of $u/u_h$ for various values of $\Delta \tilde{\pi}_{\xi}$ for the NG model. One finds $R$ is regular when $\Delta \tilde{\pi}_{\xi}=0$. $R$ asymptotes to the scalar curvature of AdS$_2$ with an unit radius, $R=-2$, in the vicinity of $u=0$.


For the $(\partial x)^4$-truncated model, the behavior of the scalar curvature is more complicated.
From Eq.~(\ref{eq:pi_truncated}), we obtain three different $\xi^{\prime}(u)$ as solutions 
for given $\pi_{\xi}$, and they give different $R$ each other.
Figure \ref{fig:patchwork_dx4-model}(b), (c) show $R$ as a function of $u/u_h$ for several $\Delta \tilde{\pi}_{\xi}$.
The red, blue, and green families of curves correspond to the three roots $\xi'(u)$ of Eq.~(\ref{eq:pi_truncated}).
In the region of $u<u_*$, the red curves correspond to the solutions satisfying the desired boundary condition at $u=0$.
We see that the red curves diverge at $u=u_*$ when $\Delta\tilde{\pi}_{\xi}\neq 0$.
When $\Delta\tilde{\pi}_{\xi}=0$, the red curve in the region of $u<u_*$ reaches $u=u_*$ with a finite value of $R$, and it connects 
smoothly to the blue curve that belongs to the different branch of the solution.
The smooth switching between the different roots is the same as what happened in the models studied in the former sections. Note that the blue smooth curves in Fig.~\ref{fig:patchwork_dx4-model}(b), (c) do not satisfy the right boundary condition at $u=0$.
After all, we find that $R$ is regular everywhere in $0<u<u_h$ only when $\pi_{\xi}=\pi_{\text{phys}}$ for the solutions with the appropriate boundary condition at $u=0$.

\section{Anisotropy of effective temperature}
\label{sec:effective-temp}
In this section, we consider the effective horizon and the effective temperature of the $\mathcal{F}^2$ model. 
We apply the patchwork condition for finding the location of the effective horizon. We find that the $\mathcal{F}^2$ model exhibits different effective temperatures for different modes.
For simplicity, we use the Schwarzschild coordinates in this section.\footnote{Note that the definition of $u$ coordinate is common between the ingoing Eddington-Finkelstein coordinates and the Schwarzschild coordinates.}

Let us consider an effective temperature of $\mathcal{F}^2$ model (\ref{eq:action_F2}) that is detected by a fluctuation of the vector potential $a_M$ around the background configuration. 
In general, the Lagrangian density can be expanded as
\begin{eqnarray}
\mathcal{L}= \mathcal{L}_{(0)}+\frac{1}{2}\mathcal{M}^{\mu\nu\rho\lambda} f_{\mu\nu} f_{\rho\lambda}+O(f_{\mu\nu}^3),  
\end{eqnarray}
where $f_{\mu\nu} = \partial_{\mu} a_{\nu} - \partial_{\nu} a_{\mu}$ is the fluctuation of the field-strength around the background configuration $F_{MN}$. $\mathcal{L}_{(0)}$ is the zeroth order of $f_{\mu\nu}$, and $\mathcal{M}^{\mu\nu\rho\lambda} \equiv \pdv[2]{\mathcal{L}(F,u)}{F_{\mu\nu}}{F_{\rho\lambda}}$.
By definition, this tensor satisfies $\mathcal{M}^{\mu\nu\rho\lambda}=\mathcal{M}^{\rho\lambda\mu\nu}=-\mathcal{M}^{\nu\mu\rho\lambda}=-\mathcal{M}^{\mu\nu\lambda\rho}$.
The linearized equation of motion for the perturbations is given by
\begin{equation}
 \partial_{\mu}\left( \mathcal{M}^{\mu\nu\rho\lambda} f_{\rho\lambda}
	\right) = 0.
\end{equation}
Using the Lagrangian density of Eq.~(\ref{eq:action_F2}), we obtain
\begin{equation}
	-\frac{1}{\mathcal{N}\sqrt{-g}}\mathcal{M}^{\mu\nu\rho\lambda}
	=
    \frac{1}{2}
	(1+c \mathcal{F})g^{\mu[\rho}g^{\lambda]\nu}
	- \frac{c}{4} F^{\mu\nu} F^{\rho\lambda}.
\end{equation}

\subsection{Effective temperature for longitudinal modes}
We consider a perturbation $a_x$ that is longitudinal to the background current. We assume that $a_x$ depends only on $t$ and $u$. In this case, $a_x(t,u)$ decouples from the other fluctuation modes.
The Lagrangian density of the quadratic order of the perturbation is given by
\begin{equation}
	\mathcal{L}_{(2)}^x
	=
	-\frac{\mathcal{N}}{2}\sqrt{-g}\left(
		\gamma^{tt;xx} f_{tx}^2 + 2 \gamma^{tu;xx} f_{tx} f_{ux} + \gamma^{uu;xx} f_{ux}^2 
	\right),
\end{equation}
where
\begin{subequations}
\begin{align}
	\gamma^{tt;xx}
	=&
	- \frac{1}{2}|g^{tt}|g^{xx}\left(
		2 + 3 |g^{tt}| g^{xx} c E^2 - g^{uu} g^{xx} c h'(u)^2
	\right),\\
	\gamma^{tu;xx}
	=&
	- |g^{tt}| g^{uu} (g^{xx})^2 c E h'(u),\\
	\gamma^{uu;xx}
	=&
	\frac{1}{2}g^{xx} g^{uu}
	\left(
		2 + |g^{tt}| g^{xx} c E^2 - 3 g^{uu} g^{xx} c h'(u)^2
	\right).
\end{align}
\end{subequations}
$\gamma^{MN;xx}$ are related to $\mathcal{M}^{MxNx}$ as $-\mathcal{N}\sqrt{-g} \:\gamma^{MN;xx} = 4\mathcal{M}^{MxNx}$.
$\gamma^{MN;xx}$ are proportional to the $M$--$N$ component of the inverse effective metric for $a_x$.
Remarkably, the $u$--$u$ component can be also written as
\begin{equation}
	\gamma^{uu;xx} = - \frac{1/\mathcal{N}}{\sqrt{-g}}\pdv[2]{\mathcal{L}(F, u)}{F_{ux}}.
\end{equation}
Since Eq.~(\ref{eq:patchL-2}) holds at $u=u_{*}$, $\gamma^{uu;xx}$ vanishes at $u=u_{*}$.
Therefore, $u=u_{*}$ is the effective horizon for $a_x$.
 
In the vicinity of $u = u_{*}$, we obtain
\begin{subequations}
\begin{align}
	\gamma^{tt;xx}=&
	- 2 u_h^4 u_{*}^4 \frac{5 u_h^4 - u_{*}^4}{(u_h^4-u_{*}^4)(3 u_h^4 + u_{*}^4)} + \cdots,\\
	\gamma^{tu;xx} =&
	- 2 u_h^2 u_{*}^2 \frac{\sqrt{6 u_h^4 - 2 u_{*}^4}}{3 u_h^4 + u_{*}^4} + \cdots,\\
	\gamma^{uu;xx} =&
	4\sqrt{2} u_{*}^4
	\frac{\sqrt{9 u_h^8 + 6 u_h^4 u_{*}^4 - 7 u_{*}^8}}{3 u_h^4 + u_{*}^4}
	\left(1-\frac{u}{u_{*}}\right) + \cdots,
\end{align}
\end{subequations}
where $\cdots$ denotes the higher order terms of $(u-u_*)$.
In the derivation of the above expansions, we have used $0<u_*<u_h$. Note that $c$-dependence accommodates in the expression of $u_*$.
We obtain the effective temperature as
\begin{equation}
	T_{*}
	= \frac{1}{4\pi} \left.\frac{\partial_{u} \gamma^{uu;xx}}{\gamma^{tu;xx}}\right|_{u=u_{*}}
	=
	\frac{T}{2 u_{*} u_h}
	\sqrt{\frac{9 u_h^8 + 6 u_{*}^4 u_h^4 - 7 u_{*}^8}{3 u_h^4 - u_{*}^4}}.
	\label{eq:Teff_F2}
\end{equation}
By using Eq.~(\ref{eq:uStar_F2}), the ratio of the effective temperature and the Hawking temperature of the black hole geometry as a function of $e$ is given by
\begin{equation}
\begin{aligned}
	\frac{T_{*}}{T}
	=&
	\frac{2^{3/4}}{\sqrt{5 ce^2+\sqrt{9 \left(ce^2+8\right) ce^2+16}+4}}\\
	&\times\left[
		\frac{\left(ce^2-2\right) \left(9 \left(ce^2+8\right) ce^2+16\right)}{-3 ce^2+\sqrt{9 \left(ce^2+8\right) ce^2+16}-8}
	\right]^{1/4}\\
	=&
	1 + \frac{ce^2}{2} - \frac{5}{4}c^2e^4 + \order{e^5}.
\end{aligned}
\end{equation}
If we set $c=1$, the result agrees with the DBI case (\ref{eq:effectiveT_DBI}) to the order of $e^2$.

\subsection{Effective temperature for transverse modes}
Next, let us consider a fluctuation $a_y$ that is transverse to the background current. We assume that $a_y$ depends only on $t$ and $u$. In this case, $a_y(t,u)$ decouples from the other perturbations.
The Lagrangian density that contains the quadratic order of $a_y(t,u)$ is given by
\begin{equation}
	\mathcal{L}_{(2)}^y = - \frac{\mathcal{N}}{2}\sqrt{-g}(1+c\mathcal{F})g^{xx} \left[
		- |g^{tt}| ( \partial_{t} a_y )^2 + g^{uu} (\partial_{u} a_y)^2
	\right].
\end{equation}
We immediately see that the effective metric for $a_y(t,u)$ is proportional to the metric of the background geometry, in this case. This means that the location of the effective horizon and the effective temperature for $a_y$ are $u_h$ and $T$, respectively: they are the same as that of the event horizon and the temperature of the background geometry.

We have found that the effective temperature for the longitudinal modes and that for the transverse modes do not agree in general for the $\mathcal{F}^2$ model. On the other hand, we can check that the model which is obtained by the derivative expansion of the DBI theory to the quadratic order of the fluctuations has a common effective temperature and the effective horizon for the longitudinal modes and the transverse modes. This occurs because it is different from the $\mathcal{F}^2$ model even for $c=1$. One can see the difference between them by comparing (\ref{eq:action_F2}) (with $c=1$) and (\ref{eq:DBI_expansion}).\footnote{Note that this difference does not affect the computation of the nonlinear conductivity in Sec.~\ref{sec:higher-order}. This is why the conductivity obtained from the $\mathcal{F}^2$ model with $c=1$ agrees with that from the DBI theory to the order of $e^2$.} This suggests that the terms that are contained in the derivative expansion of the DBI action to the quadratic order and that are absent from the $\mathcal{F}^2$ model play an important role in the isotropization of the effective temperatures.\footnote{In general, a nonlinear Maxwell theory shows birefringence in the presence of a background flux. Conditions for the absence of the birefringence and its relationship to the DBI theory have been considered on the Minkowski spacetime in \cite{deMelo:2014isa} and the references therein.}

\section{Conclusion and Discussions}\label{sec:conclusion}
In this paper, we have proposed a new method to compute nonlinear transport coefficients such as nonlinear DC conductivity and nonlinear friction coefficient in holography. Our method works even for the models without the square-root structure, thus having a wider application than the already-existing conventional method. 

Our basic idea is to request regularity which is a standard philosophy in holography. However, the point is that we conjecture that the regularity condition is equivalent to the patchwork condition (with the appropriate boundary conditions at the boundary). We have checked that our conjecture works for several concrete examples of models.
Since the patchwork point plays a role of an effective horizon, the presence of the notion of the effective temperature is suggested to be quite common and important in the realization of the nonequilibrium steady states in holography. 

Our method enables us to compute not only the nonlinear responses but also the effective temperatures in nonequilibrium steady states for a wider range of holographic models. We found that the effective temperature for the longitudinal modes can be different from that for the transverse modes in some models such as the $\mathcal{F}^2$ model. These analyses thus may shed some light on the mechanism of the isotropization process of the (effective) temperatures.

In the present work, we have analyzed only the models where the Lagrangian is given in terms of the powers of $\mathcal{F} = -\frac{1}{4} F_{\mu\nu}F^{\mu\nu}$ for conductors and in terms of the powers of $(\partial_{\alpha}x)^{2}$ for test objects that undergo frictional force. It is interesting to generalize our work to more general models.
The above-mentioned points should be investigated in further studies.

\section*{Acknowledgements}
The authors thank S.~Kinoshita for his valuable comments and discussions on the first version of the paper.
The authors acknowledge K.~Izumi, N.~Tanahashi for discussions and comments.
The authors are grateful to the Yukawa Institute for Theoretical Physics at Kyoto University. Discussions during the YITP workshop YITP-W-22-09 on ``Strings and Fields 2022'' were useful to complete this work. Hospitality at Asia Pacific Center for Theoretical Physics during the program ``QCD and gauge/gravity duality'' is kindly acknowledged. The authors also thank RIKEN iTHEMS NEW working group for fruitful discussions.
The work of S.\,I.~is supported by National Natural Science Foundation of China with Grant No.~12147158.
The work of S.\,N.~is supported in part by JSPS KAKENHI Grants No.~JP19K03659, No.~JP19H05821, and the Chuo University Personal Research Grant.
The work of K.\,T.~is supported by JSPS KAKENHI Grant No.~JP22K20350 and No.~JP23K17664, and by JST PRESTO Grant No.~JPMJPR2256.

\appendix
\section{Method of Ref.~\cite{Kim:2011zd}}\label{sec:Kim-Pang}

Here, we overview the
method proposed in Ref.~\cite{Kim:2011zd}.
The computational procedure 
proposed there is the following.
\begin{enumerate}[label=\roman*)]
    \item Derive the expression of the \emph{linear} conductivity from the membrane paradigm
    as a function of $u_h$, where $u_h$ is the location of the horizon of the bulk black hole.
    \item Obtain the locus of the effective horizon (which they call singular shell) $u_*$ as a function of $E$.
    \item Replace $u_h$ with $u_*$ in the expression of the linear conductivity. Then the obtained expression gives the
    nonlinear conductivity.
\end{enumerate}
The last procedure is based on the observation that the $E$-dependence of the conductivity in the models with the DBI action is encoded only in the $E$-dependence of $u_*$. 

First, we demonstrate how it works for the models with the DBI action.
For instance, the locus of the effective horizon is given by $u_* = (1+e^2)^{-1/4} u_h$ for the model
we have considered in Sec.~\ref{sec:DBI}.
The linear conductivity for this model is given by $\sigma_{\text{linear}} = 1/u_h$.
Applying the above procedure, we obtain the nonlinear conductivity
\begin{equation}
    \sigma_{u_h\to u_{*}}^{\text{DBI}} = \frac{1}{u_*}  = \frac{1}{u_h} (1+e^2)^{1/4},
\end{equation}
which agrees with the result of the real-action method.

When we attempt to apply this method to more general models, we find our approach is necessary.
%
Let us consider
the $\mathcal{F}^2$ model, for example.
The linear conductivity for this model is again given by $\sigma_{\text{linear}} = 1/u_h$.
If we follow the procedure i), ii), iii) mentioned above,
we obtain
\begin{equation}
\begin{aligned}
    \sigma_{u_h\to u_{*}}^{\mathcal{F}^2}
    &\equiv \frac{1}{u_*}\\
    &=
    \frac{1}{u_h}
    \left(
        \frac{
            2(2-ce^2)
        }{
            3ce^2 + 8 - \sqrt{9c^2 e^4 + 72 c e^2 + 16}
        }
    \right)^{1/4}\\
    &=
    \frac{1}{u_h}
    \left(
        1 + \frac{c}{4} e^2 - \frac{9}{128} (5 c - 1) c e^4 + \order{e^6}
    \right),
\end{aligned}
\label{eq:sigma_OSM_F2}
\end{equation}
where we have employed $u_*$ given in (\ref{eq:uStar_F2}).\footnote{
The method for calculating the location of the effective horizon employed in \cite{Kim:2011zd} is valid only for the models with the DBI action. For more general models considered in this paper, we need the patchwork condition to obtain $u_*$.}
The $\order{e^4}$ contribution differs from the correct conductivity $J/E=u_{h}^{-1}j/e$ with $j$ given by (\ref{eq:current_F2}) from the regularity. 



\section{$T=0$ limit of $\mathcal{F}^2$ model}
\label{sec:T=0}
In the main part of this paper, we have considered the nonlinear conductivity at finite temperatures. We make a comment for the zero-temperature cases.
The D3-D7 model has a nonlinear conductivity even at zero temperature which is given by $J = E^{3/2}$. We obtain this result by taking the $e\to\infty$ limit in (\ref{eq:current_DBI}).

Let us consider the zero temperature limit in the $\mathcal{F}^2$ model.\footnote{We show only for the $\mathcal{F}^2$ model here, since the analysis for the $\mathcal{F}^3$ model is technically difficult.}
In the limit of $T=0$, Eqs.~(\ref{eq:uStar_F2}), (\ref{eq:J_F2})  and (\ref{eq:Teff_F2}) become
\begin{equation}
	u_{*} = \frac{2^{1/4}}{c^{1/4}\sqrt{E}},~
	J = \frac{4}{3}\sqrt{\frac{\sqrt{2}}{3}} c^{1/4}E^{3/2},~
	T_{*} = \frac{\sqrt{3/2}}{\pi}c^{1/4}\sqrt{E},
	\label{eq:us_J_T=0}
\end{equation}
respectively. Here we have chosen the positive sign for $J$.
\begin{figure}[htbp!]
	\centering
	\subfloat[]{\includegraphics[width=7.5cm]{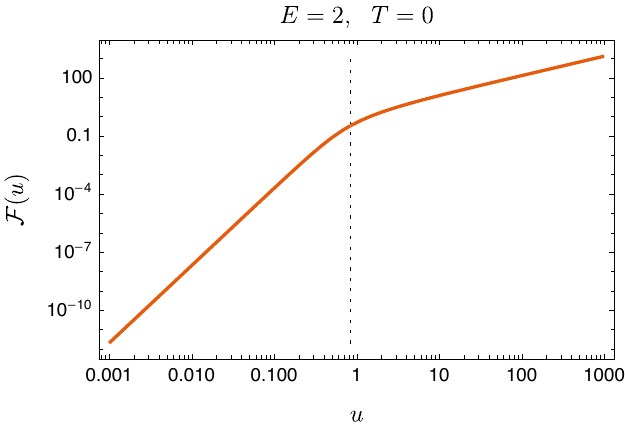}\label{fig:calF-u_Tzero-a}}
	\subfloat[]{\includegraphics[width=7.5cm]{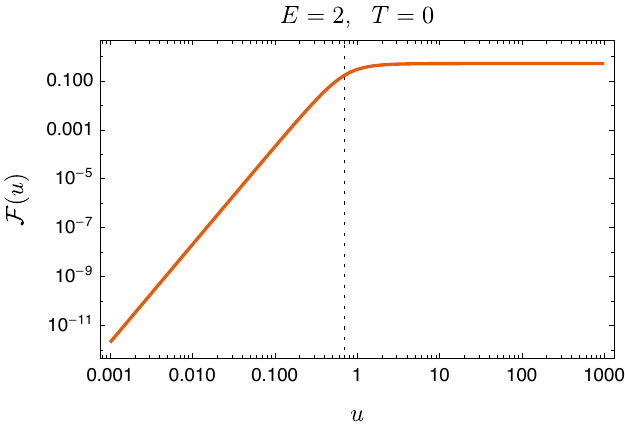}\label{fig:calF-u_Tzero-b}}
	\caption{
		(a)
		$\mathcal{F}(u)$ for $E=2$ in the $\mathcal{F}^2$ model with $c=1$
  at $T=0$.
		The dotted line shows $u=u_{*}$.
		(b)
		$\mathcal{F}(u)$ for $E=2$ in the D3-D7 model 
  at $T=0$.
		The dotted line shows $u=u_{*} = 1/\sqrt{E}$.
		Both plots are shown in a log-log scale.
	}
	\label{fig:calF-u_Tzero}
\end{figure}
Although the $E$-dependencies are determined by the scaling properties at $T=0$ completely,
the coefficients are different from those of the DBI theory.

One can see some qualitative difference in $\mathcal{F}(u)$ between the $\mathcal{F}^2$ model and the DBI model for $E\neq 0$.
$\mathcal{F}(u)$ for the $\mathcal{F}^2$ model with $E\neq 0$
diverges at $u=\infty$ as is shown in Fig.~\ref{fig:calF-u_Tzero}(a).
This behavior can be also understood from Eq.~(\ref{eq:scF(uh)_F2-1}).
At finite temperatures, $\mathcal{F}(u)$ at $u=u_h$ is written as
\begin{equation}
	\mathcal{F}(u_h) = - \frac{1}{c}\left(
		1 - \frac{1}{\pi T} \frac{J}{E}
	\right),
\end{equation}
and $\mathcal{F}(u_h)$ diverges in the limit of $u_h \to \infty$ where $T\to 0$.
However, one should note that this singularity is hidden by the effective horizon located at $u=u_{*}$ for $E\neq 0$.

For the DBI theory, $\mathcal{F}(u_h)$ is evaluated by substituting (\ref{eq:hPrime_DBI}) into (\ref{eq:calF_def}) as
\begin{equation}
	\mathcal{F}(u_h) = \frac{1}{2}\left(
		1 - \pi^{2} T^{2}\frac{E^2}{J^2}
	\right),
\end{equation}
and $\mathcal{F}(u_h)$ goes to $1/2$ in the limit of $u_h \to \infty$ where $T\to 0$. We can see this behavior in Fig.~\ref{fig:calF-u_Tzero}(b).

\section{The higher-order models in the Schwarzschild coordinates}
We present computations in the Schwarzschild coordinates where the metric is given by (\ref{eq:bulk_metric}).

\label{sec:Schwarzschild}
\subsection{$\mathcal{F}^2$ model}
We consider the action given in (\ref{eq:action_F2}).
The Lagrangian density with our ansatz (\ref{eq:ansatz_of_A_DBI}) is explicitly written as
\begin{equation}
\begin{aligned}
	\mathcal{L} =&
	\frac{\mathcal{N}}{2}\sqrt{|g_{tt}|g_{xx}g_{uu}}\Big[
		\left(
			|g^{tt}| F_{tx}^2
			- g^{uu} F_{ux}^2
		\right)\\
		&+
		\frac{c}{4}g^{xx}\left(
			|g^{tt}| F_{tx}^2
			- g^{uu} F_{ux}^2
		\right)^2
	\Big].
\end{aligned}
	\label{eq:Lagrangian_F2_Sch}
\end{equation}
The equation of motion for $A_x$ is given by
\begin{equation}
\begin{aligned}
	J =& - \frac{1}{\mathcal{N}}\pdv{\mathcal{L}}{A_x'(u)}\\
	=&
	\sqrt{\frac{|g_{tt}|g_{xx}}{g_{uu}}}\left[
		a'(u)
		+
		\frac{c}{2}
		g^{xx}
		\left(
			|g^{tt}| E^2
			- g^{uu} a'(u)^2
		\right)
		a'(u)
	\right]\\
	\equiv& J(u, a'(u)).
\end{aligned}
	\label{eq:eom_F2_Sch}
\end{equation}
This is a cubic equation of $a'(u)$ that has three roots.
The explicit expressions of these roots are very complicated, and we do not exhibit them here.

Let us see the asymptotic behaviors of the three roots.
In the vicinity of the AdS boundary, the solutions have series expansions as follows:
\begin{eqnarray}
    	a'(u) &=&
	J u + \order{u^2},
 	\label{eq:hPrime_at_zero-1_Sch}\\
	a'(u) &=&
	\pm \frac{\sqrt{2}}{u^2} - \frac{1}{2} J u  + \order{u^2}.
	\label{eq:hPrime_at_zero-2_Sch}
\end{eqnarray}
Only the first solution (\ref{eq:hPrime_at_zero-1_Sch}) matches with our boundary condition $a(u)=0$ at $u\to0$.
On the other hand, the solutions in the vicinity of the event horizon are expanded as follows:
\begin{eqnarray}
	a'(u) &=&
	u_h^2 \left(
		\pm \frac{1}{4}\frac{e}{1-u/u_h}
		\pm \frac{4 e + 3 c e^3 \mp 4 j}{8 c e^2}
	\right)\nonumber\\
    & &+ \order{u - u_h},
 	\label{eq:hPrime_at_uh-1_Sch}\\
 a'(u) &=&
	u_h^2 \frac{2}{c} \frac{j}{e^2} + \order{u - u_h}.
	\label{eq:hPrime_at_uh-2_Sch}
\end{eqnarray}
For these solutions, we obtain
\begin{eqnarray}
	\mathcal{F}(u) &=&
	- \frac{1}{c}\left(1 \pm \frac{j}{e}\right) + \order{u-u_h},
 	\label{eq:scF(uh)_F2-1_Sch}\\
 \mathcal{F}(u) &=&
	\frac{e^2}{8} \frac{1}{1-u/u_h} - \frac{5}{16} e^2 + \order{u-u_h},
	\label{eq:scF(uh)_F2-2_Sch}
\end{eqnarray}
respectively.
If we choose the first two solutions (\ref{eq:hPrime_at_uh-1_Sch}), $\mathcal{F}(u)$ remain finite at $u=u_h$ whatever combinations of $j$ and $e$ we have: the regularity of $\mathcal{F}(u)$ at $u=u_h$ alone does not fix the conductivity at all.
However, these solutions are not necessarily smoothly connected to (\ref{eq:hPrime_at_zero-1_Sch}) that satisfies the right boundary condition.

\begin{figure*}[htbp!]
	\centering
	\subfloat[]{\includegraphics[width=8.0cm]{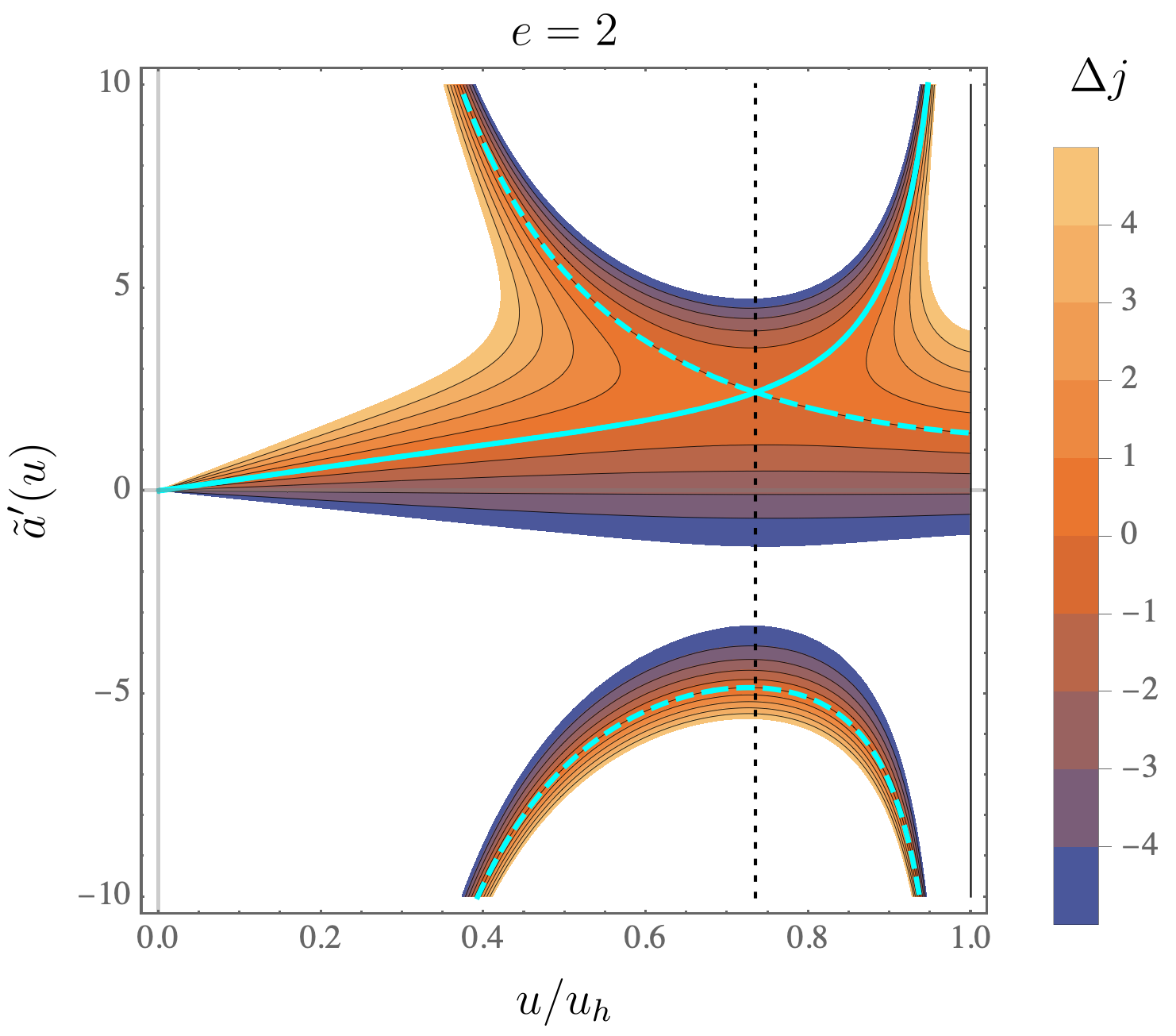}\label{fig:j_contours_F2_Sch-a}}
	\subfloat[]{\includegraphics[width=7.0cm]{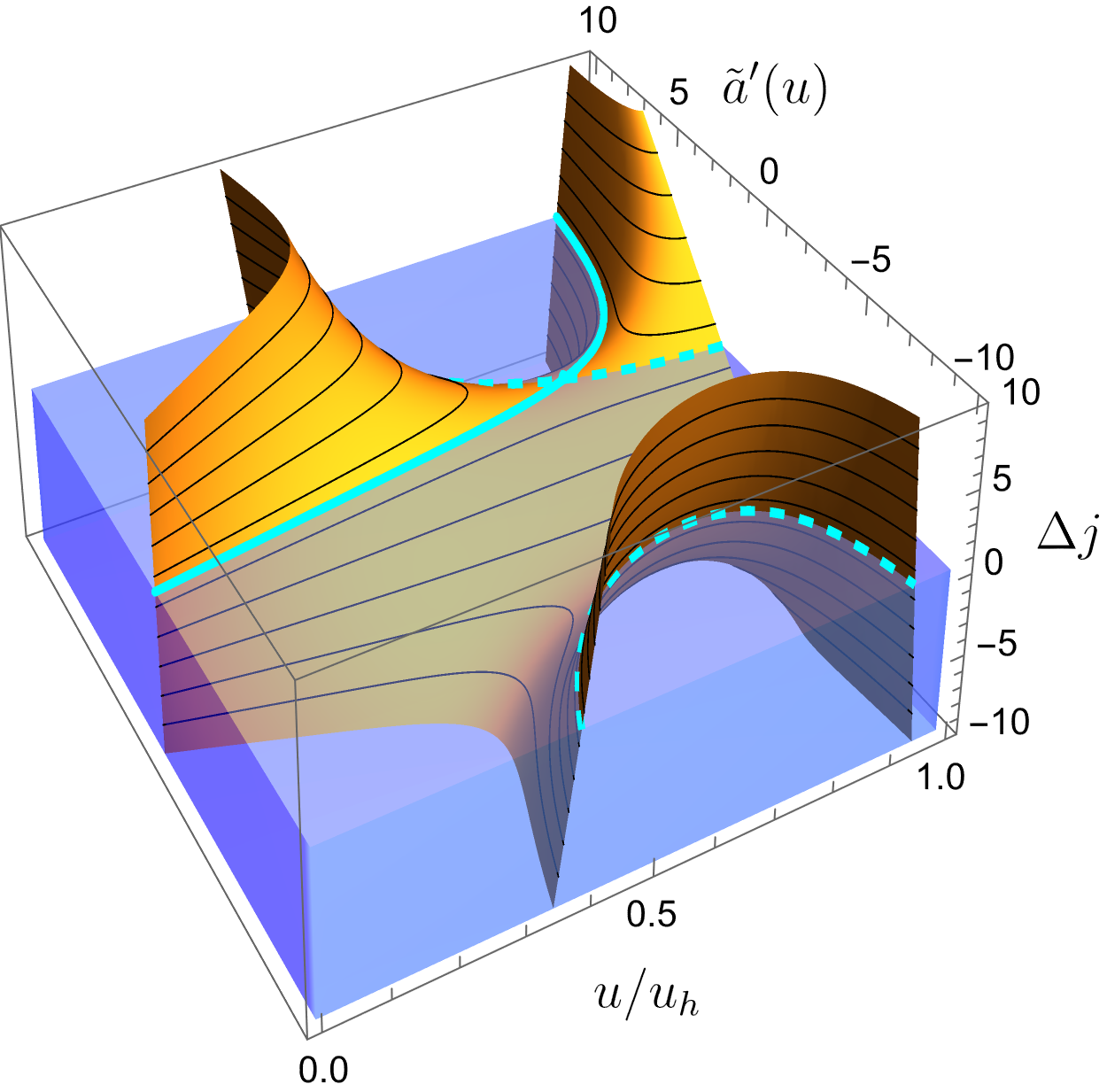}\label{fig:j_contours_F2_Sch-b}}
	\caption{
        (a)
        Contour plot of $J(u,h')$ for the $\mathcal{F}^2$ model in the Schwarzshild coordinates. Each contour represents the solution $h'(u)$ for given $J$. Here, $J$ and $h^{\prime}$ are rescaled as $j=u_h^3 J$ and $\tilde{h}'=u_h^2 h'$, respectively, and $\Delta j = j-j_\text{phys}$. The solid and dashed cyan curves are the contours at $j=j_{\text{phys}}=2.8424$ that go through the patchwork point. The solid cyan curve is the regular physical solution and satisfies the right boundary condition at $u=0$.
        The vertical dotted and solid lines indicate the locations of the saddle point (effective horizon) $u=u_{*}$ and the black hole horizon $u=u_h$, respectively.
        (b)
        Surface plot of $J(u,h')$ for the $\mathcal{F}^2$ model in the Schwarzshild coordinates.
        The orange surface represents $\Delta j$, while the top plane of the translucent blue box represents $\Delta j=0$.
        The solid and dashed cyan curves are the section of these two surfaces. The crossing point of these curves is the saddle point of the orange surface. Here, we set $e=2$ and $c=1$ in these panels.
	}
	\label{fig:j_contours_F2_Sch}
\end{figure*}

Let us survey the behavior of the solutions in the entire region of $0\le u \le u_h$.
The numerical results are shown in Fig.~\ref{fig:j_contours_F2_Sch}.
In Fig.~\ref{fig:j_contours_F2_Sch}, $\tilde{a}^{\prime}(u)\equiv u_{h}^{2} a^{\prime}(u)$ as a function $u$ is given for various values of $j=u_h^3 J(u,a'(u))$, for $e=2$ and $c=1$. 
We find that the patchwork occurs when $j=j_{\text{phys}}=2.8424$ and $a'(u)$ satisfies the desired behavior at both $u=0$ and $u=u_h$. Indeed, the location of the patchwork is the saddle point of the landform if we regard $j=u_h^3 J(u,a'(u))$ as the ``height.'' 
Note that $\mathcal{F}(u)$ is regular at the horizon although $a'(u)$ is divergent there in this solution.

The location of the saddle point $(u_*, a'(u_*))$ is given by the patchwork condition.
(\ref{eq:patchJ-1}) and (\ref{eq:patchJ-2}) for the present case are explicitly given as
\begin{subequations}
\begin{align}
	0 =& \left(\x^2 \left(3-\frac{3 c e^2}{2}\right)+\frac{1}{\x^2}\right) \ya \nonumber\\
	&+\frac{1}{2} c \left(11 \x^8-14 \x^4+3\right) \x^2 \ya^3,\\
	0 =& \x^3 \left(1- \frac{c}{2} e^2 \right) - \frac{1}{\x} +\frac{3}{2} c (\x^4-1)^2 \x^3 \ya^2,
\end{align}
\end{subequations}
respectively, where $\x \equiv u_{*}/u_h$.
Solving these equations, we obtain
\begin{equation}
	\x = \left(
		\frac{\sqrt{9 c^2 e^4+72 c e^2+16}-3 c e^2-8}{2 c e^2-4}
	\right)^{1/4},
	\label{eq:uStar_F2_Sch}
\end{equation}
and
\begin{equation}
	\ya = \pm\left(
		\frac{\x^4 \left(c e^2-2\right)+2}{3 c (\x^4-1)^2 \x^4}
	\right)^{1/2}.
\end{equation}
The current density is obtained by substituting these results into Eq.~(\ref{eq:eom_F2_Sch}):
\begin{equation}
	j = \pm\frac{\left(\x^4 \left(c e^2-2\right)+2\right)^{3/2}}{3 \sqrt{3} \sqrt{c} \x^3 \left(\x^4-1\right)}.
	\label{eq:J_F2_Sch}
\end{equation}
The sign should be chosen in such a way that $j\cdot e>0$.
This agrees with the result (\ref{eq:J_F2}) we have obtained in the ingoing Eddington-Finkelstein coordinates.

\subsection{$\mathcal{F}^3$ model}

Let us consider the $\mathcal{F}^3$ model given by (\ref{eq:action-F3}).
The equation of motion is explicitly given by
\begin{equation}
\begin{aligned}
	J =& - \frac{1}{\mathcal{N}}\pdv{\mathcal{L}}{A_x'(u)}\\
	=&
	\sqrt{\frac{|g_{tt}|g_{xx}}{g_{uu}}}
	\Bigg[
		a'(u) + \frac{1}{2} g^{xx}(|g^{tt}| E^2 - g^{uu} a'(u)^2) a'(u)\\
				&+ \frac{3}{8}b (g^{xx})^2 (|g^{tt}| E^2 - g^{uu} a'(u)^2)^2 a'(u)
	\Bigg]\\
	 \equiv& J(u,a'(u)),
\end{aligned}
\end{equation}
within the ansatz (\ref{eq:ansatz_of_A_DBI}).
In this case, the equation of motion is a quintic equation of $a'(u)$.
Let us examine the behavior of the solutions numerically since there is no algebraic formula for the roots of the quintic equation.

\begin{figure*}[htbp!]
	\centering
	\subfloat[]{\includegraphics[width=8.0cm]{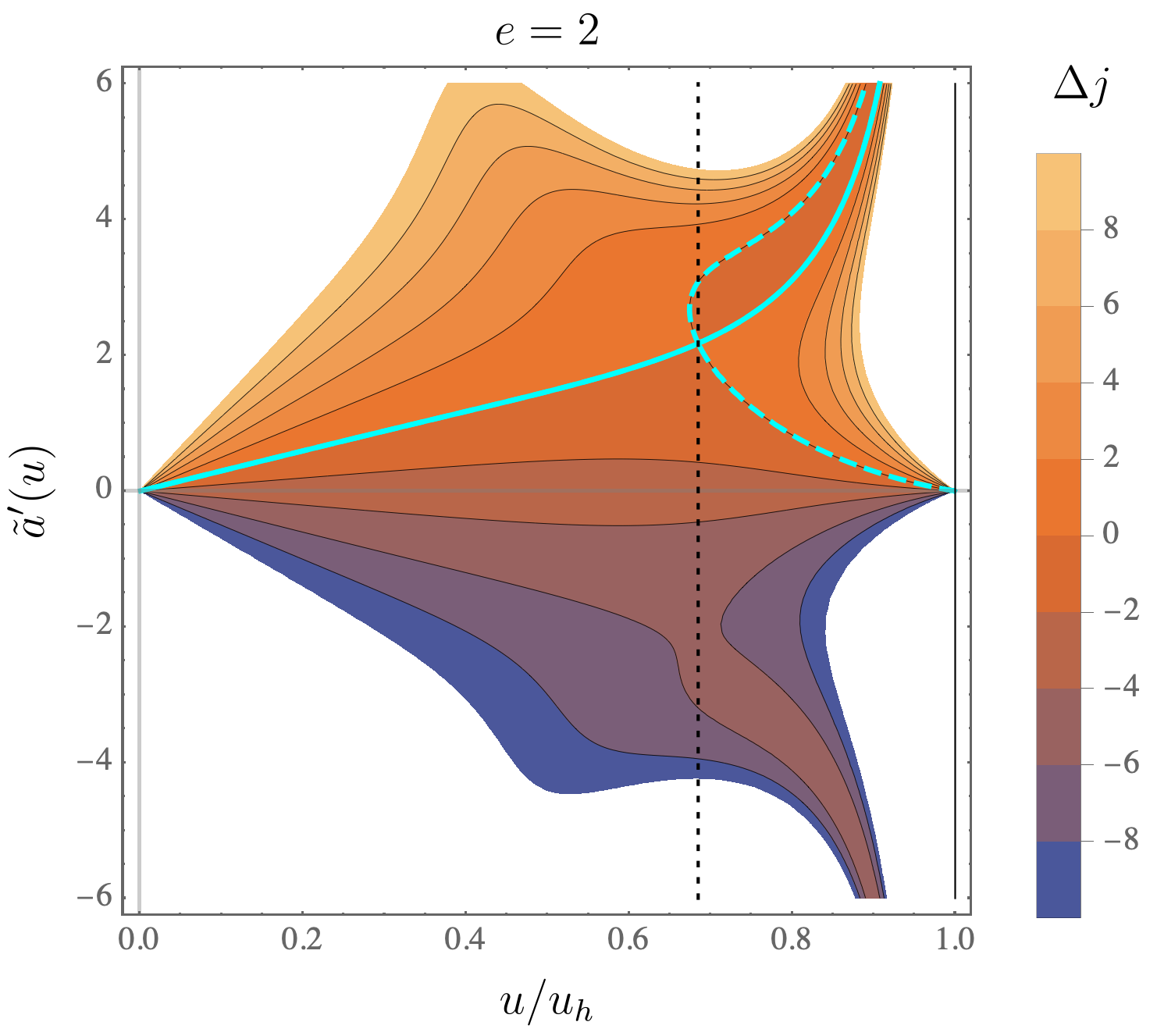}\label{fig:j_contours_F3_Sch-a}}
	\subfloat[]{\includegraphics[width=7.0cm]{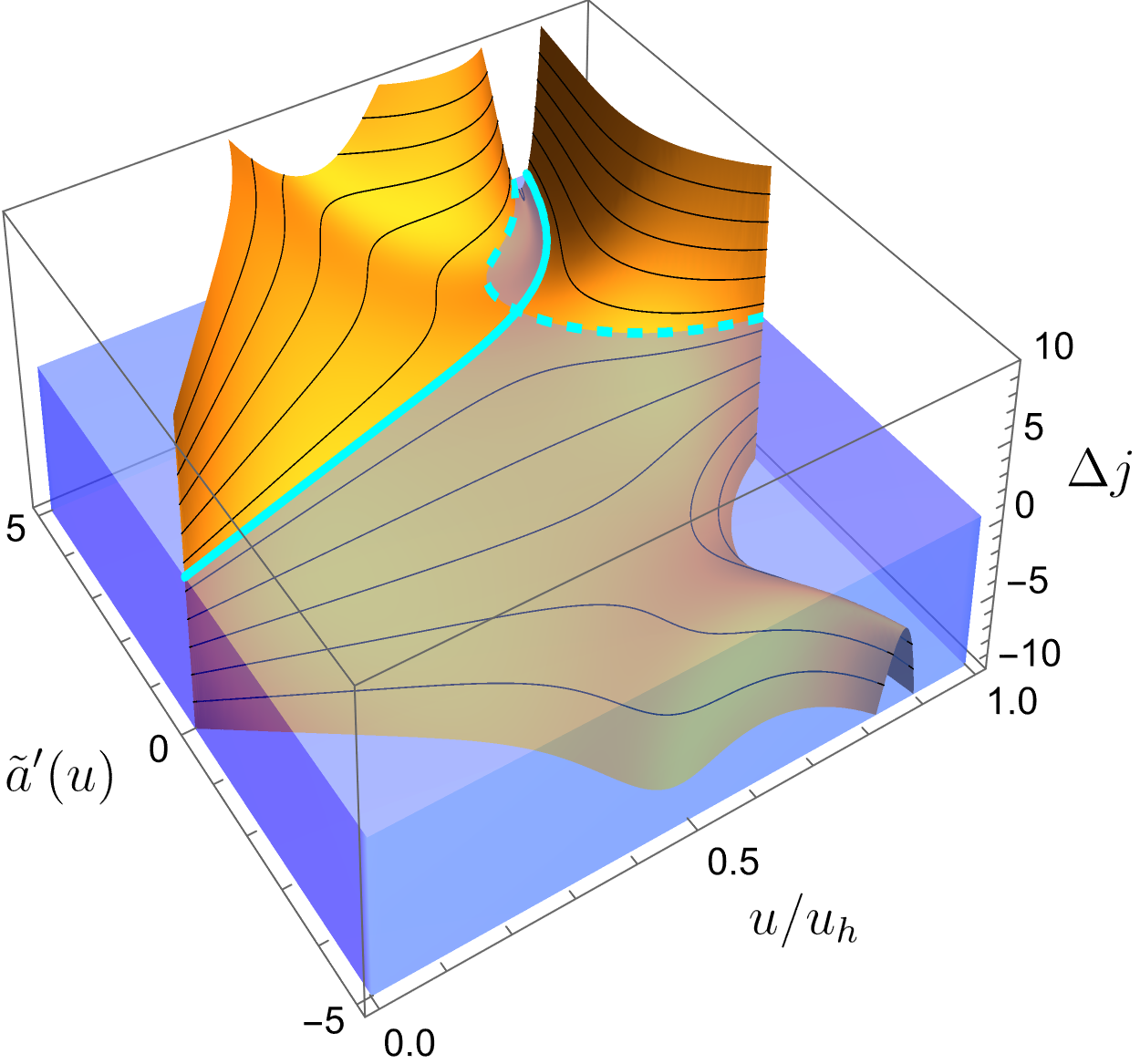}\label{fig:j_contours_F3_Sch-b}}
	\caption{
        (a)
        Contour plot of $J(u,h')$ for the $\mathcal{F}^3$ model in the Schwarzshild coordinates. Each contour represents the solution $h'(u)$ for given $J$. Here, $J$ and $h^{\prime}$ are rescaled as $j=u_h^3 J$ and $\tilde{h}'=u_h^2 h'$, respectively, and $\Delta j = j-j_\text{phys}$. The solid and dashed cyan curves are the contours at $j=j_{\text{phys}}=2.9621$ that go through the patchwork point. The solid cyan curve is the regular physical solution and satisfies the right boundary condition at $u=0$.
        The vertical dotted and solid lines indicate the locations of the saddle point (effective horizon) $u=u_{*}$ and the black hole horizon $u=u_h$, respectively.
        (b)
        Surface plot of $J(u,h')$ for the $\mathcal{F}^3$ model in the Schwarzshild coordinates.
        The orange surface represents $\Delta j$, while the top plane of the translucent blue box represents $\Delta j=0$.
        The solid and dashed cyan curves are the section of these two surfaces. The crossing point of these curves is the saddle point of the orange surface. Here, we set $e=2$ and $b=1$ in these panels.
	}
	\label{fig:j_contours_F3_Sch}
\end{figure*}
We show the solutions for various values of $j$ in Fig.~\ref{fig:j_contours_F3_Sch}.
We find that the patchwork occurs when $ (j_{\text{phys}}, u_*) = (2.9621, 0.68475 u_h) $ for $e=2$.
We can employ the small-$e$ approximation as we did in Sec.~\ref{subsec:F3-model}. We express $\tilde{u}_*$ and $\tilde{a}(u_*)$ in powers of $e$ and substitute them into the patchwork condition. Then we obtain $j$ in powers of $e$ order by order. The results agree with Eq.~(\ref{eq:current_F3}).

\bibliography{main}

\providecommand{\href}[2]{#2}\begingroup\raggedright\begin{thebibliography}{10}

\bibitem{Maldacena:1998}
J.~M. Maldacena, {\slshape {The Large N limit of superconformal field theories
  and supergravity},} \href{http://dx.doi.org/10.1023/A:1026654312961}{{\em
  Adv. Theor. Math. Phys.} {\bfseries 2} (1998) 231--252},
  \href{http://arxiv.org/abs/hep-th/9711200}{{ arXiv:hep-th/9711200}}.

\bibitem{Gubser:1998}
S.~S. Gubser, I.~R. Klebanov, and A.~M. Polyakov, {\slshape {Gauge theory
  correlators from noncritical string theory},}
  \href{http://dx.doi.org/10.1016/S0370-2693(98)00377-3}{{\em Phys. Lett. B}
  {\bfseries 428} (1998) 105--114},
  \href{http://arxiv.org/abs/hep-th/9802109}{{ arXiv:hep-th/9802109}}.

\bibitem{Witten:1998}
E.~Witten, {\slshape {Anti-de Sitter space and holography},}
  \href{http://dx.doi.org/10.4310/ATMP.1998.v2.n2.a2}{{\em Adv. Theor. Math.
  Phys.} {\bfseries 2} (1998) 253--291},
  \href{http://arxiv.org/abs/hep-th/9802150}{{ arXiv:hep-th/9802150}}.

\bibitem{Son:2002sd}
D.~T. Son and A.~O. Starinets, {\slshape {Minkowski space correlators in AdS /
  CFT correspondence: Recipe and applications},}
  \href{http://dx.doi.org/10.1088/1126-6708/2002/09/042}{{\em JHEP} {\bfseries
  09} (2002) 042}, \href{http://arxiv.org/abs/hep-th/0205051}{{
  arXiv:hep-th/0205051}}.

\bibitem{Yee:2009vw}
H.-U. Yee, {\slshape {Holographic Chiral Magnetic Conductivity},}
  \href{http://dx.doi.org/10.1088/1126-6708/2009/11/085}{{\em JHEP} {\bfseries
  11} (2009) 085}, \href{http://arxiv.org/abs/0908.4189}{{
  arXiv:0908.4189~[hep-th]}}.

\bibitem{Herzog:2006gh}
C.~P. Herzog, A.~Karch, P.~Kovtun, C.~Kozcaz, and L.~G. Yaffe, {\slshape
  {Energy loss of a heavy quark moving through N=4 supersymmetric Yang-Mills
  plasma},} \href{http://dx.doi.org/10.1088/1126-6708/2006/07/013}{{\em JHEP}
  {\bfseries 07} (2006) 013}, \href{http://arxiv.org/abs/hep-th/0605158}{{
  arXiv:hep-th/0605158}}.

\bibitem{Gubser:2006bz}
S.~S. Gubser, {\slshape {Drag force in AdS/CFT},}
  \href{http://dx.doi.org/10.1103/PhysRevD.74.126005}{{\em Phys. Rev. D}
  {\bfseries 74} (2006) 126005}, \href{http://arxiv.org/abs/hep-th/0605182}{{
  arXiv:hep-th/0605182}}.

\bibitem{Karch:2007pd}
A.~Karch and A.~O'Bannon, {\slshape {Metallic AdS/CFT},}
  \href{http://dx.doi.org/10.1088/1126-6708/2007/09/024}{{\em JHEP} {\bfseries
  09} (2007) 024}, \href{http://arxiv.org/abs/0705.3870}{{
  arXiv:0705.3870~[hep-th]}}.

\bibitem{Kim:2011zd}
K.-Y. Kim and D.-W. Pang, {\slshape {Holographic DC conductivities from the
  open string metric},} \href{http://dx.doi.org/10.1007/JHEP09(2011)051}{{\em
  JHEP} {\bfseries 09} (2011) 051}, \href{http://arxiv.org/abs/1108.3791}{{
  arXiv:1108.3791~[hep-th]}}.

\bibitem{Kim:2011qh}
K.-Y. Kim, J.~P. Shock, and J.~Tarrio, {\slshape {The open string membrane
  paradigm with external electromagnetic fields},}
  \href{http://dx.doi.org/10.1007/JHEP06(2011)017}{{\em JHEP} {\bfseries 06}
  (2011) 017}, \href{http://arxiv.org/abs/1103.4581}{{
  arXiv:1103.4581~[hep-th]}}.

\bibitem{Sonner:2012if}
J.~Sonner and A.~G. Green, {\slshape {Hawking Radiation and Non-equilibrium
  Quantum Critical Current Noise},}
  \href{http://dx.doi.org/10.1103/PhysRevLett.109.091601}{{\em Phys. Rev.
  Lett.} {\bfseries 109} (2012) 091601},
  \href{http://arxiv.org/abs/1203.4908}{{ arXiv:1203.4908~[cond-mat.str-el]}}.

\bibitem{Gubser:2006nz}
S.~S. Gubser, {\slshape {Momentum fluctuations of heavy quarks in the
  gauge-string duality},}
  \href{http://dx.doi.org/10.1016/j.nuclphysb.2007.09.017}{{\em Nucl. Phys. B}
  {\bfseries 790} (2008) 175--199},
  \href{http://arxiv.org/abs/hep-th/0612143}{{ arXiv:hep-th/0612143}}.

\bibitem{Casalderrey-Solana:2007ahi}
J.~Casalderrey-Solana and D.~Teaney, {\slshape {Transverse Momentum Broadening
  of a Fast Quark in a N=4 Yang Mills Plasma},}
  \href{http://dx.doi.org/10.1088/1126-6708/2007/04/039}{{\em JHEP} {\bfseries
  04} (2007) 039}, \href{http://arxiv.org/abs/hep-th/0701123}{{
  arXiv:hep-th/0701123}}.

\bibitem{Gursoy:2010aa}
U.~Gursoy, E.~Kiritsis, L.~Mazzanti, and F.~Nitti, {\slshape {Langevin
  diffusion of heavy quarks in non-conformal holographic backgrounds},}
  \href{http://dx.doi.org/10.1007/JHEP12(2010)088}{{\em JHEP} {\bfseries 12}
  (2010) 088}, \href{http://arxiv.org/abs/1006.3261}{{
  arXiv:1006.3261~[hep-th]}}.

\bibitem{Nakamura:2013yqa}
S.~Nakamura and H.~Ooguri, {\slshape {Out of Equilibrium Temperature from
  Holography},} \href{http://dx.doi.org/10.1103/PhysRevD.88.126003}{{\em Phys.
  Rev. D} {\bfseries 88} (2013) 126003},
  \href{http://arxiv.org/abs/1309.4089}{{ arXiv:1309.4089~[hep-th]}}.

\bibitem{Hartnoll:2008kx}
S.~A. Hartnoll, C.~P. Herzog, and G.~T. Horowitz, {\slshape {Holographic
  Superconductors},}
  \href{http://dx.doi.org/10.1088/1126-6708/2008/12/015}{{\em JHEP} {\bfseries
  12} (2008) 015}, \href{http://arxiv.org/abs/0810.1563}{{
  arXiv:0810.1563~[hep-th]}}.

\bibitem{Hartnoll:2008vx}
S.~A. Hartnoll, C.~P. Herzog, and G.~T. Horowitz, {\slshape {Building a
  Holographic Superconductor},}
  \href{http://dx.doi.org/10.1103/PhysRevLett.101.031601}{{\em Phys. Rev.
  Lett.} {\bfseries 101} (2008) 031601},
  \href{http://arxiv.org/abs/0803.3295}{{ arXiv:0803.3295~[hep-th]}}.

\bibitem{Gubser:2008zu}
S.~S. Gubser, {\slshape {Colorful horizons with charge in anti-de Sitter
  space},} \href{http://dx.doi.org/10.1103/PhysRevLett.101.191601}{{\em Phys.
  Rev. Lett.} {\bfseries 101} (2008) 191601},
  \href{http://arxiv.org/abs/0803.3483}{{ arXiv:0803.3483~[hep-th]}}.

\bibitem{Gubser:2008wv}
S.~S. Gubser and S.~S. Pufu, {\slshape {The Gravity dual of a p-wave
  superconductor},} \href{http://dx.doi.org/10.1088/1126-6708/2008/11/033}{{\em
  JHEP} {\bfseries 11} (2008) 033}, \href{http://arxiv.org/abs/0805.2960}{{
  arXiv:0805.2960~[hep-th]}}.

\bibitem{Roberts:2008ns}
M.~M. Roberts and S.~A. Hartnoll, {\slshape {Pseudogap and time reversal
  breaking in a holographic superconductor},}
  \href{http://dx.doi.org/10.1088/1126-6708/2008/08/035}{{\em JHEP} {\bfseries
  08} (2008) 035}, \href{http://arxiv.org/abs/0805.3898}{{
  arXiv:0805.3898~[hep-th]}}.

\bibitem{Ammon:2008fc}
M.~Ammon, J.~Erdmenger, M.~Kaminski, and P.~Kerner, {\slshape
  {Superconductivity from gauge/gravity duality with flavor},}
  \href{http://dx.doi.org/10.1016/j.physletb.2009.09.029}{{\em Phys. Lett. B}
  {\bfseries 680} (2009) 516--520}, \href{http://arxiv.org/abs/0810.2316}{{
  arXiv:0810.2316~[hep-th]}}.

\bibitem{Ammon:2009fe}
M.~Ammon, J.~Erdmenger, M.~Kaminski, and P.~Kerner, {\slshape {Flavor
  Superconductivity from Gauge/Gravity Duality},}
  \href{http://dx.doi.org/10.1088/1126-6708/2009/10/067}{{\em JHEP} {\bfseries
  10} (2009) 067}, \href{http://arxiv.org/abs/0903.1864}{{
  arXiv:0903.1864~[hep-th]}}.

\bibitem{Tseytlin:1997csa}
A.~A. Tseytlin, {\slshape {On nonAbelian generalization of Born-Infeld action
  in string theory},}
  \href{http://dx.doi.org/10.1016/S0550-3213(97)00354-4}{{\em Nucl. Phys. B}
  {\bfseries 501} (1997) 41--52}, \href{http://arxiv.org/abs/hep-th/9701125}{{
  arXiv:hep-th/9701125}}.

\bibitem{Hashimoto:1997gm}
A.~Hashimoto and W.~Taylor, {\slshape {Fluctuation spectra of tilted and
  intersecting D-branes from the Born-Infeld action},}
  \href{http://dx.doi.org/10.1016/S0550-3213(97)00399-4}{{\em Nucl. Phys. B}
  {\bfseries 503} (1997) 193--219},
  \href{http://arxiv.org/abs/hep-th/9703217}{{ arXiv:hep-th/9703217}}.

\bibitem{Boyd_book}
R.~W. Boyd, {\em Nonlinear Optics, Third Edition}.
\newblock Academic Press, London, 3rd~ed., 2008.

\bibitem{Huttner2017}
U.~Huttner, M.~Kira, and S.~W. Koch, {\slshape Ultrahigh off-resonant field
  effects in semiconductors,}
  \href{http://dx.doi.org/https://doi.org/10.1002/lpor.201700049}{{\em Laser \&
  Photonics Reviews} {\bfseries 11} (2017) 1700049}.

\bibitem{Watanabe2020}
H.~Watanabe, Y.~Liu, and M.~Oshikawa, {\slshape On the general properties of
  non-linear optical conductivities,}
  \href{https://doi.org/10.1007/s10955-020-02654-5}{{\em Journal of Statistical
  Physics} {\bfseries 181} (Dec, 2020) 2050--2070}.

\bibitem{Takasan2023}
K.~Takasan, M.~Oshikawa, and H.~Watanabe, {\slshape Drude weights in
  one-dimensional systems with a single defect,}
  \href{https://link.aps.org/doi/10.1103/PhysRevB.107.075141}{{\em Phys. Rev.
  B} {\bfseries 107} (Feb, 2023) 075141}.

\bibitem{Michishita2021}
Y.~Michishita and R.~Peters, {\slshape Effects of renormalization and
  non-hermiticity on nonlinear responses in strongly correlated electron
  systems,} \href{https://link.aps.org/doi/10.1103/PhysRevB.103.195133}{{\em
  Phys. Rev. B} {\bfseries 103} (May, 2021) 195133}.

\bibitem{Nakamura:2010zd}
S.~Nakamura, {\slshape {Negative Differential Resistivity from Holography},}
  \href{http://dx.doi.org/10.1143/PTP.124.1105}{{\em Prog. Theor. Phys.}
  {\bfseries 124} (2010) 1105--1114}, \href{http://arxiv.org/abs/1006.4105}{{
  arXiv:1006.4105~[hep-th]}}.

\bibitem{Ishigaki:2020coe}
S.~Ishigaki and S.~Nakamura, {\slshape {Mechanism for negative differential
  conductivity in holographic conductors},}
  \href{http://dx.doi.org/10.1007/JHEP12(2020)124}{{\em JHEP} {\bfseries 12}
  (2020) 124}, \href{http://arxiv.org/abs/2008.00904}{{
  arXiv:2008.00904~[hep-th]}}.

\bibitem{Nakamura:2012ae}
S.~Nakamura, {\slshape {Nonequilibrium Phase Transitions and Nonequilibrium
  Critical Point from AdS/CFT},}
  \href{http://dx.doi.org/10.1103/PhysRevLett.109.120602}{{\em Phys. Rev.
  Lett.} {\bfseries 109} (2012) 120602},
  \href{http://arxiv.org/abs/1204.1971}{{ arXiv:1204.1971~[hep-th]}}.

\bibitem{Ali-Akbari:2013hba}
M.~Ali-Akbari and A.~Vahedi, {\slshape {Non-equilibrium Phase Transition from
  AdS/CFT},} \href{http://dx.doi.org/10.1016/j.nuclphysb.2013.09.008}{{\em
  Nucl. Phys. B} {\bfseries 877} (2013) 95--106},
  \href{http://arxiv.org/abs/1305.3713}{{ arXiv:1305.3713~[hep-th]}}.

\bibitem{Zeng:2018ero}
H.-B. Zeng and H.-Q. Zhang, {\slshape {Universal critical exponents of
  nonequilibrium phase transitions from holography},}
  \href{http://dx.doi.org/10.1103/PhysRevD.98.106024}{{\em Phys. Rev. D}
  {\bfseries 98} (2018) 106024}, \href{http://arxiv.org/abs/1807.11881}{{
  arXiv:1807.11881~[hep-th]}}.

\bibitem{Vahedi:2018gvn}
A.~Vahedi and M.~Shakeri, {\slshape {Non-Equilibrium Critical Phenomena From
  Probe Brane Holography in Schr\"odinger Spacetime},}
  \href{http://dx.doi.org/10.1007/JHEP01(2019)047}{{\em JHEP} {\bfseries 01}
  (2019) 047}, \href{http://arxiv.org/abs/1811.05823}{{
  arXiv:1811.05823~[hep-th]}}.

\bibitem{Imaizumi:2019byu}
T.~Imaizumi, M.~Matsumoto, and S.~Nakamura, {\slshape {Current Driven
  Tricritical Point in Large- $N_c$ Gauge Theory},}
  \href{http://dx.doi.org/10.1103/PhysRevLett.124.191603}{{\em Phys. Rev.
  Lett.} {\bfseries 124} (2020) 191603},
  \href{http://arxiv.org/abs/1911.06262}{{ arXiv:1911.06262~[hep-th]}}.

\bibitem{Endo:2023vov}
D.~Endo, Y.~Fukazawa, M.~Matsumoto, and S.~Nakamura, {\slshape {Electric-field
  driven nonequilibrium phase transitions in AdS/CFT},}
  \href{http://dx.doi.org/10.1007/JHEP03(2023)173}{{\em JHEP} {\bfseries 03}
  (2023) 173}, \href{http://arxiv.org/abs/2302.13535}{{
  arXiv:2302.13535~[hep-th]}}.

\bibitem{Ghimire2019}
S.~Ghimire and D.~A. Reis, {\slshape High-harmonic generation from solids,}
  \href{https://doi.org/10.1038/s41567-018-0315-5}{{\em Nat. Phys.} {\bfseries
  15} (Jan, 2019) 10--16}.

\bibitem{Dzsaber2021}
S.~Dzsaber, X.~Yan, M.~Taupin, G.~Eguchi, A.~Prokofiev, T.~Shiroka, P.~Blaha,
  O.~Rubel, S.~E. Grefe, H.-H. Lai, Q.~Si, and S.~Paschen, {\slshape Giant
  spontaneous hall effect in a nonmagnetic weyl–kondo semimetal,}
  \href{http://dx.doi.org/10.1073/pnas.2013386118}{{\em Proceedings of the
  National Academy of Sciences} {\bfseries 118} (2021) e2013386118}.

\bibitem{Babington:2003vm}
J.~Babington, J.~Erdmenger, N.~J. Evans, Z.~Guralnik, and I.~Kirsch, {\slshape
  {Chiral symmetry breaking and pions in nonsupersymmetric gauge / gravity
  duals},} \href{http://dx.doi.org/10.1103/PhysRevD.69.066007}{{\em Phys. Rev.
  D} {\bfseries 69} (2004) 066007},
  \href{http://arxiv.org/abs/hep-th/0306018}{{ arXiv:hep-th/0306018}}.

\bibitem{Albash:2007bq}
T.~Albash, V.~G. Filev, C.~V. Johnson, and A.~Kundu, {\slshape {Quarks in an
  external electric field in finite temperature large N gauge theory},}
  \href{http://dx.doi.org/10.1088/1126-6708/2008/08/092}{{\em JHEP} {\bfseries
  08} (2008) 092}, \href{http://arxiv.org/abs/0709.1554}{{
  arXiv:0709.1554~[hep-th]}}.

\bibitem{Christensen:1998hg}
M.~Christensen, V.~P. Frolov, and A.~L. Larsen, {\slshape {Soap bubbles in
  outer space: Interaction of a domain wall with a black hole},}
  \href{http://dx.doi.org/10.1103/PhysRevD.58.085008}{{\em Phys. Rev. D}
  {\bfseries 58} (1998) 085008}, \href{http://arxiv.org/abs/hep-th/9803158}{{
  arXiv:hep-th/9803158}}.

\bibitem{Hashimoto:2014yza}
K.~Hashimoto, S.~Kinoshita, K.~Murata, and T.~Oka, {\slshape {Electric Field
  Quench in AdS/CFT},} \href{http://dx.doi.org/10.1007/JHEP09(2014)126}{{\em
  JHEP} {\bfseries 09} (2014) 126}, \href{http://arxiv.org/abs/1407.0798}{{
  arXiv:1407.0798~[hep-th]}}.

\bibitem{Kinoshita:2016lqd}
S.~Kinoshita, T.~Igata, and K.~Tanabe, {\slshape {Energy extraction from Kerr
  black holes by rigidly rotating strings},}
  \href{http://dx.doi.org/10.1103/PhysRevD.94.124039}{{\em Phys. Rev. D}
  {\bfseries 94} (2016) 124039}, \href{http://arxiv.org/abs/1610.08006}{{
  arXiv:1610.08006~[gr-qc]}}.

\bibitem{Kinoshita:2017mio}
S.~Kinoshita and T.~Igata, {\slshape {The essence of the
  Blandford\textendash{}Znajek process},}
  \href{http://dx.doi.org/10.1093/ptep/pty024}{{\em PTEP} {\bfseries 2018}
  (2018) 033E02}, \href{http://arxiv.org/abs/1710.09152}{{
  arXiv:1710.09152~[gr-qc]}}.

\bibitem{Herzog:2006se}
C.~P. Herzog, {\slshape {Energy Loss of Heavy Quarks from Asymptotically AdS
  Geometries},} \href{http://dx.doi.org/10.1088/1126-6708/2006/09/032}{{\em
  JHEP} {\bfseries 09} (2006) 032},
  \href{http://arxiv.org/abs/hep-th/0605191}{{ arXiv:hep-th/0605191}}.

\bibitem{Caceres:2006as}
E.~Caceres and A.~Guijosa, {\slshape {On Drag Forces and Jet Quenching in
  Strongly Coupled Plasmas},}
  \href{http://dx.doi.org/10.1088/1126-6708/2006/12/068}{{\em JHEP} {\bfseries
  12} (2006) 068}, \href{http://arxiv.org/abs/hep-th/0606134}{{
  arXiv:hep-th/0606134}}.

\bibitem{Seiberg:1999vs}
N.~Seiberg and E.~Witten, {\slshape {String theory and noncommutative
  geometry},} \href{http://dx.doi.org/10.1088/1126-6708/1999/09/032}{{\em JHEP}
  {\bfseries 09} (1999) 032}, \href{http://arxiv.org/abs/hep-th/9908142}{{
  arXiv:hep-th/9908142}}.

\bibitem{Gibbons:2000xe}
G.~W. Gibbons and C.~A.~R. Herdeiro, {\slshape {Born-Infeld theory and stringy
  causality},} \href{http://dx.doi.org/10.1103/PhysRevD.63.064006}{{\em Phys.
  Rev. D} {\bfseries 63} (2001) 064006},
  \href{http://arxiv.org/abs/hep-th/0008052}{{ arXiv:hep-th/0008052}}.

\bibitem{Baggioli:2016oju}
M.~Baggioli and O.~Pujolas, {\slshape {On Effective Holographic Mott
  Insulators},} \href{http://dx.doi.org/10.1007/JHEP12(2016)107}{{\em JHEP}
  {\bfseries 12} (2016) 107}, \href{http://arxiv.org/abs/1604.08915}{{
  arXiv:1604.08915~[hep-th]}}.

\bibitem{Izumi:2014loa}
K.~Izumi, {\slshape {Causal Structures in Gauss-Bonnet gravity},}
  \href{http://dx.doi.org/10.1103/PhysRevD.90.044037}{{\em Phys. Rev. D}
  {\bfseries 90} (2014) 044037}, \href{http://arxiv.org/abs/1406.0677}{{
  arXiv:1406.0677~[gr-qc]}}.

\bibitem{Ishigaki:2021vyv}
S.~Ishigaki, S.~Kinoshita, and M.~Matsumoto, {\slshape {Dynamical stability and
  filamentary instability in holographic conductors},}
  \href{http://dx.doi.org/10.1007/JHEP04(2022)173}{{\em JHEP} {\bfseries 04}
  (2022) 173}, \href{http://arxiv.org/abs/2112.11677}{{
  arXiv:2112.11677~[hep-th]}}.

\bibitem{deMelo:2014isa}
C.~A.~M. de~Melo, L.~G. Medeiros, and P.~J. Pompeia, {\slshape {Causal
  Structure and Birefringence in Nonlinear Electrodynamics},}
  \href{http://dx.doi.org/10.1142/S021773231550025X}{{\em Mod. Phys. Lett. A}
  {\bfseries 30} (2015) 1550025}, \href{http://arxiv.org/abs/1407.0567}{{
  arXiv:1407.0567~[hep-th]}}.

\end{thebibliography}\endgroup
\bibliographystyle{ytphys}
\end{document}